\documentstyle[12pt,epsfig]{article}

\def\shiftdown#1{#1\llap{\lower.04ex\hbox{#1}}}

\begin{document}




\vspace*{0.3cm}

\begin{center}
{\bf {\large Field-theoretical three-body relativistic equations for the
multichannel $\pi N\leftrightarrow \gamma N\leftrightarrow \pi \pi
N\leftrightarrow \gamma \pi N$ reactions \footnotemark }}

\footnotetext{
Supported by DFG 436 GEO.}

\vspace{0.7cm}
\end{center}

\vspace{0.7cm}

\begin{center}
{\large {\it A.\ I.\ Machavariani$^{\dag\diamond \ast}$\ and\ Amand\
Faessler $^{\dag}$}}

\vspace{0.6cm} {\em $^{\dag }$ Institut f\"{u}r Theoretische Physik,
Universit\"{a}t T\"{u}bingen, Auf der Morgenstelle 14, D-72076 T\"{u}bingen,
Germany,}\\[0pt]
{\em $^{\diamond }$ Joint\ Institute\ for\ Nuclear\ Research,\ Moscow
Region \\
141980 Dubna,\ Russia}\\[0pt]
{\em $^{*}$ High Energy Physics Institute of Tbilisi State University,
University Str. 9 }\\[0pt]
{\em 380086 Tbilisi, Georgia }
\end{center}

\vspace{0.5cm} \medskip

\begin{abstract}
A new kind of the relativistic three-body equations for the coupled $\pi N$
and $\gamma N$ scattering reactions with the $\pi \pi N$ and $\gamma \pi N$
three particle final states are suggested. These equations are derived in
the framework of the standard field-theoretical $S$-matrix approach in the
time-ordered three dimensional form. Therefore corresponding relativistic
covariant equations are three-dimensional from the beginning and  the
considered formulation is free of the ambiguities which appear due to a
three dimensional reduction of the four dimensional Bethe-Salpeter
equations. The solutions of the considered equations satisfy the unitarity
condition and are exactly gauge invariant even after the truncation of the
of the multiparticle ($n>3$) intermediate states. Moreover the form of these
three-body equations does not depend on the choice of the model Lagrangian
and it is the same for the formulations with and without quark degrees of
freedom. The effective potential of the suggested equations is defined by
the vertex functions with two on-mass shell particles. It is emphasized that
these INPUT vertex functions can be constructed from experimental data.

Special attention is given to the construction of the intermediate on shell
and off shell $\Delta$ resonance states. These intermediate $\Delta$ states
are obtained after separation of the $\Delta$ resonance pole contributions
in the intermediate $\pi N$ Green function. The resulting amplitudes for the 
$\Delta\Longleftrightarrow N\pi$; $\Delta\Longleftrightarrow N\gamma;$ $%
\Delta^{\prime}\Longleftrightarrow \Delta\gamma$ transition have the same
structure as the vertex functions for transitions between the on mass shell
particle states with spin $1/2$ and $3/2$. Therefore it is possible to
introduce the real value for the magnetic momenta for the $%
\Delta^{\prime}\Longleftrightarrow \Delta\gamma$ transition amplitudes  in
the same way as it is done for the $N^{\prime}\Longleftrightarrow N\gamma$
vertex function.

\end{abstract}

\newpage

\begin{center}
{\bf 1. INTRODUCTION}
\end{center}

\medskip

The problem of the relativistic description of an particle interactions in
the framework of a potential picture is usually solved by relativistic
generalization of the Lippmann-Schwinger type equation of the
nonrelativistic collision theory \cite{New,Gold}. In quantum field theory
potentials of the relativistic Lippmann-Schwinger type equation are
constructed from more simple vertex functions which can be determined by a
Lagrangian. Depending on the initial general relations in quantum field
theory, one can distinguish three essentially different derivations of the
field-theoretical generalizations of the Lippmann-Schwinger type equations
which have different off mass shell behavior of the amplitudes,the
potentials, the vertex functions and the propagators. The most popular
source of the derivation of the such type field-theoretical equations is the
Bethe-Salpeter equation \cite{IZ,Brown,GrossB,Klein,Briodo}. The effective
potential of these four-dimensional equations consists of the sum of the
Feynman diagrams. The well known quasipotential method was applied \cite
{LT,BlanS,Brown,GrossB} in order to obtain the equivalent three-dimensional
equations. The different three-dimensional representations of the
Bethe-Salpeter equations, derived in the framework of the different
quasipotential methods, have different free Green functions and diverse
quasipotentials $W$ which are constructed from the Bethe-Salpeter potential $%
V$. For practical calculations the quasipotential is usually taken in the
Born approximation $W\approx V$. Therefore the results of these
approximations depend on the form of the three-dimensional reduction. In
addition for construction of $V$ or $W$ three-variable vertex functions are
required as the ``input''  functions\footnotemark . Therefore, in the
practical calculations based on the Bethe-Salpeter equations or their
quasipotential reductions the off-mass shell variables in the vertex
functions are usually neglected or a separable form for all three variables
is introduced \cite{Pearce,Lahiff}.

\footnotetext{
The only exception is the quasipotential equation derived by F. Gross\cite
{Gross}, where the input vertex functions are the two-variable vertices,
because one particle in these vertices is on mass shell.}

Another derivation of the relativistic Lippmann-Schwinger equation is based
on the field-theoretical generalization of the Schr\"odinger equation in the
form of the functional Tomonaga-Schwinger equations \cite{Bog,Schweber}. In
the framework of this method a covariant Hamiltonian theory for construction
of the relativistic three-dimensional equations was given in Refs. \cite
{Kadysh1,Kadysh2}. In this approach the covariant equations are
three-dimensional from the beginning, and therefore they are free from the
ambiguities of the three-dimensional reduction. The potential in these
equations is constructed from the vertex functions with all particles on
mass shell, i.e., the related ``input'' vertex functions are used dependent
also on three variables. Apart from this problem in these equations one has
the non-physical $``spurious^{\prime\prime}$ degrees of freedom. Light-front
reformulation of these equations was done in Refs. \cite{Karmanov1,Karmanov2}%
. The generalization of the Faddeev equations in this formulation was done
in Ref.\cite{Vinog}.

The third way of the derivation of the three-dimensional Lippmann-Schwinger
type integral equations in the quantum field theory proceeds from the
field-theoretical generalizations of the off-shell unitarity conditions.
These conditions can be treated as a field-theoretical spectral
decomposition of the scattering amplitude over the asymptotic $^{\prime
\prime }in(out)^{\prime \prime }$ states. This formulation, suggested first
by F. Low \cite{Low} in the framework of the old perturbation theory, was
developed afterwards in Refs. \cite{Schweber,Chew,Ban,Ther,MR,M1,MCH,MFB}.
In particular, in Refs.\cite{Schweber,Chew,Ban,Ther} these
three-dimensional, time-ordered and quadratically nonlinear equations were
used for the evaluation of the $\pi N$ scattering amplitudes. In the Refs.%
\cite{MR,MCH,M1} the explicit linearization procedures of these equations
were suggested. The linearized equations have the form of relativistic
Lippmann-Schwinger type equations and they were employed for the calculation
of the low energy $\pi N$ and $NN$ scattering reactions. The corresponding
equations are also three-dimensional and time ordered from the beginning and
their potentials are constructed from completely dressed matrix elements
with two on mass shell particles. Using the on-mass shell methods, such as
the dispersion relations, sum rules, current algebra etc. one can determine
the required ``input'' vertices with two on mass shell particles. The
effective potential of the considered equations contains also the equal-time
commutators of the two external interacting field operators. Therefore in
this approach it is necessary to use some model Lagrangian in order to
determine these equal-time commutators. The resulting operators calculated
by the equal-time commutators are sandwiched between the real asymptotic
on-mass shell states, i.e. these equal-time commutators are also determined
by the one-variable vertex functions which can be considered as ``input''
vertices.

The equal-time term of a nonrenormalizable Lagrangian produce a number of
contact terms \cite{M1}. But first of all our aim is to determine the
minimal number of the terms from equal-time commutators (in the Chew-Low
model these terms were omitted!) which are necessary for the description of
the multichannel $\pi N$ and $\gamma N$ scattering reactions. These simplest
Lagrangians can be improved to become closer to the more complete Lagrangian
by inclusion of more complicated symmetries and by adding more terms,
capable to describe different mechanisms of the interactions. In our
previous works \cite{MCH,M1,MFB} about the $NN$ scattering we have estimated
the contributions of the additional contact terms which arise after using
the $\pi N$ Lagrangian with the nonrenormalizable pseudo-vector coupling. In
papers \cite{M1,MBFE} we have considered the structure of the present
field-theoretical approach with quark degrees of freedom. We have shown,
that the structure of the present relativistic three-dimensional equations
and their potentials with and without quark degrees of freedom is the same.
Moreover for each Lagrangian (even in the formulation with quark degrees of
freedom) the terms produced by the equal-time commutators are Hermitian and
do not contribute in the unitarity condition.

Presently, the interest to investigate reactions with three-body final $%
\gamma \pi N$ states is stimulated by the proposal to determine the magnetic
moment of the $\Delta^{+}$ resonance in the $\gamma p\to \gamma^{\prime}{%
\pi^o}^{\prime}p^{\prime}$ reaction. The basic idea of this investigation 
\cite{Kond} is to separate the contribution of the $\Delta\to\gamma^{\prime}%
\Delta^{\prime}$ vertex function which in analogy to the $%
N-\gamma^{\prime}N^{\prime}$ vertex, contains the magnitude of the $\Delta$
magnetic moment at threshold. The contribution of the $\Delta^+\to\gamma^{%
\prime}{\Delta^+}^{\prime}$ vertex function in the $\gamma p\to
\gamma^{\prime}{\pi^o}^{\prime}p^{\prime}$ reaction was numerically
estimated in Refs.\cite{M3,Dr1,Dr2,M5} in order to study the dependence of
the observables on the value of the $\Delta^+$ magnetic moment. First data
about the $\gamma p\to \gamma^{\prime}{\pi^o}^{\prime}p^{\prime}$ reaction
were obtained in a recent experiment by the A2/TAPS collaboration at MAMI 
\cite{Kotull} and future experimental investigations of this reaction are
planned by using the Crystal Ball detector at MAMI \cite{BeckN}.

This paper is devoted to the three-body generalization of the two-body
field-theoretical equations \cite{Schweber,Low,Chew,Ban,Ther,M1,MBFE} for
the case of the multichannel $\pi N-\gamma N-\pi\pi N-\gamma\pi N$ 
reactions. The potential of these equations has minimal off shellness (i.e.
only two of external particles are being off mass shell) and they are
analytically connected with any other field-theoretical equations, based on
the Bethe-Salpeter or the Tomonaga-Schwinger equations. The form of these
equations do not depend on the choice of the Lagrangian. and they  do not
change their form even for the formulations with quark-gluon degrees of
freedom. Therefore this formulation can help to clarify the difference
between  a number of theoretical models which describe with quite a good
accuracy the experimental data of the two-body $\pi N$ and $\gamma N$
reactions in the framework of the different approximations without of the
reproduction of the three-body $\gamma \pi N$ and $\pi\pi N$ data. From this
point of view the unified description of the multichannel $\pi N-\gamma
N-\pi \pi N-\gamma \pi N$ reactions up to threshold of the creation of the
third pi-meson can clarify the dynamic of the two-body $\gamma N$, $\pi N$, $%
\gamma \pi$ and $\pi\pi$ interactions in the low and intermediate energy
region. In the recent investigations \cite{Lahiff,M3} the importance of the
choice of the form of the intermediate $\Delta$ propagator was demonstrated
by the description of the $\pi N$ and multichannel $\gamma N$ scattering
reactions. Therefore the unified description of the $\pi N-\gamma N-\pi \pi
N-\gamma \pi N$ reactions can be employed also for the determination of the
properties of the intermediate $\Delta$ resonance propagation.

The organization of this papers is as follows. In Sec.2 we state the
three-body spectral decomposition equations (which have the form of the off
shell unitarity conditions \cite{New}) for the amplitudes of the coupled $%
\pi N-\gamma N-\pi\pi N-\gamma\pi N$ channels. These quadratically nonlinear
three-dimensional equations are derived after extraction of the two external
particles from the asymptotic $in$ or $out$ states, performed in the
framework the standard $S$-matrix reduction formulas \cite
{IZ,Bog,Schweber,BD}. In this section we consider also the connected and
disconnected parts of the three-body amplitudes and the construction of the
equal-time commutators. In Sec. 3 and in Appendix A  the equivalence of the
above quadratically nonlinear equations and the Lippmann-Schwinger type
equations is outlined. The analytical expressions of the three-body
potential of the presented Lippmann-Schwinger type equations is given
Appendix B. The procedure of extraction of the intermediate on shell and off
shell $\Delta$-isobar degrees of freedom from the intermediate $\pi N$
states in the suggested three-body equation is developed in Sec. 4. In Sec.
5 we apply this procedure to the derivation of the three-body equations for
the coupled $\Delta-\pi\Delta-\gamma N-\gamma\Delta$ transition amplitudes.
The unitarity and gauge invariance for the derived three-body equations are
demonstrated in Sec. 6. Finally in Sec.7 we give some concluding remarks.

\begin{center}
{\bf 2.  Spectral decomposition method of the multichannel $\pi N$ and $%
\gamma N$ scattering amplitudes over  the complete set of the asymptotic $%
^{\prime\prime}in^{\prime\prime}$  or $^{\prime\prime}out^{\prime\prime}$
states.}
\end{center}

\medskip

Consider the $S$-matrix element $S_{\alpha,\beta}$ and the scattering
amplitude $f_{\alpha,\beta}$ for the $\alpha;\beta=1,2,3,4\equiv\pi N,\gamma
N,\pi\pi N,\gamma\pi N$ states

$$
S_{\alpha,\beta}=<out;\alpha|\beta;in>=
<in;\alpha|\beta;in>-(2\pi)^4i\delta^{(4)}(P_\alpha-P_\beta) f_{\alpha,\beta}%
\eqno(2.1)
$$

where $P_{\alpha}\equiv (P_{\alpha}^o,{\bf P}_{\alpha})$ stands for the
complete four-momentum of the asymptotic state $\alpha$ and

\[
f_{\alpha\beta}=-<out;{\widetilde \alpha}|j_{a}(0)|\beta; in>
\]
$$
=<out;{\widetilde \alpha}|\Bigl[j_{a}(0),a_{b}^+(0)\Bigr] |{\widetilde \beta}%
;in> + i\int d^4 x e^{-ip_b x}<out;{\widetilde\alpha} |T%
\Bigl(j_{a}(0)j_{b}(x)\Bigr) |{\widetilde \beta};in>,\eqno(2.2)
$$

$a=\pi^{\prime},\gamma^{\prime}; \ b=\pi,\gamma$ denotes the one particle
pion or photon states  extracted from the $\alpha\ \ \
^{\prime\prime}out^{\prime\prime}$ and $\beta\ \ \
^{\prime\prime}in^{\prime\prime}$ states correspondingly 
$$
\alpha={\widetilde \alpha}+a;\ \ \ \beta= {\widetilde \beta}+b.\eqno(2.3)
$$
The four-momentum of the asymptotic one-particle states $a\ or\ b$ is $%
p_{\pi}=\Bigl(\sqrt{m_{\pi}^2+{\bf p}_{\pi}^2},{\bf p}_{\pi}\Bigr) \equiv
\Bigl( \omega_{\pi}({\bf p}_{\pi}),{\bf p}_{\pi}\Bigr)$ for pion with mass $%
m_{\pi}$ or $p_{\gamma}=\Bigl(|{\bf k}_{\gamma}|,{\bf k}_{\gamma}\Bigr)
\equiv \Bigl( \omega_{\gamma}({\bf p}_{\gamma}),{\bf p}_{\gamma}\Bigr)$ for
photon with the observed (physical) mass $m_{\gamma}=0$.

The expression (2.2) is defined through the meson and photon current
operators

$$
j_{\pi}(x)=\Bigr(\Box_x+m_{\pi}^2\Bigl)\Phi_{\pi}(x);\ \ \ J_{\eta{\bf k}%
}(x)= \epsilon^{\mu}_{\eta}({\bf k}) \Box_xA_{\mu}(x),\eqno(2.4a)
$$
where $\epsilon^{\eta}_{\mu}({\bf k})= \Bigl(\epsilon^{0}_{\mu}({\bf k}),{%
\stackrel{\longrightarrow}{\epsilon}}_{\mu}({\bf k})\Bigr)$ denotes the
polarization four-vector of photon with the four momentum $k=(|{\bf k}|,{\bf %
k}$). The current operators in Eq.(2.4a) is defined through  the $\pi$ meson 
${\Phi}_{\pi}(x)$ and photon $A_{\mu}(x)$ Heisenberg field operators as

$$
a_{{\bf p}_{\pi}}^+(x_0)=-i \int d^3x e^{-ip_{\pi}x} {\ \stackrel{%
\longleftrightarrow}{\partial}}_{x_o} {\Phi}_{\pi}(x),\eqno(2.4b)
$$
$$
a_{{\bf k}\eta}^+(x_0)=-i\epsilon^{\mu}_{\eta}({\bf k}) \int d^3x e^{-ikx} {%
\ \stackrel{\longleftrightarrow}{\partial}}_{x_o} A_{\mu}(x).\eqno(2.4c)
$$
These operators transforms into meson and photon creation or annihilation
operators in the asymptotic regions $x_o\to\pm\infty$. Here and afterwards
we use the definitions and normalization conditions from the Itzykson and
Zuber's book \cite{IZ}.

The $S$-matrix element $S_{\alpha,\beta}$ (2.1) and scattering amplitude $%
f_{\alpha,\beta}$ (2.2) consists of the connected $S^c_{\alpha,\beta}$,\ $%
f^c_{\alpha,\beta}$ and disconnected $S^d_{\alpha,\beta}$, $%
f^d_{\alpha,\beta}$ parts

$$
S_{\alpha,\beta}=S^d_{\alpha,\beta}+S^c_{\alpha,\beta};\ \ \ \ \ \ \ \ \ \ \
\ \ \ f_{\alpha,\beta}=f^d_{\alpha,\beta}+f^c_{\alpha,\beta}; \eqno(2.5)
$$
where for the two-body and for the three-body channels we have

$$
S^d_{a+N^{\prime},b+N}=<in;a,N^{\prime}|b,N;in>\eqno(2.6a)
$$
\[
S^d_{a+\pi^{\prime}+N^{\prime},b+N}=
-(2\pi)^4i\delta^{(4)}(P_{a+\pi^{\prime}+N^{\prime}}-P_{b+N})f^d_{a+\pi^{%
\prime}+N^{\prime},b+N}
\]
$$
=-(2\pi)^4i\delta^{(4)}(P_{a+\pi^{\prime}+N^{\prime}}-P_{b+N}) \Biggl[ %
<in;N^{\prime}|N;in>f^{1}_{a+\pi^{\prime},b}
+<in;\pi^{\prime}|b;in>f^{2}_{a+N^{\prime},N}\Biggr]
\eqno(2.6b)
$$

\[
S^d_{a+\pi^{\prime}+N^{\prime},b+\pi+N}=
-(2\pi)^4i\delta^{(4)}(P_{a+\pi^{\prime}+N^{\prime}}-P_{b+\pi+N})
f^d_{a+\pi^{\prime}+N^{\prime},b+\pi+N}=
\]
\[
<in;a,\pi^{\prime},N^{\prime}|b,\pi,N;in>
-(2\pi)^4i\delta^{(4)}(P_{a+\pi^{\prime}+N^{\prime}}-P_{b+\pi+N})
\]
$$
\Biggl[
<in;N^{\prime}|N;in>f^{1}_{a+\pi^{\prime},b+\pi}
+<in;\pi^{\prime}|b;in>f^{2}_{ a+N^{\prime},\pi+N}
+<in;\pi^{\prime}|\pi;in>f^{3}_{ a+N^{\prime}, b+N} \Biggr]\eqno(2.6c)
$$

\footnotetext{
This condition is one of the axioms in the axiomatic quantum field theory.
It can be argued by the following chain of transformations $|{\bf p}_A;
in>=\sum_{n^{\prime}} |n^{\prime}; out><out; n^{\prime}|{\bf p}_A; in>=
\sum_{{\bf p^{\prime}}_A}|{\bf p^{\prime}}_A; out><out;{\bf p}_A|{\bf %
p^{\prime}}_A; in> =|{\bf p}_A; out>$, where we have taken into account that
the $S$ matrix of the $1\to n^{\prime}$ transition does not disappear only
for the transition $1\to 1^{\prime}$. Therefore $<out;{\bf p^{\prime}}_A|%
{\bf p}_A;in>=<in;{\bf p^{\prime}}_A|{\bf p}_A;in>$.}



\begin{figure}[htb]
\centerline{\epsfysize=145mm\epsfbox{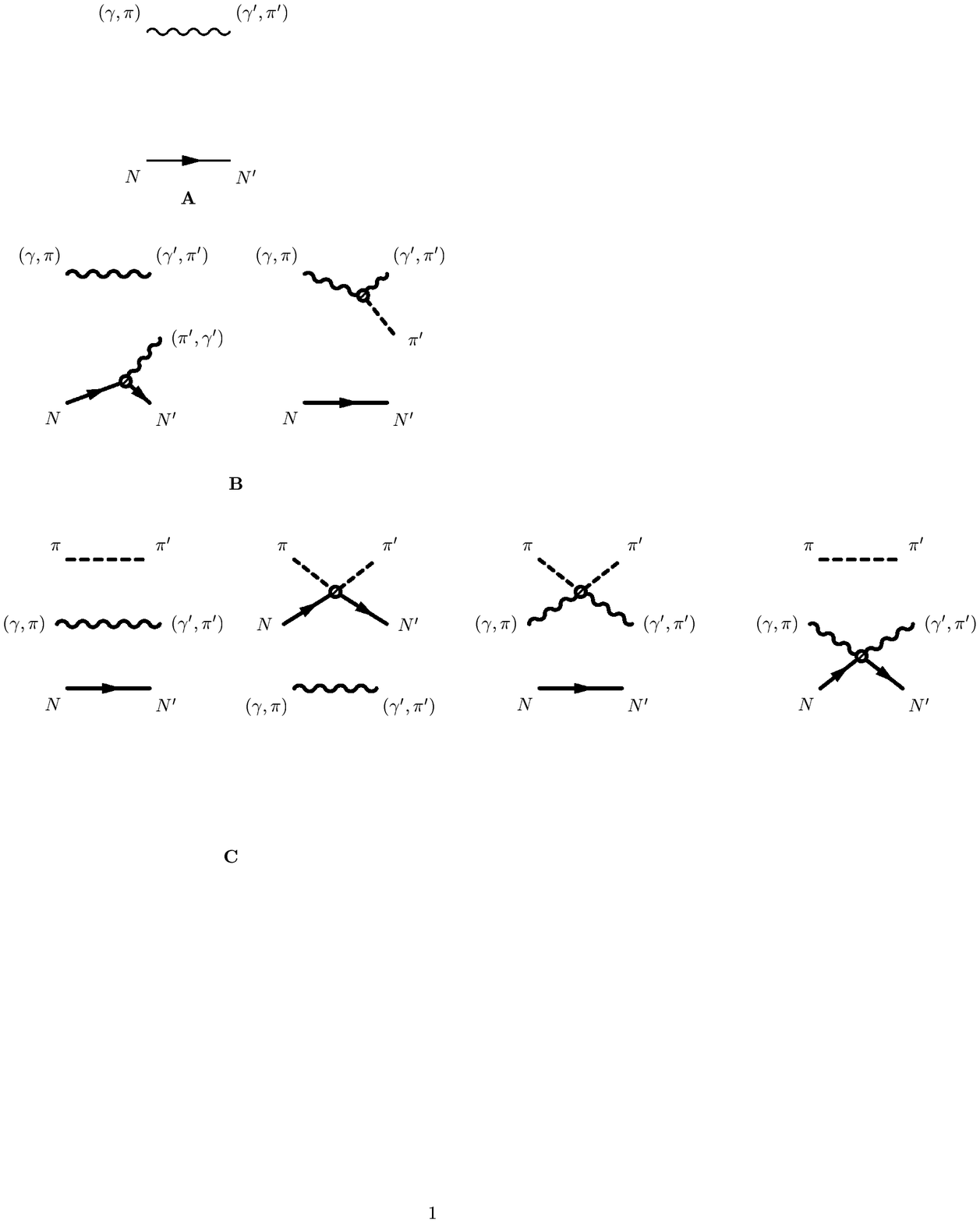}} 
\vspace{-4.0cm} 
\caption{{\protect\footnotesize {\it The disconnected parts of the two-body $%
S$-matrix elements (2.6a) and three-body $S$-matrix (2.6b), (2.6c)
correspondingly in Fig.1A and Fig.1B, Fig.1C. The curled line relates to the 
$a=\gamma^{\prime}\ or\ \pi^{\prime}$ and $b=\gamma\ or\ \pi$ asymptotic
states, the dashed line describes pion and the solid line stands for the
nucleon. The shaded circle corresponds to the vertex function. }}}
\label{fig:one}
\end{figure}
\vspace{5mm}

where $f^i$ stands for the following connected amplitudes

$$
f^{1}_{a+\pi^{\prime},b}=-<out;\pi^{\prime}|j_{a}(0)|b; in>;\ \ \ \ \ \ \ \
\ \ \ f^{2}_{a+N^{\prime},N}=-<out;N^{\prime}|j_{a}(0)|N; in> \eqno(2.7a)
$$

\[
f^{1}_{a+\pi^{\prime},b+\pi}= -<out;\pi^{\prime}|j_{a}(0)|b,\pi; in>;\ \ \ \
\  \ \ \ \ \ \ \ f^{2}_{a+N^{\prime},b+N}=-<out;N^{\prime}|j_{a}(0)|b,N; in>
\]
$$
f^{3}_{a+N^{\prime},\pi+N}=-<out;N^{\prime}|j_{a}(0)|\pi,N; in>. \eqno(2.7b)
$$

The graphical representation of Eq. (2.6a,b,c) is given in Fig. 1. In the
derivation of Eq. (2.6a,b,c) and Eq. (2.7a,b,c) we have used  the one
particle stability condition $|A; in>\equiv|{\bf p}_A; in>=|{\bf p}_A; out>$ 
\footnotemark and the condition $<0|j_{a}(0)|a; in>=0$. It must be noted,
that $S^d_{a+\pi^{\prime}+N^{\prime},b+N}=0$  due to the energy-momentum
conservation rule and the asymptotic particle stability condition.

We shall find now the equations for the connected part of the scattering
amplitude $f^c_{\alpha,\beta}$ (2.5). For this aim we insert the complete
set of the asymptotic $^{\prime\prime}in^{\prime\prime}$ states $\sum_n
|n;in><in;n|={\widehat 1}$ between the current operators in expression (2.2)
and after integration over $x$ we get

$$
f_{\alpha\beta}= W_{\alpha\beta}+(2\pi)^3\sum_{\gamma=1}^4 f_{\alpha\gamma} {%
\frac{{\delta^{(3)}( {\bf p}_b+{\bf P_{{\widetilde \beta}}-P_{\gamma} })}}{{%
\omega_b({\bf p_b})+{\ P_{{\widetilde \beta}}^o-P_{\gamma}^o+i\epsilon }} }} 
{F^{\ast}}_{\beta \gamma} \eqno(2.8a)
$$

where $p^o_b=\omega_b({\bf p_b})$ is the energy of the incoming $\pi$ meson
or photon, $W_{\alpha\beta}$ contains all contributions of the intermediate
states of the $\beta\to \alpha$ reaction except the $s$-channel $\gamma=\pi
N,\gamma N,\pi \pi N, \gamma\pi N$ exchange diagrams which are included  in
the second term of the equation (2.8a).

\[
W_{\alpha\beta}= -<out;{\widetilde \alpha}|\Bigl[j_{a}(0),a_{b}^+(0)\Bigr]|{%
\widetilde \beta};in>
\]
$$
+(2\pi)^3\sum_{n=N,3\pi N,...} <out;{\widetilde \alpha}|j_{a}(0)|n;in> {%
\frac{{\delta^{(3)}( {\bf p}_b+{\bf P_{{\widetilde \beta}}-P_{n} })}}{{%
\omega_b({\bf p_b})+{\ P_{{\widetilde \beta}}^o-P_{n}^o+i\epsilon }} }}
<in;n|j_{b}(0)|{\widetilde \beta};in>\eqno(2.9a)
$$
\[
+(2\pi)^3\sum_{n=N,\pi N,2\pi N,3\pi N,...} <out;{\widetilde \alpha}%
|j_{b}(0)|n;in> {\frac{{\delta^{(3)}( {-{\bf p}}_b+{\bf P_{{\widetilde \alpha%
}}-P_{n} })}}{{-\omega_b({\bf p_b})+ P_{{\widetilde \alpha}}^o-P_{n}^o } }}
<in;n|j_{a}(0)|{\widetilde \beta};in>.
\]

This term consist also from the disconnected and connected parts

$$
W_{\alpha\beta}=W^c_{\alpha\beta}+W^d_{\alpha\beta}\eqno(2.9b)
$$

The consistent procedure of extraction of the complete set of a connected
terms for the three-dimensional expressions like (2.9a) is well known as
cluster decomposition \cite{alf,Ban}. In Appendix B this procedure is
applied to the three-body potential (2.9a), where $<out;{\widetilde \alpha}|$
are replaced by $<in;{\widetilde \alpha}|$. The explicit formula for the
two-body $b+N\to a+N^{\prime}$ potential are given in Appendix B by
Eq.(B.4a)-(B.4h) which are depicted in Fig.2A-Fig.2H respectively. These
diagrams have different chronological sequences of the absorption of the
initial emission of the final particles. In particular, the $s$-channel
diagram 2A corresponds to the chain of reactions, where firstly the initial
nucleon and $b=\pi\ or\ \gamma$ transforms into intermediate $%
N^{\prime\prime},3\pi^{\prime\prime}N^{\prime\prime},...$ states which
afterwards produces the final nucleon and $a=\pi^{\prime}\ or\
\gamma^{\prime}$-particle states. On the diagram 2B at first the final $%
N^{\prime}$  and the intermediate states $2\pi^{\prime\prime},3\pi^{\prime%
\prime},..$ are generated from the initial $bN$ states and next we obtain
final $a$ particle from the intermediate $2\pi^{\prime\prime},3\pi^{\prime%
\prime},..$ states. Unlike the diagram 2B, on the diagram 2C the
intermediate $2\pi^{\prime\prime},3\pi^{\prime\prime},..$ states arise from
the initial pion and afterwards these intermediate states generates the
final $aN^{\prime\prime}$ state together with the initial nucleon. In Fig.
2D we have first the creation of the final nucleon $N^{\prime}$ with the
following absorption of the initial nucleon $N$. Therefore, in this, so
called $Z$ diagram \cite{alf}, the antinucleon intermediate states are
appearing. The combination of this $Z$ diagrams with the corresponding
nucleon exchange diagrams produces the one nucleon exchange Feynman diagram.
Thus the numbers of the particle and antiparticle exchange diagrams in the
time-ordered formulations coincides.


\vspace{5mm}


\begin{figure}[htb]
\centerline{\epsfysize=145mm\epsfbox{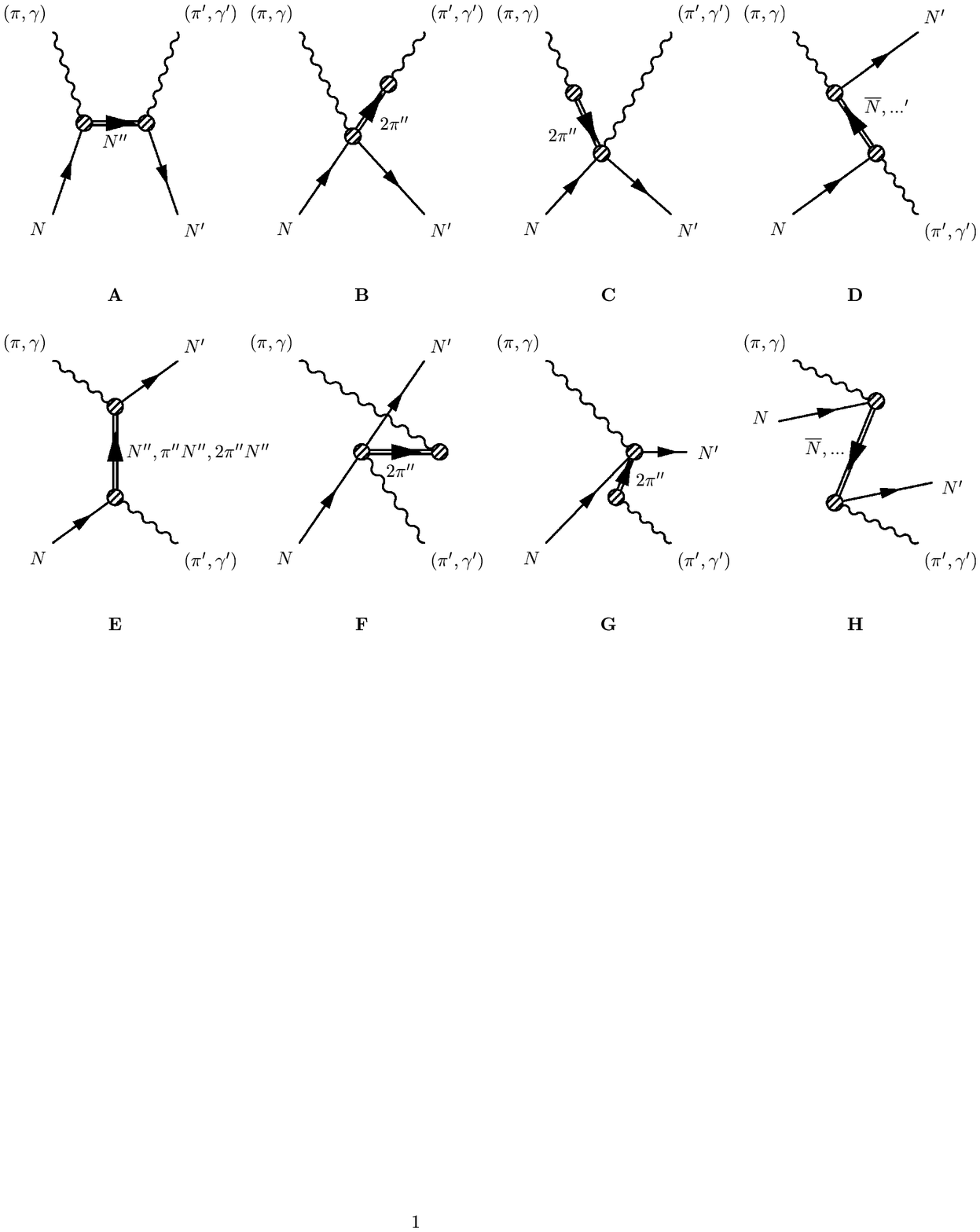}} 
\vspace{-5.0cm} 
\caption{{\protect\footnotesize {\it The simplest on-mass shell $N,{%
\overline N},\pi N, 2\pi, 3\pi,...$ exchange diagrams which are taken into
account in the second and the third term of effective potential (2.9a) for
the binary reactions $b N\to a N^{\prime}$. The intermediate $2\pi$ and $3\pi
$ states can be replaced with the effective $\sigma,\rho,\omega,...$ heavy
meson states. All considered diagrams have the three-dimensional
time-ordered form with the $^{\prime\prime}dressed^{\prime\prime}$
renormalizable vertices. Therefore they differ from the Feynman diagrams. }}}
\label{fig:two}
\end{figure}
\vspace{5mm}

The diagrams in Fig. 3 describe the effective potentials $W_{\alpha \beta
}^{c}$ (B.8a)-(B.h) for the reactions with the three-body final states $%
a+\pi ^{\prime }+N^{\prime }$. The remaining $8$ diagram describing $b+N\to
a+\pi ^{\prime }+N^{\prime }$ reaction can be obtained from the diagrams in
Fig. 3A - Fig.3H by transposition of the final $\pi ^{\prime }$ to the other
vertex function.



\begin{figure}[htb]
\centerline{\epsfysize=145mm\epsfbox{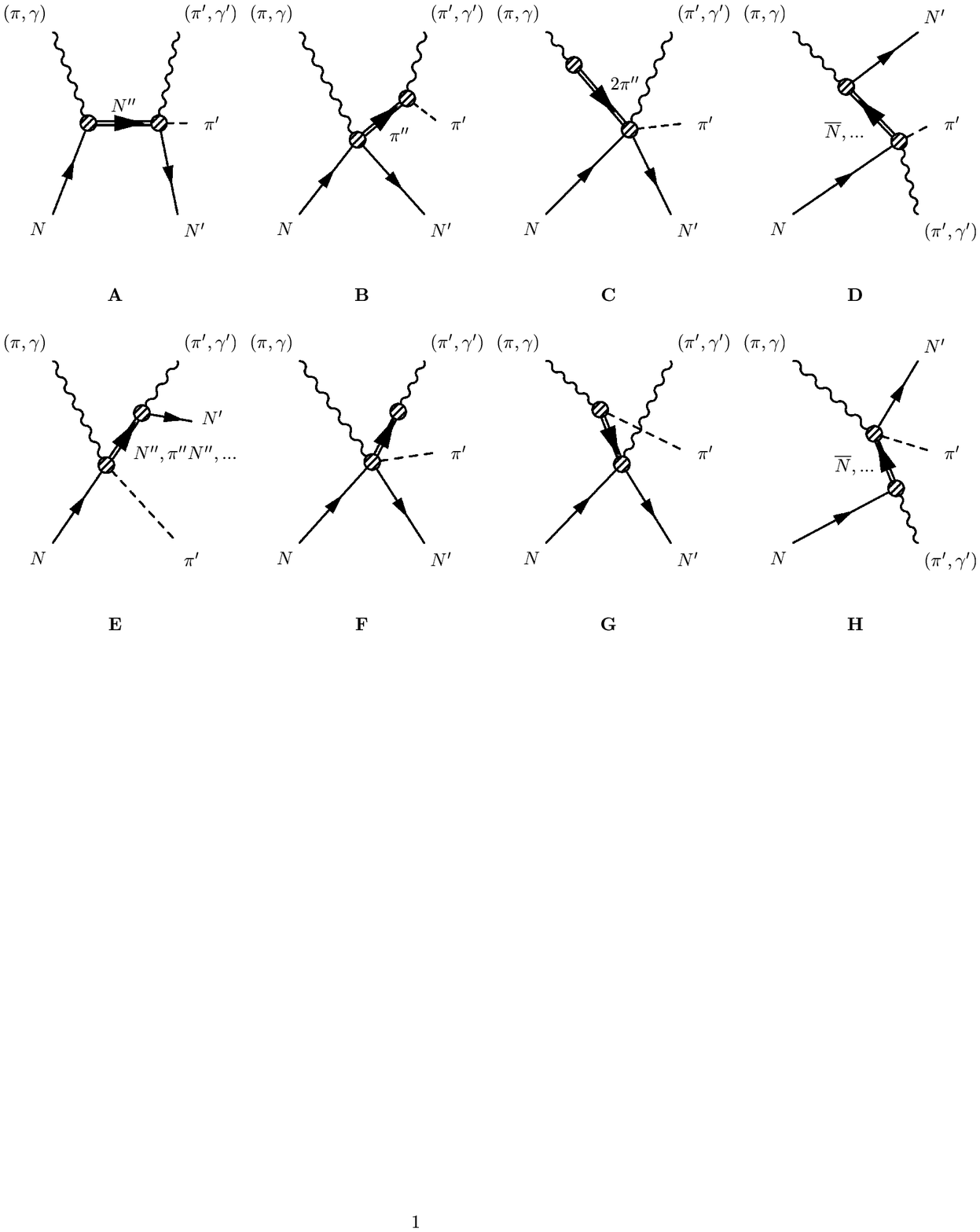}} 
\vspace{-5.0cm} 
\caption{{\protect\footnotesize {\it Same as in Fig.2, but for the $%
b+N\Longrightarrow a+\pi^{\prime}+N^{\prime}$ reaction. The next $8$
diagrams have the same form, but with $\pi^{\prime}$ emission from the first
vertex function. In the $s$-channel diagram ${\bf A}$ the $\pi N,\pi \pi
N,\gamma N,\gamma \pi N$ intermediate states are excluded, because they are
taken into account in the second, driving term of Eq.(2.8c) or Eq. (2.13). }}
}
\label{fig:three}
\end{figure}
\vspace{5mm}


The pure three-body potential $w^c_{\alpha\beta}$ of reaction $b+\pi+N\to
a+\pi^{\prime}+N^{\prime}$ is determined by Eq.(B.12a)-(B.12h) and is
depicted in  Fig.4A - Fig.4H. The other 40 time-ordered diagrams for this
reaction can be reproduced after crossing of the initial $\pi,\gamma$ and $%
\pi^{\prime},\gamma^{\prime}$-mesons.



\begin{figure}[htb]
\centerline{\epsfysize=145mm\epsfbox{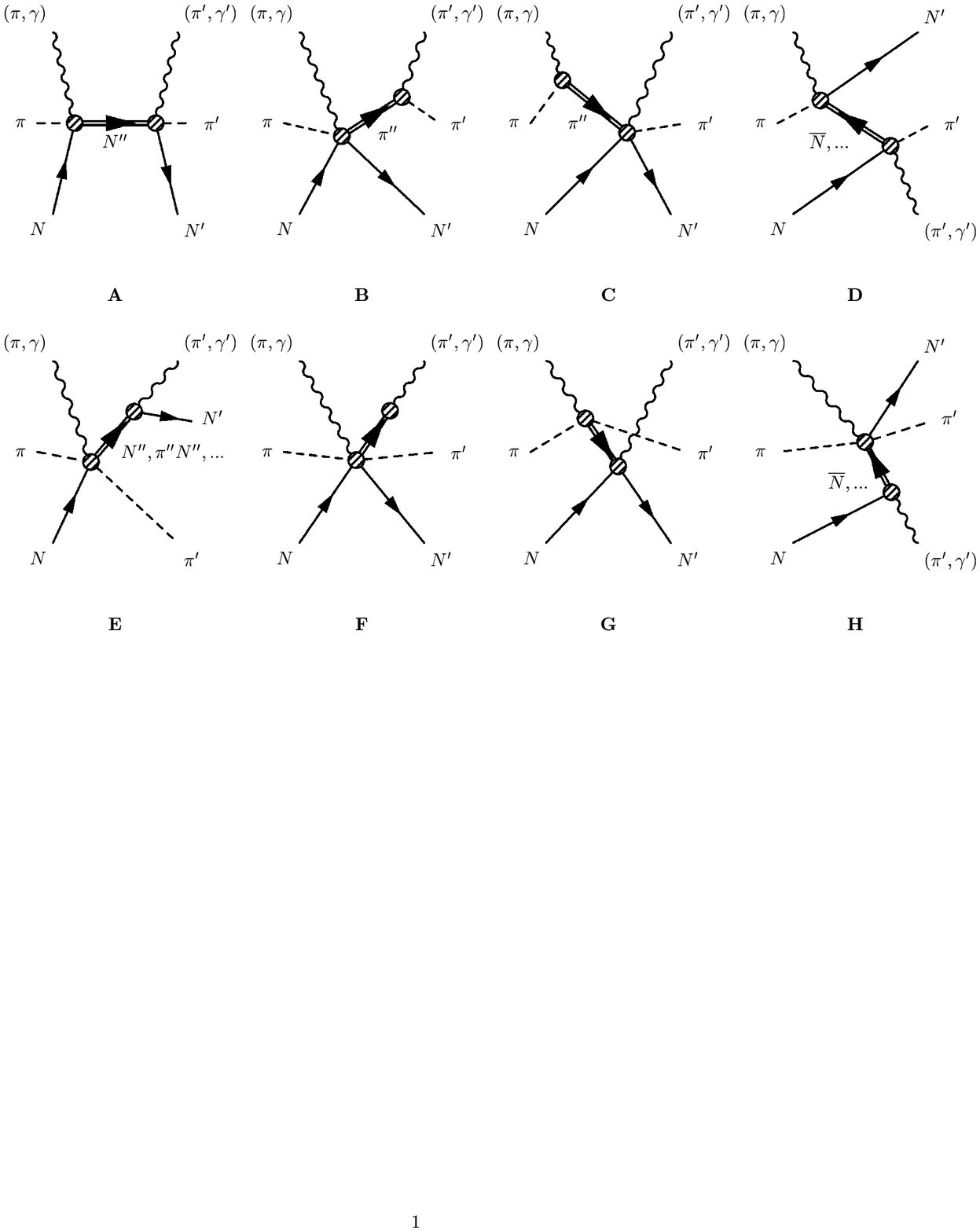}} 
\vspace{-5.0cm} 
\caption{{\protect\footnotesize {\it The first 8 diagrams for the three-body 
$b\pi N\Longleftrightarrow a \pi^{\prime}N^{\prime}$ reaction. The next 40
diagrams corresponds to the other chronological sequence of absorption of $%
\pi,\gamma$ and emission $\pi^{\prime},\gamma^{\prime}$. }}}
\label{fig:four}
\end{figure}
\vspace{5mm}


The disconnected parts of the potential (2.9a) $W^d_{\alpha\beta}$ together
with the disconnected parts of the second term on the right hand of
Eq.(2.8a) comprise the disconnected part of amplitudes (2.7a,b). For
instance, if we compare the expressions with the noninteracting pion states
in Eq.(2.7b) and in Eq.(2.8a), then we obtain

\[
<in;\pi^{\prime}|\pi;in>f^{3}_{ a+N^{\prime}, b+N}\equiv
-<in;\pi^{\prime}|\pi;in> <out;{\bf p^{\prime}}_N|j_a(0)|{\bf p}_N{\bf p}%
_b;in>
\]
$$
=<in;\pi^{\prime}|\pi;in>\Biggl\{ W^c_{ a+N^{\prime}, b+N}
+(2\pi)^3\sum_{n=N^{\prime\prime},\pi^{\prime\prime}N^{\prime\prime},...}
f_{a+N^{\prime},n} {\frac{{\delta^{(3)}( {\bf p}_N+{\bf p_{b}}-P_{n })}}{{%
\omega_b({\bf p_b})+{\ E_{{\bf p}_N}-P_{n}+i\epsilon }} }} {F_{b+N,n} }%
^{\ast}\Biggr\}
\eqno(2.8b)
$$

This means that the disconnected parts on the both side of relation (2.8a)
constitute independent equations. Therefore one can separate the connected
parts of amplitudes and effective potentials in Eq.(2.8a) as

$$
f^c_{\alpha\beta}= W^c_{\alpha\beta}+(2\pi)^3\sum_{\gamma=1}^4
f^c_{\alpha\gamma} {\frac{{\delta^{(3)}( {\bf p}_b+{\bf P_{{\widetilde \beta}%
}-P_{\gamma} })}}{{\omega_b({\bf p_b})+{\ P_{{\widetilde \beta}%
}^o-P_{\gamma}^o+i\epsilon }} }} {F^{\ast}}_{\beta \gamma}^c . \eqno(2.8c)
$$

Thus the disconnected and connected parts of the amplitudes (2.8a) and the
effective potentials (2.9a) form an independent set of equations. On the
other hand, it is well known, that the potential of a three-body Faddeev
equation contains the sum of the disconnected parts and iteration of these
disconnected parts contributes in the sought connected three-body amplitude.
The same properties have also the effective potentials of the three-body
Bethe-Salpeter equations \cite{GM,Kvin,Afnan} which are constructed in the
framework of the graphical method \cite{Taylor}. In the considered $S$%
-matrix approach the effective potential consists the product of the two
complete physical amplitudes and the last cut lemma of the graphical method
does not work for this case. Therefore here the cluster decomposition method 
\cite{alf,Ban} that separates analytically the connected and disconnected
parts of amplitudes was used. As a result of this procedure the disconnected
and connected parts of the amplitude (2.5) can be calculated independently
from each other according to the Eq.(2.8b) and Eq.(2.8c). In other words,
the contributions of the products of the disconnected amplitudes $f_{\alpha
\beta }^{d}$ (Fig.1) are already taken into account in $w_{\alpha \beta }^{c}
$ as it can be observed in Fig.3 and Fig.4. In particular, the combinations
of the disconnected parts of amplitudes depicted in Fig.1B and Fig.1C
constitute the connected term depicted in Fig.3B and in Fig.4E with the one
particle $N^{\prime \prime }\ or\ \pi ^{\prime \prime }$ intermediate states.



Eq. (2.8a) contains two type of transition amplitudes $f_{\alpha\beta}$
(2.2) and

$$
F_{\alpha\beta}=-<in;{\widetilde \alpha}|j_{a}(0)|\beta;in>.\eqno(2.10)
$$
The later includes only $^{\prime\prime}in^{\prime\prime}$ asymptotic states
in contrary to $f_{\alpha\beta}$. In particular, for the two-body states $%
\alpha=1^{\prime},2^{\prime}\equiv
\pi^{\prime}N^{\prime},\gamma^{\prime}N^{\prime}$ and an arbitrary initial $%
\beta$ state, we have

\[
F_{{\pi^{\prime}N^{\prime}},\beta}=f_{{\pi^{\prime}N^{\prime}},\beta}=-<out;%
{\bf p^{\prime}}_N|j_{\pi^{\prime}}(0)|\beta;in>;
\]
$$
F_{{\gamma^{\prime}N^{\prime}},\beta}=f_{{\gamma^{\prime}N^{\prime}}%
,\beta}=-<out;{\bf p^{\prime}}_N|J_{{\bf k}^{\prime}\eta^{\prime}}(0)|%
\beta;in>, \eqno(2.11a)
$$
because $<out;{\bf p^{\prime}}_N|= <in;{\bf p^{\prime}}_N|$, but 
$$
F_{\alpha\beta}\ne f_{\alpha\beta}\eqno(2.11b)
$$
for the three-body states $\alpha,\beta=3,4\equiv\pi\pi N,\gamma\pi N$.

For $F_{\alpha\beta}$ (2.10) we can derive similar to Eq.(2.2) relations

$$
F_{\alpha\beta} =<in;{\widetilde \alpha}|\Bigl[j_{a}(0),a_{b}^+(0)\Bigr] |{%
\widetilde \beta};in> + i\int d^4 x e^{-ip_b x}<in;{\widetilde\alpha}
|T\Bigl(j_{a}(0)j_{b}(x)\Bigr) |{\widetilde \beta};in>,\eqno(2.12)
$$
where after insertion of the completeness relation $\sum_n |n;in><in;n|={%
\widehat 1}$ between the current operators in (2.12) and subsequent
integration over $x$ and separation of the connected parts, we get

$$
F^c_{\alpha\beta}=w^c_{\alpha\beta}+(2\pi)^3\sum_{\gamma=1}^4
F^c_{\alpha\gamma} {\frac{{\delta^{(3)}( {\bf p}_b+{\bf P_{{\widetilde \beta}%
}-P_{\gamma} })}}{{\omega_b({\bf p_b})+{\ P_{{\widetilde \beta}%
}^o-P_{\gamma}^o+i\epsilon }} }} {F^c_{\beta \gamma} }^{\ast},\eqno(2.13)
$$

where $w^c_{\alpha\beta}$ can be obtained from the $W_{\alpha\beta}$ (2.6)
after replacement $<out;{\widetilde \alpha}|\Longleftarrow <in;{\widetilde
\alpha}|$.

From Eq.(2.12) one can derive also another type of relations between the ${%
F^c_{\beta \gamma} }$ and ${f^c_{\beta \gamma} }$ amplitudes. Substituting
the complete set of the $^{\prime\prime}out^{\prime\prime}$ states $\sum_n
;out><out;n|={\widehat 1}$ between the current operators in (2.12), we
obtain 
$$
F^c_{\alpha\beta}={\widetilde W}^c_{\alpha\beta}+(2\pi)^3\sum_{\gamma=1}^4 {%
f^c_{\gamma\alpha}}^{\ast} {\frac{{\delta^{(3)}( {\bf p}_b+{\bf P_{{%
\widetilde \beta}}-P_{\gamma} })}}{{\omega_b({\bf p_b})+{\ P_{{\widetilde
\beta}}^o-P_{\gamma}^o+i\epsilon }} }} {f^c_{\gamma\beta}} ,\eqno(2.14)
$$
where ${\widetilde W}^c_{\alpha\beta}$ differs from $w^c_{\alpha\beta}$
(2.6) by the intermediate $^{\prime\prime}out^{\prime\prime}$ states. 

Equations (2.8c) and (2.13), (2.14) represent the spectral decomposition
formulas (or off shell unitarity conditions) for the three-body amplitudes
in the standard quantum field theory. Such three-dimensional time-ordered
relations are often considered in the textbooks in the quantum field theory 
\cite{BD,IZ,alf} and in the nonrelativistic collision theory \cite{New,Gold}
for the two-body reactions. Therefore, we will treat Eq. (2.8c) and Eq.
(2.13), Eq. (2.14) as the three-body generalization of the field-theoretical
spectral decomposition formulas (or off shell unitarity conditions) for the
two-body amplitudes.

All of the above considered expressions are three-dimensional and
time-ordered from the beginning, and the corresponding relativistic
equations  are often called as the equations of the "old perturbation
theory". Equations (2.8c) and (2.13), (2.14) are formulated for the matrix
representation of the physical (i.e. renormalized) current operators (2.4a)
and their equal-time commutators with the Heisenberg field operators
(2.4b,c). These expressions and the left hand sides of Eqs. (2.2), (2.8a)
and (2.13) $<out(in);{\widetilde \alpha}|j_{a}(0)|\beta; in>$ are
Lorentz-invariant, but due to presence of the step functions $\theta(\pm x^o)
$ in the time-ordered expression (2.2) the propagators of the considered
equations (2.8a,b,c) and (2.9a) violated the manifestly Lorentz-covariance
form of the considered equations. A time-ordered product is a
Lorentz-covariant object, since a Lorentz-transformation cannot change the
order of the operators in the time-ordered product and the resulting
expressions for the $S$-matrix \cite{Bog}. In order to restore of the
manifestly Lorentz-covariance one can introduce the ``covariant time'', i.e.
instead of $x_o$ and $y_o$ in (2.2) one can use $X_o=\lambda_{\mu}x^{\mu}$
and $Y_o=\lambda_{\mu}y^{\mu}$, where in the c.m. frame ${\bf p}_{b}+{\bf p}%
_{\widetilde \beta}=0$ four-vector $\lambda_{\mu}$ is time-like unit vector $%
\lambda_{\mu}=(1,{\bf 0})$ and in an arbitrary system $\lambda_{%
\mu}=(p_{b}+p_{\widetilde \beta})_{\mu}/ |(p_{b}+p_{\widetilde \beta})|$.
Then in the c.m. system we obtain the expression(2.2) again and in any
arbitrary system we will have the covariant propagator with $\omega_{\gamma}(%
{\bf k}_{\gamma})\Rightarrow \lambda_{\mu} {k_{\gamma}}^{\mu}$, $%
\omega_{\pi}({\bf p}_{\pi})\Rightarrow \lambda_{\mu} {p_{\pi}}^{\mu}$ and $%
E_{p_{N}}\Rightarrow \lambda_{\mu} {p_{N}}^{\mu}\ etc.$ \cite{Kadysh2,M1}. 
This procedure restores the explicit form of Lorentz-invariance of the
considered equations.

Unlike the two-body case, in the three-body formulation it is necessary to
operate with the two kind amplitudes $f_{\gamma\beta}$ (2.2) and $%
F_{\gamma\beta}$ (2.10). The advantage of Eq. (2.13) is that it contains
only $F_{\gamma\beta}$. Using the $T$-invariance and relations (2.11a), we
see, that for the reactions with the two-body initial states $\beta\equiv
\pi N,\gamma N$ $f_{\gamma\beta}$ (2.2) and $F_{\gamma\beta}$ (2.10)
coincides.


The important part of the effective potential $w^c_{\alpha\beta}$ is the
equal-time commutator

$$
Y_{\alpha\beta}=<out;{\widetilde \alpha}| \Bigl[j_{a}(0),a_{b}^+(0)\Bigr]|{%
\widetilde \beta};in>_c.\eqno(2.15)
$$

The explicit form of this expression can be determined from the $a\ priori$
given Lagrangian and equal-time commutations relation between the Heisenberg
field operators. In the case of renormalized Lagrangian models or for
nonrenormalizable simple phenomenological Lagrangians the equal-time
commutators are easy to calculate \cite{M1}. In that case potential (2.15)
consists of the off shell ${\bf internal}$ one particle exchange potentials
(see Fig.5A, Fig.5C and Fig.5E) and of the contact (overlapping) terms
(Fig.5B, Fig.5D and Fig.5F). This is the only part of Eq.(2.13) which
contains ${\bf explicitely}$ the ${\bf internal}$ particle exchange
diagrams, since other terms in the effective potential (2.9a,b) and in
Eq.(2.13) consists of the matrix elements of the source operators of the $%
{\bf external}$ particle operators $j_{a}(x),\ and\ j_{b}(x)$. In order to
clarify the structure of the equal-time terms, we will consider Lagrangian
of the linear $\sigma $ model with the electromagnetic fields \cite
{IZ,GrossB,alf}

\[
{\cal L}_I= -e{\overline \psi}\gamma^{\mu}{\frac{{1+\tau_3}}{2}}\psi A_{\mu}
-g_{\pi}{\overline \psi}[\Phi_{\sigma}+i\gamma_5{\bf \tau \Phi_{\pi} } ]\psi
-ie A_{\mu}\epsilon^{3ij}\Phi_{\pi}^i\partial_{\mu}\Phi_{\pi}^j +{\frac{{e^2}%
}{2}}A_{\mu}A^{\mu}\Phi_{\pi}^2
\]

$$
-{\frac{ {g_{\pi}^2(m_{\sigma}^2-m_{\pi}^2)}}{{4m_N^2} }} (\Phi_{\sigma}^2+%
\Phi_{\pi}^2)^2 -{\frac{ {g_{\pi}(m_{\sigma}^2-m_{\pi}^2)}}{{2m_N} }}
\Phi_{\sigma}(\Phi_{\sigma}^2+\Phi_{\pi}^2),\eqno(2.16)
$$

where $\Phi_{\sigma}$ is the auxiliary scalar field operator with mass $%
m_{\sigma}$ and $g_{\pi}$ is defined thought the pion decay constant $f_{\pi}
$ as $g_{\pi}f_{\pi}=m_N$.

The corresponding current operator and the equation of motion are

$$
\partial_{\nu}\partial^{\nu}A_{\mu}=J_{\mu}= -e{\overline \psi}\gamma^{\mu}{%
\frac{{1+\tau_3}}{2}}\psi
-ie\epsilon^{3ij}\Phi_{\pi}^i\partial_{\mu}\Phi_{\pi}^j
+e^2A_{\mu}\Phi_{\pi}^2.\eqno(2.17a)
$$

\[
(\partial_{\nu}\partial^{\nu}+m_{\pi}^2)\Phi_{\pi}^i=j_{\pi}^i= -g_{\pi}{%
\overline \psi}i\gamma_5 \tau^i\psi
-2ieA_{\mu}\epsilon^{3ij}\partial^{\mu}\Phi_{\pi}^j
+e\partial^{\mu}A_{\mu}\epsilon^{3ij}\Phi_{\pi}^j
+e^2A_{\mu}A^{\mu}\Phi_{\pi}^i
\]
$$
-{\frac{ {g_{\pi}^2(m_{\sigma}^2-m_{\pi}^2)}}{{m_N^2} }} (\Phi_{\sigma}^2+%
\Phi_{\pi}^2)\Phi_{\pi}^i -{\frac{ {g_{\pi}(m_{\sigma}^2-m_{\pi}^2)}}{{m_N} }%
} \Phi_{\sigma}\Phi_{\pi}^i. \eqno(2.17b)
$$

Using Eq. (2.4b,c) and the equal-time commutation relation between the pion
fields $\Bigl[\partial _{o}\Phi _{\pi }^{i}(x),\Phi _{\pi }^{j}(y)\Bigr%
]\delta (x_{o}-y_{o})=-i\delta ^{ij}\delta ^{(4)}(x-y)$, for $a=\pi ^{\prime
}$ and $b=\pi $ we get from Eq.(2.15)

\[
Y_{\alpha\beta}=2ep_{\pi}^{\mu}\epsilon^{3ij} <out;{\widetilde \alpha}%
|A_{\mu}(0)|{\widetilde \beta};in>_c +<out;{\widetilde \alpha}%
|e^2A_{\nu}(0)A^{\nu}(0)+e\partial^{\mu}A_{\mu}(0) |{\widetilde \beta};in>_c
\]

\[
-{\frac{ {g_{\pi}^2(m_{\sigma}^2-m_{\pi}^2)}}{{m_N^2} }} <out;{\widetilde
\alpha}|
\Bigl([\Phi_{\sigma}^2(0)+\Phi_{\pi}^2(0)]\delta^{ij}+2\Phi_{\pi}^j(0)
\Phi_{\pi}^i(0)\Bigr)|{\widetilde \beta};in>_c
\]
\[
-{\frac{ {g_{\pi}(m_{\sigma}^2-m_{\pi}^2)}}{{m_N} }} <out;{\widetilde \alpha}%
| \Phi_{\sigma}(0)\delta^{ij}|{\widetilde \beta};in>_c
\]

\[
\approx -{\frac{ {g_{\pi}^2(m_{\sigma}^2-m_{\pi}^2)}}{{m_N^2} }} <out;{%
\widetilde \alpha}| \Bigl([\Phi_{\sigma}^2(0)+\Phi_{\pi}^2(0)]\delta^{ij}+
2\Phi_{\pi}^j(0)\Phi_{\pi}^i(0)\Bigr)|{\widetilde \beta};in>_c
\]
$$
-{\frac{ {g_{\pi}(m_{\sigma}^2-m_{\pi}^2)}}{{m_N} }} {\frac{{\delta^{ij}<out;%
{\widetilde \alpha}| j_{\sigma}(0)|{\widetilde \beta};in>_c} }{{%
(P_{\widetilde \alpha}-P_{\widetilde \beta})^2 -m^2_{\sigma}} }}\eqno(2.18a)
$$

where ${\widetilde \alpha}=N^{\prime}\ or\ \pi^{\prime}N^{\prime}\ or\
\gamma^{\prime}N^{\prime}$; ${\widetilde \beta}=N\ or\ \pi N\ or\ \gamma N$
and we have omitted terms with the intermediate photons.

The first term of Eq.(2.18a) corresponds to the contact (overlapping)
potential which is depicted in Fig.5B, Fig.5D and Fig.5F. Second term
describes the isoscalar $\sigma$-meson exchange interaction and it is given
in Fig.5A, Fig.5C and Fig.5E. For the other kind equal-time commutators with 
$b=\pi$ $({\widetilde \beta}=N\ or\ \pi N\ or\ \gamma N)$  and $%
a=\gamma^{\prime}$ $({\widetilde \alpha}=N^{\prime}\ or\
\pi^{\prime}N^{\prime})$ we have

\[
Y_{\alpha\beta}=2\Biggl[
-e\epsilon^{3ij}<out;{\widetilde \alpha}|
\Bigl(i\partial_{\mu}-(p_{\pi})_{\mu}\Bigr) \Phi_{\pi}^j(0)|{\widetilde \beta%
};in>_c +e^2<out;{\widetilde \alpha}|A_{\mu}(0)\Phi_{\pi}^i(0) |{\widetilde
\beta};in>_c\Biggr]
\epsilon^{\mu}_{\eta}({\bf k}_{\gamma}) 
\]
$$
=-2e\epsilon^{3ij}(P_{\widetilde \alpha}-P_{ \beta})_{\mu} {\frac{{<out;{%
\widetilde \alpha}| j_{\pi}^j(0)|{\widetilde \beta};in>_c}}{{(P_{\widetilde
\alpha}-P_{ \beta})^2 -m^2_{\pi}} }} \epsilon^{\mu}_{\eta}({\bf k}_{\gamma})
+2e^2<out;{\widetilde \alpha}|A_{\mu}(0)\Phi_{\pi}^i(0)|{\widetilde \beta}%
;in>_c \epsilon^{\mu}_{\eta}({\bf k}_{\gamma}). \eqno(2.18b)
$$

For $b=\gamma$ and $a=\pi^{\prime}$

\[
Y_{\alpha\beta}= -2ie\epsilon^{3ij}<out;{\widetilde \alpha}|
\partial_{\mu}\Phi_{\pi}^j(0)|{\widetilde \beta};in>_c \epsilon^{\mu}_{\eta}(%
{\bf k}_{\gamma}) +2e^2<out;{\widetilde \alpha} |A_{\mu}(0)\Phi_{\pi}^i(0)|{%
\widetilde \beta};in>_c \epsilon^{\mu}_{\eta}({\bf k}_{\gamma}) 
\]
$$
=-2e\epsilon^{3ij}(P_{\widetilde \alpha}-P_{\widetilde \beta})_{\mu} {\frac{{%
<out;{\widetilde \alpha}| j_{\pi}^j(0)|{\widetilde \beta};in>_c} }{{%
(P_{\widetilde \alpha}-P_{\widetilde \beta})^2 -m^2_{\pi}} }}
\epsilon^{\mu}_{\eta}({\bf k}_{\gamma}) +2e^2<out;{\widetilde \alpha}%
|A_{\mu}(0)\Phi_{\pi}^i(0)|{\widetilde \beta};in>_c \epsilon^{\mu}_{\eta}(%
{\bf k}_{\gamma}), \eqno(2.18c)
$$

\vspace{5mm}



\begin{figure}[htb]
\centerline{\epsfysize=145mm\epsfbox{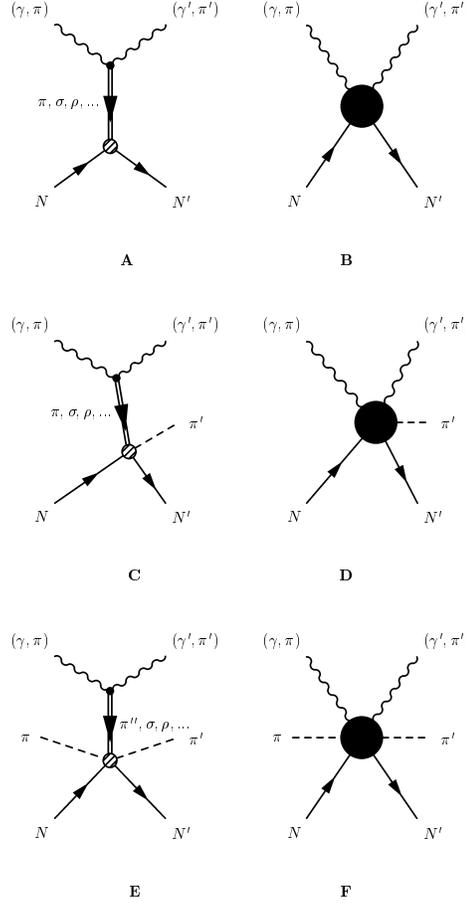}} 
\vspace{-1.0cm}
\par
\caption{{\protect\footnotesize {\it The graphical representation of the
equal-time commutators (2.15) in the effective potential (2.9a). This term
is depicted separately for the binary reactions ${\bf A,B}$, for the
reactions $2\Longrightarrow 3^{\prime}$ ${\bf C,D}$ and for the three-body
reactions $3\Longleftrightarrow 3^{\prime}$ ${\bf E,F}$. Diagrams ${\bf A,C,E%
}$ correspond to one off-mass shell particle exchange interactions. The
vertex functions between the $(\gamma,\pi)$ and $(\gamma^{\prime},\pi^{%
\prime})$ states are given in the lowest, tree approximation. Diagrams ${\bf %
B,D,F}$ describe the contact (overlapping) interaction which does not
contain the intermediate hadron propagation between hadron states.}}}
\label{fig:five}
\end{figure}
\vspace{3mm}


and finally for $b=\gamma$ and $a=\gamma^{\prime}$

$$
Y_{\alpha\beta}= e^2<out;{\widetilde \alpha}| \Phi_{\pi}(0)^2|{\widetilde
\beta};in>_c {\epsilon_{\mu}}_{\eta^{\prime}}({\bf k^{\prime}}_{\gamma})
\epsilon^{\mu}_{\eta}({\bf k}_{\gamma})\eqno(2.18d) 
$$
consists of a contact (overlapping) term only. If we replace $\Phi_{\pi}^2$
by a new type auxiliary scalar field $\sigma^{\prime}$, then  expression
(2.18d) transforms into one $\sigma^{\prime}$ scalar meson exchange diagram.
Using more complete Lagrangian, one can obtain also heavy $\rho,\omega$
meson exchange diagrams \cite{M1,MR}. Moreover, in the Ref.\cite{MCH,M1} the
One Boson Exchange (OBE) Bonn model of $NN$ interaction was exactly
reproduced from the equal-time commutator. For the three-body asymptotic
states equal-time commutators (2.15) produce more complicated contact terms
(Fig. 5C, Fig.5D and Fig. 5E, Fig.5F) which can be treated as pure
three-body forces. For these terms with the $\sigma, \rho,\omega,...$ field
operators it is necessary to use corresponding derivation of the spectral
decomposition formulas like Eq.(2.13). Note, that for the two-body reactions
equal-time commutators are determined by the one-variable vertex functions
with two on mass shell particles. For the three-body reactions the
equal-time commutators3.commutators are defined through the scattering amplitudes with
three or four on-mass shell particles.

Thus starting from the Lagrangian of the well-known linear $\sigma $-model
(2.16) we have obtained the one-particle exchange potentials (Fig. 5A, Fig.
5C and Fig. 5E) and contact (overlapping) terms. For the binary reactions
these terms contribute in the potential of the considered equations.
Unfortunately, the explicit expression of the one-variable vertices on Fig.
5A and Fig. 5B are not well determined. Using the model Lagrangians we can
extract the coupling constants i.e.e we can obtain the threshold values of
these vertex functions. For the $\gamma NN$ and $\gamma \pi \pi $ systems
these vertices are determined from the experimental data. The $\pi NN$
vertex functions can be constructed from the dispersion relations. Apart
from this, we can determine the asymptotic behavior of these vertex
functions based on the quark counting rules\cite{MMT,Brodsky} and dispersion relations
\cite{Hohler:1976ax,Krivoruchenko:1994qb,Mergell:1996bf,Dubnicka:1996sp,
Watanabe}
or on the Regge trajectories theory \cite{Collins}. Another possibility to
determine the equal-time commutators in the two-body reactions, using the
inverse scattering method , is considered in Ref.\cite{Ml}.

Presently, there exist numerous phenomenological models of the $NN$, $\pi N$
and $\gamma N$ reactions, where the pieces of Lagrangian (2.16) are used by
construction of the effective potentials by the description of the
corresponding reactions. Therefore in the framework of the considered
formulation it seems possible to achieve a unified description of the
coupled $\pi N-\gamma N-\pi \pi N-\gamma \pi N$ reactions proceeding from
the generalized Lagrangian (2.16) with the $\rho ,\omega ,...$ degrees of
freedom. On the other hand, the Lagrangian of the linear $\sigma $-model can
be reproduced from a set of more complicated and complete QCD motivated
Lagrangians. Thus the unified description of the coupled $\pi N-\gamma N-\pi
\pi N-\gamma \pi N$ reactions allows us to determine the form of the
equivalent Lagrangians for the $\gamma \pi N$ interactions which are
sufficient and necessary for a description of the experimental observables
in the low and intermediate energy region.

In the quantum field theory with the quark-gluon degrees of freedom one can
construct the hadron creation and annihilation operators as well as the
Heisenberg field operators of hadrons from the quark-gluon fields in the
framework of the Haag-Nishijima-Zimmermann \cite{H,N,Z,HW} treatment. In
this case  the form of the hadron quantum field operators is changed, but 
equations (2.8c), (2.13), (2.14) and (2.15) remain the same \cite{M1,M3,MBFE}
and  one can separate again the one off-mass shell meson exchange diagrams
in Fig.5A, Fig.5C and Fig.5E \cite{MBFE}. The contribution of the
overlapping (contact) terms in Fig.5B, Fig.5D and Fig.5F can be estimated
using the pure quark-gluon exchange or overlapping (contact) terms.


\begin{center}
{\bf 3. The relativistic  Lippmann-Schwinger type equations for the 
multichannel amplitude $F_{\alpha\beta}$ }
\end{center}

\medskip

The Lippmann-Schwinger type equations for the multichannel scattering $t$%
-matrix $T_{\alpha \beta }(E)$ with the Hermitian potential $V=V^{+}$ have
the following form \cite{New,Gold}

$$
T_{\alpha\beta}(E_{\beta})\equiv <in;\alpha|T(E_{\beta})|\beta;in>
=V_{\alpha\beta}+\sum_{\gamma}V_{\alpha\gamma}{\frac{1}{{E_{\beta}-
E_{\gamma}+i\epsilon}}}T_{\gamma\beta}(E_{\beta})\eqno(3.1a)
$$
$$
=V_{\alpha\beta}+\sum_{\gamma,\gamma^{\prime}}V_{\alpha\gamma} <in;\gamma|{%
\frac{1}{{E_{\beta}-H+i\epsilon}}}|\gamma^{\prime};in>
V_{\gamma^{\prime}\beta},\eqno(3.1b)
$$
where $<in;\alpha|$ denotes the asymptotic $\alpha$-channel wave function
with the energy $E_{\alpha}$ and quantum numbers $\alpha$, $\sum_{\gamma}$
stands for the integration over the momenta and the summation over the
quantum numbers of the complete set intermediate $\gamma$-channel states, $H$
is the full Hamiltonian which has the complete set of the eigenfunctions $%
H|\Psi_{\gamma}>= E_{\gamma}|\Psi_{\gamma}>$; $\sum_{\gamma}|\Psi_{\gamma}><%
\Psi_{\gamma}|={\widehat 1}$. Using the decomposition formula of the full
Green function $G(E)=1/(E-H+i\epsilon)$ over the complete set of the
functions $|\Psi_{\gamma}>$

$$
G(E)=\sum_{\gamma}{\frac{{|\Psi_{\gamma}><\Psi_{\gamma}|}}{{%
E-E_{\gamma}+i\epsilon}}}, \eqno(3.2)
$$

equations (3.1a) and (3,1b) can be written as the quadratically nonlinear
integral equations \cite{New,Gold}

$$
T_{\alpha\beta}(E_{\beta})=V_{\alpha\beta}+ \sum_{\gamma}T_{\alpha\gamma}{%
\frac{1}{{E_{\beta}-E_{\gamma}+i\epsilon }}} T_{\beta \gamma}^{\ast},%
\eqno(3.3)
$$
where we have taken into account the relation $T_{\alpha\beta}(E_{\beta})=
<in;\alpha|V|\Psi_{\beta}>$.

The three-dimensional equations (3.3) have the same form as equations
(2.13). In spite of the great complexity of the nonlinear equations (3.3),
they have very attractive properties, since they can be considered as the
off shell generalization of the unitarity conditions \cite{New} and, unlike
the Lippmann-Schwinger type equations (see the book of Goldberger and
Watson, ch.5, Eq. (88)-Eq.(100)), they are free from the difficulty of
construction of the orthonormal set of the intermediate channel states.
Therefore they may be considered as a basis of derivation of the special
Lippmann-Schwinger type equations which have no inherent pathological
properties. In the present formulation the nonlinear field-theoretical
equations (2.13) are considered as the origin for a derivation  of the
linearized relativistic Lippmann-Schwinger type equations.

Unlike the nonrelativistic case, potential $w_{\alpha \beta }^{c}$, is not
Hermitian due to the propagators of intermediate states. Nevertheless, in
appendix A we show that quadratically nonlinear equations (2.10) are
equivalent to the following Lippmann-Schwinger type equations

$$
{\cal T}_{\alpha\beta}(E_{\beta})=U_{\alpha\beta}(E_{\beta})+
\sum_{\gamma=1}^4 U_{\alpha\gamma}(E_{\beta}) {\frac{1}{{\omega_b({\bf p_b}%
)+ P_{{\widetilde \beta} }^o-P_{\gamma}^o+i\epsilon } }} {\cal T}_{
\gamma\beta}(E_{\beta}),\eqno(3.4)
$$

here $P_{\beta}^o\equiv E_{\beta}=\omega_b({\bf p_b})+P_{{\widetilde \beta}
}^o$ and  for the sake of simplicity we have omitted the total
three-momentum conservation $\delta$ function $(2\pi)^3\delta({\bf %
P_{\beta}-P_{\gamma} })$.


\vspace{5mm}


\begin{figure}[htb]
\centerline{\epsfysize=225mm\epsfbox{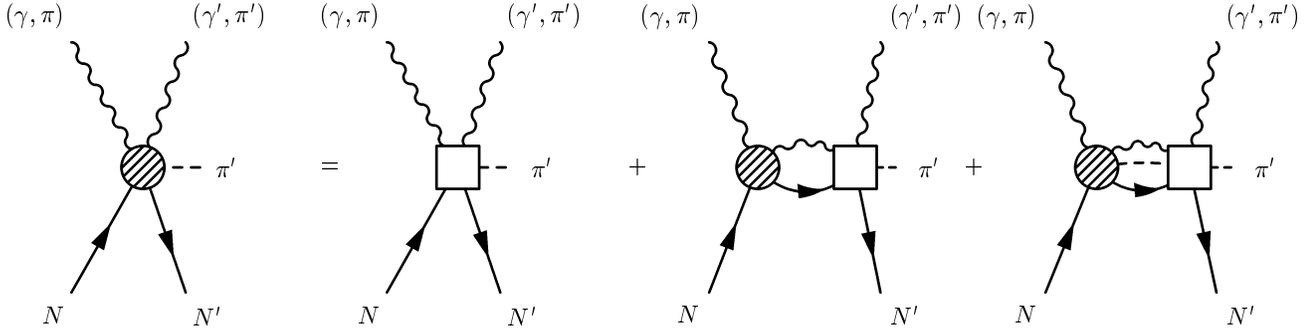}} 
\vspace{-17.0cm} 
\caption{{\protect\footnotesize {\it The graphical representation of
Eq.(3.4) for the $b+N\Longrightarrow a+\pi^{\prime}+N^{\prime}$ reaction
with $a=\pi^{\prime}\ or\ \gamma^{\prime}$ and $b=\pi\ or\ \gamma$. The
effective potential (3.5) of these equations is depicted as square and it is
represented in Fig.3 and Fig. 5C, Fig. 5D. }}}
\label{fig:six}
\end{figure}
\vspace{5mm}

The explicit form of the linear energy depending potential 
$$
U_{\alpha \beta }(E)=A_{\alpha \beta }+E\ B_{\alpha \beta }\eqno(3.5)
$$
with Hermitian $A$ and $B$ 
$$
A_{\alpha \beta }=A_{\beta \alpha }^{*};\ \ \ \ \ \ \ B_{\alpha \beta
}=B_{\beta \alpha }^{*},\eqno(3.6)
$$
is defined in the Appendix B. $U_{\alpha \beta }(E)$ is simply connected
with the $w_{\alpha \beta }^{c}$-potential (B.4a)-(B.4h), (B.8a)-(B.8h) and
(B.12a)-(B.12h) 
$$
U_{\alpha \beta }(E_{\alpha })=w_{\alpha \beta }^{c}.\eqno(3.7)
$$
Therefore, for any field-theoretical potential $w_{\alpha \beta }^{c}$ on
can unambiguously construct $U_{\alpha \beta }(E)$. The graphical form of
the potential is depicted in Fig.2, Fig.3 and Fig.4 for the two-body $\alpha
,\beta =1,2=\pi N,\gamma N$ and the three-body $\alpha ,\beta =3,4=\pi \pi
N,\gamma \pi N$ transition channels.

On energy shell solutions of the equations (2.10) and (3.4) coincide 
$$
{\cal T}_{\alpha\beta}(E_{\beta}=E_{\alpha})=F^c_{\alpha\beta}
|_{_{_{E_{\beta}=E_{\alpha} } } }\eqno(3.8)
$$
and in the half on energy shell region these amplitudes are simply connected

$$
F^c_{\alpha\beta}\equiv -<in;{\widetilde \alpha}|j_{a}(0)|%
\beta;in>_{connected}= w^c_{\alpha\beta}+\sum_{\gamma=1}^4
w^c_{\alpha\gamma} {\frac{1}{{E_{\beta}-E_{\gamma}+i\epsilon}}}{\cal T}%
_{\gamma\beta}(E_{\beta}) .\eqno(3.9)
$$

Equations (3.4) are our final equations for the multichannel $\pi
N\Longleftrightarrow\gamma N\Longleftrightarrow  \pi \pi
N\Longleftrightarrow \gamma \pi N$ scattering amplitudes, because their
solution allows us to determine $F^c_{\alpha\beta}$ (2.8c) that on energy
shell coincides with the solution of Eq.(2.13), and in the off energy shell
region $F^c_{\alpha\beta}$ is determined through ${\cal T}%
_{\gamma\beta}(E_{\beta})$ according to Eq.(3.9). For construction of the
complete $3\to 3^{\prime}$ transition amplitude it is necessary to take into
account  the disconnected parts (Fig. 1C) and to use Eq.(2.14) or (2.8).
Equation (3.4) have the form of the multichannel Lippmann-Schwinger
equations with the connected potentials $U_{\alpha\beta}(E)$ that are
single-valued determined by the potentials of the initial field-theoretical
equation (2.13) $w^c_{\alpha\beta}$. The structure of the system of
relativistic equations (3.4) is more simple as the field-theoretical
generalizations of the Faddeev equations,  based on the Bethe-Salpeter
equations \cite{Kvin,Afnan} or in the framework of the other relativistic
three-dimensional formulations (see the list of the corresponding quotations
in book\cite{GM}), because $U_{\alpha\beta}(E)$ does not contain any
disconnected parts and the complications, coming from double counting
problems (i.e. from the iterations of the disconnected parts of the
three-body potentials), does not take a part in this approach.\footnotemark 

\footnotetext{
Instead of Eq.(2.13) we can consider the quadratically-nonlinear three-body
equations for the sum of the connected and the disconnected parts of
amplitudes
\par
$$
F_{\alpha\beta}=w_{\alpha\beta}+(2\pi)^3\sum_{\gamma=1}^4 F_{\alpha\gamma} {%
\frac{{\delta^{(3)}( {\bf p}_b+{\bf P_{{\widetilde \beta}}-P_{\gamma} })}}{{%
\omega_b({\bf p_b})+{\ P_{{\widetilde \beta}}^o-P_{\gamma}^o+i\epsilon }} }} 
{F_{\beta \gamma} }^{\ast},\eqno(2.13')
$$
\par
where $F_{\alpha\beta}=F^c_{\alpha\beta}+F^d_{\alpha\beta}$ denotes the
complete $\beta\to\alpha$-transition amplitude and Eq.(2.13') have the form
of the off-shell unitarity conditions \cite{New}. Equation (2.13') can be
considered also as the basis for the derivation of the field-theoretical
generalizations of the Faddeev-type equations. For this aim we must derive
Eq.(2.13') from the Faddeev-type equations after rearrangement of the three
body amplitude $T_{\alpha=3^{\prime}\beta=3}(E)$ over the auxiliary
amplitudes $T_{\alpha=3^{\prime}\beta=3}(E)=\sum_{i=1,3}
T^i_{3^{\prime}3}(E)+  T^{123}_{3^{\prime}3}(E)= {\cal T}_{3^{\prime}3}(E)+{%
\ T}^d_{3^{\prime}3}(E)$, where ${\ T}^d_{\alpha\beta}(E)$ denotes the
disconnected part of amplitude and $T^{123}_{3^{\prime}3}(E)$ is introduced
for the pure three-body interactions \cite{New}. This means that we must
construct the complete operator ${\cal U}_{\alpha\beta}(E)$ with the
connected and the disconnected parts from $w_{3^{\prime}3}$. ${\cal U}%
_{\alpha\beta}(E)$  can contain only the three-body irreducible diagrams,
i.e. diagrams which does not include the 1,2,3-particle exchange
contributions in the $s$-channel. As result we will reproduce the
Lippmann-Schwinger-type equations 
$$
T_{\alpha\beta}(E_{\beta})={\cal U}_{\alpha\beta}(E_{\beta}) +\sum_{\gamma} 
{\cal U}_{\alpha\gamma}(E_{\beta}) {\frac{1}{{\omega_b({\bf p_b})+ P_{{%
\widetilde \beta} }^o-P_{\gamma}^o+i\epsilon } }} {\ T}_{
\gamma\beta}(E_{\beta}),\eqno(3.4')
$$
The detailed investigation of this equation and their comparison with
Eq.(3.4) we plane in the forthcoming papers.}


The essential difference between the nonrelativistic Lippmann-Schwinger
equation \cite{New,Gold} their relativistic field-theoretical generalization
(3.4)  is the nonlinearity of the field-theoretical equations. The most
famous nonlinearity was investigated in the nonlinear Chew-Low equations 
\cite{Schweber,Chew,Ban,Ther,MR,M1} for the $\pi N$ scattering problem. It
was demonstrated, that even the $s$-channel nonlinearity in Eq.(3.3)
generates the infinite sum of the Castilleho-Dalitz-Dyson (CDD) poles \cite
{CDD} in the $\pi N$ amplitude. Subsequently,  the position of these poles
was determined in the dispersion theory with the quark-gluon degrees of
freedom \cite{Saf} and in Ref.\cite{Ther} one of the CDD poles was used to
reproduce the $\Delta$ resonance. In the book of Goldberger and Watson \cite
{Gold} it was argued the uniqueness of the solution of nonlinear (3.3) and
linear (3.1a) equations. The argumentation of such type assertion were given
also in Ref.\cite{Haag2}, where it was shown, that if the spectrum of the
free Hamiltonian does not contain any compound or exited states, then the
conditions of completeness and orthonormality of the wave functions, that
are the solution of the Lippmann-Schwinger equations (3.1a), are compatible
with the solution of the nonlinear Chew-Low equations without CDD poles i.e.
solutions of linear Lippmann-Schwinger equations (3.4) and nonlinear
equations (3.3) are identical. Also in Ref.\cite{Warnock} for the Chew-Low
equations it was shown, that under special conditions imposed on the
coupling constants and vertex functions the Chew-Low type equations have a
unique solution.

From the explicit form of the $w^c_{\alpha\beta}$-potentials (B4a)-(B.4h), 
(B.8a)-(B.8h) and (B.12a)-(B.12h) and their graphical representation in Fig.
2, Fig.3, Fig.4 and Fig.5 we see, that besides the $u$-channel crossing
terms in Fig.2E, Fig.3E and Fig.4E, the unknown amplitudes placed also in
all other potential terms except the diagrams with the antinucleon exchange.
Even the terms with $\pi$-meson intermediate states from the equal-time
commutators in Fig.5C, Fig.5D, Fig.5E and Fig.4G generate the nonlinearity.
In Ref.\cite{MR} it was shown, that also the potential in the Bethe-Salpeter
equation contains the nonlinearity in the  crossed $u$-channel diagram for
the $\pi N$ scattering reactions. One can show, that the other kind of
nonlinearities, depicted in Fig.2, Fig.3, Fig.4 and Fig.5 arise in the
Bethe-Salpeter  equation too. The source of such type nonlinearities is the
field-theoretical nature of the considered approaches, that originate the
essential  non-linear functional equations \cite{Bog,IZ}, which are the base
of the above integral equations. Therefore the problem of the role of these
nonlinearities in the considered reactions is the important problem for the
future investigations.

\medskip

\begin{center}
{\bf 4. $\Delta$ degrees of freedom.}
\end{center}

\medskip

The relativistic field-theoretical equations (2.13) and (3.4)  are the
multichannel generalization of the equations for the elastic $\pi N$
scattering from ref.\cite{MR,M1}, where  all of $\pi N$ partial waves up to $%
300$ MeV, including the resonance $\pi N$ $P33$ partial wave, were
described. The effective $\pi N$ potentials there was constructed from the
one-variable phenomenological $\pi N-N$, $\sigma N-N$, $\rho N-N$ and $\rho
-\pi\pi$ vertex functions and corresponding simplest phenomenological
Lagrangians. In the considered generalization of the Chew-Low model \cite
{Low,Chew,Ban,Ther} the contributions of the intermediate $\Delta$ resonance
in the low-energy $\pi N$ scattering reaction was reproduced without any
additional assumptions.

Another way to taken into account the intermediate $\Delta$-resonance
effects is to introduce $\Delta$-degrees of freedom, i. e. to construct the
intermediate $\Delta$-resonance in the same manner as the usual one-particle
hadron states. Unfortunately, it is not possible to construct the
representation of the Poincare group for the unstable states, because the
asymptotic $^{\prime\prime}out^{\prime\prime}$ or $^{\prime\prime}in^{\prime%
\prime}$ states (i. e. the Fock space) is determined for the stable
particles in the asymptotic region $x_o\to \pm\infty$.  Therefore, we will
treat the $\Delta$ degrees of freedom as intermediate $\pi N$ cluster state
using the procedure of the separation of the $\Delta$ resonance poles from
the intermediate $\pi N$ Green functions in the $P33$ partial states. In
particular, following our previous papers \cite{M3,M5}, we firstly separate
the intermediate full $\pi N$ Green functions in the matrix elements (2.2)
for the transition between the ${\widetilde \alpha}+a$ and ${\widetilde \beta%
}+b$ states. For instance the matrix element $<out;{\widetilde\alpha}%
|j_{a}(0)\theta(-x_o)j_{b}(x)|{\widetilde \beta};in>$  with the $\pi N$
intermediate state can be transformed as 
\[
\sum_{\pi N}<out;{\widetilde \alpha}|j_a(0)|{\bf p}_{\pi}{\bf p}_N;in> {%
\frac{{\delta ( {\bf p}_b+{\bf P}_{\widetilde \beta}- {\bf p}_{\pi}-{\bf p}%
_N ) }}{{{P_{\widetilde \beta}}^o+\omega_b({\bf p}_b) - \omega({\bf p}%
_{\pi})-{E_{{\bf p}}}_N+io}}} <in;{\bf p}_{\pi}{\bf p}_N|j_b(0)|{\widetilde
\beta};in>
\]
\[
=\sum_{\pi N}<out;{\widetilde \alpha}|j_a(0)|{\bf p}_{\pi}{\bf p}_N;in>_
{\pi N\ irreducable}
\]
$$
{\cal G}^{\pi N}\Bigl(E={P_{\widetilde \beta}}^o+\omega_b({\bf p}_b)\Bigr)
<in;{\bf p}_{\pi}{\bf p}_N|j_b(0)|{\widetilde \beta};in>_ {\pi N\
irreducable},\eqno(4.1)
$$
where we have to distinguish the full $\pi N$ wave function $|\Psi^{\pi N}>$
and full $\pi N$ Green function

$$
{\cal G}^{\pi N}(E,{\bf P})= \int d^3{\bf p}_Nd^3{\bf p}_{\pi}(2\pi)^3\
\delta({\bf P}- {\bf p}_N-{\bf p}_{\pi}) {\frac{{|\Psi^{\pi N}_{{\bf p}_N%
{\bf p}_{\pi} }> <{\widetilde \Psi}^{\pi N}_{{\bf p}_N{\bf p}_{\pi} }|} }{{%
E-E_{{\bf p}_N}-\omega_{\pi}({\bf p}_{\pi})+io}}}\eqno(4.2a)
$$

using the $^{\prime\prime}\pi N\ irreducibe^{\prime\prime}$  matrix elements
which does not contain the intermediate $\pi N$ states 
$$
<out;{\widetilde \alpha}|j_a(0)|{\bf p}_{\pi}{\bf p}_N;in>= \Bigl\{<out;{%
\widetilde \alpha}|j_a(0)\Bigr\}_{\pi N\ irreducable}  |\Psi^{\pi N}_{{\bf p}%
_N{\bf p}_{\pi} }>;\eqno(4.3a)
$$

$$
<{\bf p}_{\pi}{\bf p}_N|j_b(0)|{\widetilde \beta};in>= <{\widetilde \Psi}%
^{\pi N}_{{\bf p}_N{\bf p}_{\pi} }|\Bigl\{ j_b(0)|{\widetilde \beta}%
;in>\Bigr\}_{\pi N\ irreducable}. \eqno(4.3b)
$$

The wave functions $|\Psi^{\pi N}>$ are simply connected with the $t$%
-matrices ${\cal T}_{\alpha\beta}(E)$ (3.4) 
$$
{\cal T}_{\pi^{\prime}N^{\prime};\pi N}(E_{\pi N})\equiv {\cal T}%
_{1^{\prime}1}(E_{1})= \sum_{\gamma=1}^4 <{\bf p^{\prime}}_N{\bf p^{\prime}}%
_{\pi}|U_{1^{\prime}\gamma}(E_{\pi N})|\gamma;in> <in;\gamma|\Psi^{\pi N}_{%
{\bf p}_N{\bf p}_{\pi} }>. \eqno(4.4)
$$
where $E_{\pi N}= E_{{\bf p}_N}+\omega_{\pi}({\bf p}_{\pi})$.

According to Eq.(A.1) and normalization condition (A.5b) the wave function $%
<in;\beta|\Psi^{\alpha=1=\pi N}>$ satisfies the following equation of motion

\[
\Bigl(E_{{\bf p^{\prime}}_N}+\omega_{\pi}({\bf p^{\prime}}_{\pi}) -E_{{\bf p}%
_N}-\omega_{\pi}({\bf p}_{\pi})\Bigr) <in;{\bf p^{\prime}}_N{\bf p^{\prime}}%
_{\pi}|\Psi^{\pi N}_{{\bf p}_N{\bf p}_{\pi} }>=
\]
$$
\sum_{\gamma=1}^4 <in;{\bf p^{\prime}}_N{\bf p^{\prime}}_{\pi}|U_{1^{\prime}%
\gamma} \bigl( E_{{\bf p}_N}+\omega_{\pi}({\bf p}_{\pi})\bigr)
|\gamma;in> <in;\gamma|\Psi^{\pi N}_{{\bf p}_N{\bf p}_{\pi} }> \eqno(4.5a)
$$

and the normalization condition

$$
<\Psi^{\pi^{\prime}N^{\prime}}|(1-B)|\Psi^{\pi N}>\equiv <{\widetilde \Psi}%
^{\pi^{\prime}N^{\prime}}|\Psi^{\pi N}>=\delta^{\ {\pi^{\prime}N^{\prime}},{%
\pi N}}. \eqno(4.6)
$$

Next we consider two kinds of extensions (or projections) the equation
(2.5a) for the initial $\pi N$ state energy in the complex region as

$$
E_{\pi N}= E_{{\bf p}_N}+\omega_{\pi}({\bf p}_{\pi})\Longrightarrow E_{{\bf P%
}_{\Delta}}\eqno(4.7a)
$$

and

$$
E_{\pi N}= E_{{\bf p}_N}+\omega_{\pi}({\bf p}_{\pi})\Longrightarrow E^o_{%
{\bf P}_{\Delta}}+\Sigma_{\Delta}(E_{{\bf P}_{\Delta}},{\bf P}_{\Delta}) %
\eqno(4.7b)
$$

where ${\bf P}_{\Delta}={\bf p}_{N}+{\bf p}_{\pi}$, $E^o_{{\bf P}_{\Delta}}=%
\sqrt{{M^o_{\Delta}}^2+{\bf P}_{\Delta}^2}$ and $E_{{\bf P}_{\Delta}}=\sqrt{{%
M_{\Delta}}^2+{\bf P}_{\Delta}^2}$ are the energies of the $bare$ and
observed energies of $\Delta$ with the $bare$ $M^o_{\Delta}$ and the
Breit-Wigner (physical) mass $M_{\Delta}=m_{\Delta}+i\Gamma_{\Delta}/2$ mass
of $\Delta$ respectively, where $m_{\Delta}=1232MeV$ and $%
\Gamma_{\Delta}=120MeV$ are the observed mass and full width of $\Delta$
resonances, $\Sigma_{\Delta}(E,{\bf P}_{\Delta})$ is the complex function of 
$E$ which satisfies the conditions

$$
Re\biggl[
E-E^o_{{\bf P}_{\Delta}}-\Sigma_{\Delta}(E,{\bf P}_{\Delta}) \biggr]_{E=E_{%
{\bf P}_{\Delta}} }=0\eqno(4.8a)
$$
$$
Im\biggl[
E-E^o_{{\bf P}_{\Delta}}- \Sigma_{\Delta}(E,{\bf P}_{\Delta}) \biggr]_{E=E_{%
{\bf P}_{\Delta}};\ \ \ {\bf P}_{\Delta}=0} =\Gamma_{\Delta}/2\eqno(4.8b)
$$

Using the extension (4.7a) we obtain the wave function of the on shell $%
\Delta$

$$
|\Psi^{\pi N}_{{\bf p}_N{\bf p}_{\pi} }> {\stackrel{ E_{\pi N}\to E_{{\bf P}%
_{\Delta}} }{\Longrightarrow}}  |\Psi^{\Delta}_{{\bf P}_{\Delta} }> %
\eqno(4.9a)
$$

which obeys the equation

\[
\Bigl(E_{ N}({\bf p^{\prime}}_N)+\omega_{\pi}({\bf p^{\prime}}_{\pi}) -E_{%
{\bf P}_{\Delta}}\Bigr) <in;{\bf p^{\prime}}_N{\bf p^{\prime}}%
_{\pi}|\Psi^{\Delta}_{{\bf P}_{\Delta} }>=
\]
$$
\sum_{\gamma=1}^4 <in;{\bf p^{\prime}}_N{\bf p^{\prime}}_{\pi}|U_{1^{\prime}%
\gamma} \bigl( E_{{\bf P}_{\Delta}}\bigr)
|\gamma;in> <in;\gamma|\Psi^{\Delta}_{{\bf P}_{\Delta} }> \eqno(4.5b)
$$


Similarly, the off shell projection of Eq.(4.5a) according to Eq.(4.7b)
allows us to determine the explicit form of the wave function $%
|\Psi^{\Delta}_{{\bf P}_{\Delta} }>$

$$
|\Psi^{\pi N}_{{\bf p}_N{\bf p}_{\pi} }> {\stackrel{E_{\pi N}\to E^o_{\pi
N}+\Sigma_{\Delta}(E,{\bf P}_{\Delta})}{\Longrightarrow}}  |\Psi^{\Delta}_{%
{\bf P}_{\Delta} }(E)>. \eqno(4.9b)
$$

as the solution of the extended equation

\[
\Bigl(E-E^o_{{\bf P}_{\Delta}}-\Sigma_{\Delta}(E,{\bf P}_{\Delta})\Bigr) <in;%
{\bf p^{\prime}}_N{\bf p^{\prime}}_{\pi}|\Psi^{\Delta}_{{\bf P}_{\Delta}
}(E)>=
\]
$$
\sum_{\gamma=1}^4 <in;{\bf p^{\prime}}_N{\bf p^{\prime}}_{\pi}|U_{1^{\prime}%
\gamma} \Bigl( E^o_{{\bf P}_{\Delta}}+ \Sigma_{\Delta}(E,{\bf P}%
_{\Delta})\Bigr) |\gamma;in> <in;\gamma|\Psi^{\Delta}_{{\bf P}_{\Delta}
}(E)>. \eqno(4.5c)
$$

Now one can pick out the $\Delta $ resonance part from the fully $\pi N$
Green function (4.2a) using the determination of the resonances as poles on
the complex $\pi N$ energy sheet. For this aim it is enough to separate the
contribution of the $\Delta $ resonance singularity at $E_{{\bf P}_{\Delta }}
$ (4.7a) or at $E_{\pi N}^{o}+\Sigma _{\Delta }(E,{\bf P}_{\Delta })$ (4.7b)
in the integral (4.2a). The residue of this pole consists of the product of
the two $\pi N-\Delta $ wave functions which are analytically continued with
the $\pi N$ wave function (4.5a) according to projections (4.7a) or (4.9b).
Thus from the Eq.(4.2a) we obtain

$$
{\cal G}^{\pi N}(E,{\bf P})= \int d^3{\bf P}_{\Delta}(2\pi)^3\ \delta({\bf P}%
-{\bf P}_{\Delta}) \sum_{\Delta}{\frac{{|\Psi^{\Delta}_{{\bf P}_{\Delta}}> <{%
\widetilde \Psi}^{\Delta}_{{\bf P}_{\Delta}}|} }{{E-E_{{\bf P}_{\Delta}}} }}%
+ \ nonresonant\ part.\ \eqno(4.2b)
$$

and 

$$
{\cal G}^{\pi N}(E,{\bf P})= \int d^3{\bf P}_{\Delta}(2\pi)^3\ \delta({\bf P}%
-{\bf P}_{\Delta}) \sum_{\Delta}{\frac{{|\Psi^{\Delta}_{{\bf P}%
_{\Delta}}(E)> <{\widetilde \Psi}^{\Delta}_{{\bf P}_{\Delta}}(E)|} }{{E-E^o_{%
{\bf P}_{\Delta}}-\Sigma_{\Delta}(E,{\bf P}_{\Delta})}}} + \ nonresonant\
part.\ \eqno(4.2c)
$$

The essential feature of the different representation of the $\pi N$ Green
function (4.2b) and (4.2c) is that they differently take into account $\Delta
$ degrees of freedom in the intermediate states. For instance, the general
formula (4.1) for the $s$-channel $b+{\widetilde \beta}\to a+{\widetilde
\alpha}$ transitions, after replacement of the fully $\pi N$ Green function 
(4.2a) with their resonance part (4.2b), takes the form

$$
\sum_{\Delta}<out;{\widetilde \alpha}| j_a(0)|\Psi^{\Delta}_{{\bf P}%
_{\Delta} }>_{\pi N\ irreducable} {\frac{{\delta ( {\bf p}_b+{\bf P}%
_{\widetilde \beta}- {\bf P}_{\Delta} ) }}{{{\ P_{\widetilde \beta}}%
^o+\omega_b({\bf p}_{b}) - {E_{{\bf P}}}_{\Delta}} }} <{\widetilde \Psi}%
^{\Delta}_{{\bf P}_{\Delta}}|j_b(0) |{\widetilde \beta};in>_ {\pi N\
irreducable}\eqno(4.10a)
$$

and from Eq.(4.2c) we obtain

\[
\sum_{\Delta}<out;{\widetilde \alpha}| j_a(0)|\Psi^{\Delta}_{{\bf P}%
_{\Delta} } \Bigl(P_{\widetilde \beta}^o+\omega_b({\bf p}_{b})\Bigr) >_{\pi
N\ irreducable}
\]
$$
{\frac{{\delta ( {\bf p}_b+{\bf P}_{\widetilde \beta}- {\bf P}_{\Delta} ) }}{%
{{\ P_{\widetilde \beta}}^o+\omega_b({\bf p}_{b}) - {E^o_{{\bf P}}}%
_{\Delta}- \Sigma_{\Delta} \Bigl(P_{\widetilde \beta}^o+\omega_b({\bf p}%
_{b}),{\bf P}_{\Delta}\Bigr) } }} <{\widetilde \Psi}^{\Delta}_{{\bf P}%
_{\Delta}} \Bigl(P_{\widetilde \beta}^o+\omega_b({\bf p}_{b})\Bigr)|j_b(0) |{%
\widetilde \beta};in>_ {\pi N\ irreducible}.\eqno(4.10b)
$$

In the both expressions (4.10a) and (4.10b) we have neglected the
nonresonant part of the $P_{33}$ $\pi N$ partial wave contributions.

In Eq.(4.10a) intermediate $\Delta$ propagators can be considered as on
shell propagators, because only the Breit-Wigner mass and width are taken
into account in the energy of $\Delta$. The more general expression (4.10b)
includes the renormalization effects in the mass operator $\Sigma_{\Delta}(E,%
{\bf P}_{\Delta})$ which satisfies the conditions (4.8a,b). The explicit
form of $\Sigma_{\Delta}(E,{\bf P}_{\Delta})$  depends on the model of
interaction of the ingredient quark-gluon and pion-nucleon fields. An
overview of these models is out of the scope of the present paper.

In order to demonstrate the suggested recipe of the construction of the
intermediate $\Delta$ propagators we consider the separable model for the $%
\pi N$ potential $V$ and $t$-matrix for the $P33$ partial wave in the c.m.
frame ${\bf p_N=-p_{\pi}=p}$

$$
V(p^{\prime},p)=\lambda g(p^{\prime})g(p);\ \ \  t(p^{\prime},p,E)={\frac{{%
\lambda g(p^{\prime})g(p)}}{{D(E)}}};\eqno(4.11a)
$$

$$
D(E)=\lambda\int {\frac{ {|g(p)|^2p^2dp}}{{E+i\epsilon - E_{{\bf p}%
_N}-\omega_{\pi}({\bf p}) } }},\eqno(4.11b)
$$
where the $t$-matrix satisfies the usual Lippmann-Schwinger equation which
is valid also for the $\pi N$  wave function

$$
<p^{\prime}|\Psi_p>={\frac{{\delta(p^{\prime}-p)}}{{p^{\prime}p} }}+\int {%
\frac{{V(p^{\prime},q) q^2dq}}{{E_{{\bf p}_N}+\omega_{\pi}({\bf p}%
)+i\epsilon- E_{{\bf q}_N}-\omega_{\pi}({\bf q}) }}} <q|\Psi_p>\eqno(4.12a)
$$

and has the following solution

$$
<p^{\prime}|\Psi_{p}>={\frac{{\delta(p^{\prime}-p)}}{{p^{\prime}p}}}+
\lambda {\frac{{g(p^{\prime})g(p)}}{{D\Bigl(E_{{\bf p}_N}+\omega_{\pi}({\bf p%
})\Bigr)} }}.\eqno(4.12b)
$$

Using the procedure (4.7a,b) we get

$$
|\Psi^{\pi N}_{{\bf p}_N{\bf p}_{\pi} }> {\stackrel{ E_{\pi N}\to M_{\Delta} 
}{\Longrightarrow}} |\Psi^{\Delta} >= \lambda {\frac{{g(p^{\prime})g(M_{%
\Delta})}}{{D(M_{\Delta }) } }}\eqno(4.13a)
$$

$$
|\Psi^{\pi N}_{{\bf p}_N{\bf p}_{\pi} }> {\stackrel{ E_{\pi N}\to
M^o_{\Delta}+\Sigma_{\Delta}(E) }{\Longrightarrow}} |\Psi^{\Delta}%
\Bigl(M^o_{\Delta}+\Sigma_{\Delta}(E)\Bigr) >= \lambda {\frac{ {%
g(p^{\prime})g(M^o_{\Delta}+\Sigma_{\Delta}(E) ) }}{{D\Bigl(M^o_{\Delta}+%
\Sigma_{\Delta}(E)\Bigr) } }}\eqno(4.13b)
$$

and instead of (4.11a) we obtain

\[
t(p^{\prime},p;E) \simeq <p^{\prime}|\biggl[ V+\sum_{\Delta}
V|\Psi_{\Delta}> {\frac{1}{{E-E_{\Delta} }}}<\Psi_{\Delta}|V\biggr]|p>
\]
$$
=\lambda g(p^{\prime})g(p)+\lambda^2 {\frac{{g(p^{\prime})g(p)|g(M_{%
\Delta})|^2 }}{{|D(M_{\Delta})|^2 } }} {\frac{1}{{E-M_{\Delta} } }}, %
\eqno(4.14a)
$$

or

$$
t(p^{\prime},p,E)\simeq \lambda g(p^{\prime})g(p)+\lambda^2 {\frac{{%
g(p^{\prime})g(p)|g(M^o_{\Delta}+\Sigma_{\Delta}(E))|^2}}{{%
|D(M^o_{\Delta}+\Sigma_{\Delta}(E))|^2 } }} {\frac{1}{{E-M^o_{\Delta}-%
\Sigma_{\Delta}(E) }}}, \eqno(4.14b)
$$

Comparing (4.14a) with (4.14b) we see that they have sufficiently different
form. The advantage of the representations (4.14a) and (4.14b) is that in
these expressions the propagation of the intermediate $\Delta$ is taken into
account exactly. In addition it is important to note, that the $g(p)$ and $%
D(E)$ functions can be constructed directly from the $\pi N$ phase shifts 
\cite{GM}. In the Section 6 we will consider the restrictions which
generates the unitarity condition for the $\Sigma_{\Delta}(E)$.

\vspace{5mm}

\begin{center}
{\bf 5. Three-body equation with the $\Delta$ degrees of freedom}
\end{center}

\vspace{5mm}

The main result of the previous section is the recipe of construction the
amplitudes sandwiched by the $\Delta$ resonance states. In particular, for
the amplitude of the reaction $\pi +N\to
\pi^{\prime}+N^{\prime}+\gamma^{\prime}$, according to Eq.(3.9) and
Eq.(4.5a), we get

$$
F^c_{4^{\prime},1}\equiv F^c_{\gamma^{\prime}\pi^{\prime}N^{\prime};\pi
N}\equiv -<in;{\bf p^{\prime}}_N{\bf p^{\prime}}_{\pi}|J_{\mu}(0) |{\bf p}_N%
{\bf p}_{\pi};in>_{connected}= \sum_{\sigma=1}^4
w^c_{\gamma^{\prime}\pi^{\prime}N^{\prime},\sigma} <in;\sigma|\Psi^{\pi N}_{%
{\bf p}_N{\bf p}_{\pi} }> .\eqno(5.1)
$$
In virtue of the extensions (4.9a,b) and Eq.(4.5b,c) we can obtain following
expressions for the $\Delta\Longleftrightarrow
\gamma^{\prime}\pi^{\prime}N^{\prime}$ transition amplitudes with on shell $%
\Delta$

$$
F^c_{\gamma^{\prime}\pi^{\prime}N^{\prime};\Delta}= -<in;{\bf p^{\prime}}_N%
{\bf p^{\prime}}_{\pi}|J_{\mu}(0) |\Psi^{\Delta}_{{\bf P}_{\Delta} }>=
\sum_{\sigma=1}^4 w^c_{\gamma^{\prime}\pi^{\prime}N^{\prime},\sigma}
<in;\sigma|\Psi^{\Delta}_{{\bf P}_{\Delta} }>, \eqno(5.2a)
$$

and for off shell $\Delta$

$$
F^c_{\gamma^{\prime}\pi^{\prime}N^{\prime};\Delta}(E)= -<in;{\bf p^{\prime}}%
_N{\bf p^{\prime}}_{\pi}|J_{\mu}(0) |\Psi^{\Delta}_{{\bf P}_{\Delta} }(E)>=
\sum_{\sigma=1}^4 w^c_{\gamma^{\prime}\pi^{\prime}N^{\prime},\sigma}
<in;\sigma|\Psi^{\Delta}_{{\bf P}_{\Delta} }(E)>, \eqno(5.2b)
$$

where instead of the two three-momenta ${\bf p}_N$ and ${\bf p}_{\pi}$ in
Eq.(5.1) we have only one three momentum ${\bf P}_{\Delta}={\bf p}_N+{\bf p}%
_{\pi}$, since the extensions (4.9a,b) imply the following transformations:

1. Projection of the $P33$ partial wave states which remove the dependence
on the relative $\pi N$ angles in Eq.(5.1).

2. Replacement of the relative $\pi N$ momenta by the expression $%
p^2=\Bigl((s-(m_N^2-m_{\pi}^2)(s-(m_N^2+m_{\pi}^2)\Bigr)/4s$, where $%
s=(m_{\Delta}+i\Gamma_{\Delta}/2)^2$ or $s=\Bigl[M^o_{\Delta}+\Sigma_{%
\Delta}(E,{\bf P}_{\Delta}={\bf 0})\Bigr]^2$. in the c.m. frame of the $\pi N
$ system. It is convenient to suppose that the $s$ variables in the above
extensions (4.7a,b) are real

$$
s=(m_{\Delta}+i\Gamma_{\Delta}/2)(m_{\Delta}+i\Gamma_{\Delta}/2)^*
=m_{\Delta}^2+{\frac{{\Gamma_{\Delta}^2}}{4}}\eqno(5.3a)
$$
$$
s=\Bigl[M^o_{\Delta}+\Sigma_{\Delta}(E,{\bf P}_{\Delta}={\bf 0})\Bigr] \Bigl[%
M^o_{\Delta}+\Sigma_{\Delta}(E,{\bf P}_{\Delta}={\bf 0})\Bigr]^* \eqno(5.3b)
$$
The advantage of the projection procedure with the real variables (5.3a,b)
is that the $u$, ${\overline u}$, ${\overline s}$ and other channel parts of
the potentials $U_{\alpha\beta}$ (3.4) or $w^c_{\alpha\beta}$ (2.13) do not
transform into complex potentials after this type projection and there will
not appear the nonphysical contribute in the unitarity condition.

From the $\Delta\to\gamma^{\prime}\pi^{\prime}N^{\prime}$ transition
amplitudes (5.2a) and (5.2b)  we can construct the $\Delta\to\gamma^{\prime}%
\Delta^{\prime}$ transition amplitude

$$
F_{\Delta^{\prime}\gamma^{\prime},\Delta}= -lim_{
E_{\pi^{\prime}N^{\prime}}\to E_{{\bf P^{\prime}}_{\Delta}} } <in;{\bf %
p^{\prime}}_N{\bf p^{\prime}}_{\pi}|J_{\mu}(0)|\Psi^{\Delta}_{{\bf P}%
_{\Delta} }> =lim_{ E_{\pi^{\prime}N^{\prime}}\to E_{{\bf P^{\prime}}%
_{\Delta}} } \sum_{\sigma=1}^4
w^c_{\gamma^{\prime}\pi^{\prime}N^{\prime},\sigma} <in;\sigma|\Psi^{\Delta}_{%
{\bf P}_{\Delta} }> \eqno(5.4a)
$$

\[
F_{\Delta^{\prime}\gamma^{\prime},\Delta}(E^{\prime},E)= -lim_{
E_{\pi^{\prime}N^{\prime}}\to \Bigl( E^o_{{\bf P^{\prime}}_{\Delta}}
+\Sigma_{\Delta}(E^{\prime},{\bf P^{\prime}}_{\Delta})\Bigr)} <in;{\bf %
p^{\prime}}_N{\bf p^{\prime}}_{\pi}|J_{\mu}(0) |\Psi^{\Delta}_{{\bf P}%
_{\Delta} }(E)>
\]
$$
= lim_{E_{\pi^{\prime}N^{\prime}}\to \Bigl( E^o_{{\bf P^{\prime}}_{\Delta}}
+\Sigma_{\Delta}(E^{\prime},{\bf P^{\prime}}_{\Delta})\Bigr) }
\sum_{\sigma=1}^4 w^c_{\gamma^{\prime}\pi^{\prime}N^{\prime},\sigma}
<in;\sigma|\Psi^{\Delta}_{{\bf P}_{\Delta} }(E)>, \eqno(5.4b)
$$
where also the projection on the $P33$ partial wave states is assumed.

Expression (5.4a) depends only on the two four momenta $P_{\Delta}=\Bigl(E_{%
{\bf P}_{\Delta}},{\bf P}_{\Delta}\Bigr)$ and $P_{\Delta}^{\prime}=\Bigl(E_{%
{\bf P^{\prime}}_{\Delta}},{\bf P^{\prime}}_{\Delta}\Bigr)$  and in
expressions (5.3b) there are also only two independent variables $%
P_{\Delta}=\Bigl(E^o_{{\bf P}_{\Delta}}+\Sigma_{\Delta}(E,{\bf P}_{\Delta}), 
{\bf P}_{\Delta}\Bigr)$ and $P_{\Delta}^{\prime}=\Bigl(E^o_{{\bf P^{\prime}}%
_{\Delta}}+ \Sigma_{\Delta}(E^{\prime},{\bf P^{\prime}}_{\Delta}),{\bf %
P^{\prime}}_{\Delta}\Bigr)$. From Eq.(5.4a,b) we see, that there exists two
ways of construction of the $\Delta^{\prime}-\gamma\Delta$ vertex functions:
the projection of the $\pi^{\prime}N^{\prime}-\gamma\pi N$ transition matrix
or the projection of the corresponding multichannel equation (3.4).

The functions $F_{\Delta^{\prime}\gamma^{\prime},\Delta}$ (5.4a) and $%
F_{\Delta^{\prime}\gamma^{\prime},\Delta}(E^{\prime},E)$ (5.4b) can be
treated as the covariant vertex functions with two on mass shell particles
which have masses $s$ and spin $3/2$. Therefore we can use the
representation of the $\Delta^{\prime}\gamma^{\prime}-\Delta$ vertices \cite
{Nozawa}

$$
F_{\Delta^{\prime}\gamma^{\prime},\Delta}({P^{\prime}}_{\Delta},{P}%
_{\Delta})= {\overline u}^{\sigma}(s^{\prime},{\bf P^{\prime}}_{\Delta})
V_{\sigma\mu\rho}(P^{\prime}_{\Delta}, P_{\Delta}) u^{\rho}(s,{\bf P}%
_{\Delta}) \eqno(5.5a)
$$

$$
V_{\sigma\mu\rho}( P^{\prime}_{\Delta}, P_{\Delta})= g_{\rho\sigma}\Bigl[
F_1(Q^2)\gamma_{\mu}+{\frac{{F_2(Q^2)}}{{2M_{\Delta} }}}R_{\mu}\Bigr]
+Q_{\sigma}Q_{\rho}\Bigl[{\frac{{F_3(Q^2)}}{{M_{\Delta}^2}}}\gamma_{\mu}+ {%
\frac{{F_4(Q^2)}}{{2 M_{\Delta}^3}}}R_{\mu}\Bigr],\eqno(5.5b)
$$
where $g_{\rho\sigma}$ is the metric tensor, $u^{\rho}(s,{\bf P}_{\Delta})$ 
denotes the spinor for the spin $3/2$ particle with mass $s$, $%
Q=P^{\prime}_{\Delta}-P_{\Delta}$ and $R=P^{\prime}_{\Delta}+P_{\Delta}$.
The form factors $F_i(Q^2)$ are simply connected with the charge monopole $%
G_{C0}(Q^2)$, the magnetic dipole $G_{M1}(Q^2)$, the electric quadrupole $%
G_{E2}(Q^2)$ and the magnetic octupole $G_{M3}(Q^2)$ form factors of the $%
\Delta$ resonance. The threshold values of these formfactors are determined
by the physical constants. For instance $G_{C0}(0)=e$ stands for the
electric charge and $G_{M1}(0)=\mu_{\Delta}$ can be used for the
determination of the magnetic moment of $\Delta$.

The relativistic field-theoretical construction of the $\Delta-\gamma^{%
\prime}\Delta^{\prime}$ vertex function is  significant by definition of the
magnetic moment of $\Delta$. In particular, if we use the quantum-mechanical
definition of $\mu_{\Delta}$ through the space components of the photon
current operator sandwiched by the $\Delta$ wave functions $|\Psi_{\Delta}>$ 
\cite{Heller}, then we obtain the complex magnitude for $\mu_{\Delta}$.
Moreover, if we calculate the magnetic moment of the nucleon in the
framework of this quantum-mechanical method, where we treat nucleon as the $%
\pi N$ cluster state for the $P11$ partial wave, then the resulting
effective $\mu_{N}$ will be complex too.

We emphasize that the considered field-theoretical definition of $%
\mu_{\Delta}$ through the vertex function $F_{\Delta^{\prime}\gamma^{%
\prime},\Delta}({P^{\prime}}_{\Delta},{P}_{\Delta})$  (5.5a) is analogue to
the accepted definition of the magnetic moment of nucleon \cite
{Schweber,BD,IZ,GrossB}. In particular, if we include the $s$-channel
one-nucleon exchange diagram in the second term of Eq.(2.8c) and Eq.(2.13)
(i.e. we take five intermediate states $\sigma=N,1,2,3,4$ instead of four $s$%
-channel terms), then we obtain the analog to $\Delta^{\prime}-\gamma\Delta$
vertex function $F_{N^{\prime}\gamma^{\prime},N}({P^{\prime}}_{N},{P}_{N})=-<%
{\bf p^{\prime}}|J_{\mu}(0)|{\bf p}>$  and $F_{N^{\prime}\gamma^{\prime},N}({%
P^{\prime}}_{N},{P}_{N})=\sum_{\sigma^{\prime}=N^{\prime\prime},\sigma}
w^c_{
N^{\prime}\gamma^{\prime},\sigma^{\prime}}<in;\sigma^{\prime}|\Psi^{N}_{{\bf %
P}_{N} }>$. Therefore, the other important difference between considered and
quantum-mechanical \cite{Heller} definitions of the $\Delta$ magnetic moment
is that in the considered definition are included the contributions of the
intermediate pion and nucleon  magnetic momenta.

Certainly, in the considered formulation the ambiguity by determination of
the explicit form of the intermediate $\Delta$ propagators in Eq.(4.2b) and
in Eq.(4.2c) according to the off shell extensions (4.7a,b) arises. For
instance, we can take $E_{{\bf P}_{\Delta}}= \sqrt{m^2_{\Delta}+{\bf P}%
_{\Delta}^2}+i\Gamma_{\Delta}/2$ for the on shell $\Delta$ and for the off
shell $\Delta$ we can choose $E_{{\bf P}_{\Delta}}=\sqrt{\Bigl(M^o_{\Delta}+
i\Sigma_{\Delta}(E,{\bf P}_{\Delta})/2 \Bigr)^2+ {\bf P}^2_{\Delta} }$. In
Ref.\cite{Lahiff} it was demonstrated, that the sufficient different values
for the coupling constants and cut-off parameters are necessary to use for
the description of the $\pi N$ phase shifts up to 360 MeV pion Laboratory
energy in the framework of the Bethe-Salpeter equations with the different
form of the $\Delta$ propagators. The strong sensitivity of the description
of the $\gamma p-\gamma p$, $\gamma p-\pi^o p$ and $\gamma p-\gamma\pi^o p$
reaction on the choice of the form of the intermediate $\Delta$ propagator
was shown also in our previous paper \cite{M5}, where the calculations were
performed in the Born approximation of the analogous to that considered here
a  three-dimensional field-theoretical equations. One can hope, that the
unified description of the multichannel $\gamma p-\pi N-\pi\pi N-\gamma \pi N
$ processes allows us to determine the form of $\Delta$ propagator.

Finally in this section we consider the representation of equation (3.4) in
the framework of the isobar model, where instead of the intermediate $\pi N$
and multimeson states we will take $\Delta$ and heavy meson $%
h=\sigma,\rho,\omega,..$ states. This means, that  the $\pi\pi N$ states are
replaced by $\Delta+\pi$ and $N+h$ nucleon and heavy meson states. As the
initial states we have $\beta=1,2\equiv\pi N, \gamma p$ states.

$$
{\cal T}_{\alpha\beta}(E_{\beta})=U_{\alpha\beta}(E_{\beta})+
\sum_{\Lambda=\Delta^{\prime\prime},\pi^{\prime\prime}\Delta^{\prime%
\prime},h^{\prime\prime}N^{\prime\prime}} U_{\alpha\Lambda}(E_{\beta}) {%
\frac{1}{{E_{\beta}-{\cal E}_{\Lambda}(E_{\beta}) } }} {\cal T}_{
\Lambda\beta} (E_{\beta}),\eqno(5.6)
$$

where we omitted intermediate states $\gamma N$ and $\gamma \pi N$ with
photon, because they are of higher order in $e^2$. ${\cal E}%
_{\Lambda}(E_{\beta})$ denotes the full energy of the intermediate $\Lambda$
cluster states

$$
{\cal E}_{\Lambda}(E_{\beta})= \left\{
\begin{array}{lll}
E^o_{{\bf P^{\prime\prime}}_{\Delta}}+\Sigma_{\Delta}(E_{\beta},{\bf %
P^{\prime\prime}}_{\Delta}) & \mbox{if $\Lambda=\Delta''$} &  \\ 
\omega_{\pi}({\bf p^{\prime\prime}}_{\pi})+ E^o_{{\bf P^{\prime\prime}}%
_{\Delta}}+\Sigma_{\Delta}(E_{\beta},{\bf P^{\prime\prime}}_{\Delta}) & %
\mbox{if $\Lambda=\pi''+\Delta''$} &  \\ 
E_{{\bf P^{\prime\prime}}_{N}}+ E^o_{{\bf P^{\prime\prime}}%
_{h}}+\Sigma_{h}(E_{\beta},{\bf P^{\prime\prime}}_{h}) & \mbox{if
$\Lambda=h''+N''\ (h=\sigma,\rho,\omega,...)$} & 
\end{array}
\right. \eqno(5.7)
$$

Now in order to derive the two-body equation for the ${\cal T}_{
\Lambda\beta}$ (4.6) transition amplitudes between the $\beta=\pi N,\gamma N$
and $\Lambda=\Delta,\pi \Delta,h+N$ states, we will use again the projection
procedure (4.7a,b) of the resonances $\Delta^{\prime}$ and $h^{\prime}$ from
the $\pi^{\prime}N^{\prime}$ and $\pi^{\prime}\pi^{\prime}$ final states.
Then $\alpha=1^{\prime},2^{\prime},3^{\prime},4^{\prime}\equiv
\pi^{\prime}N^{\prime},\gamma^{\prime}N^{\prime},\pi^{\prime}\pi^{\prime}N^{%
\prime},\gamma^{\prime}\pi^{\prime}N^{\prime}$  states will be replaced by

$$
\alpha=1^{\prime}\equiv \pi^{\prime}N^{\prime}\Longrightarrow
\Delta^{\prime};\ \ \ \ \  \alpha=3^{\prime}\equiv
\pi^{\prime}\pi^{\prime}N^{\prime}\Longrightarrow
\Bigl\{\pi^{\prime}\Delta^{\prime};\ h^{\prime}N^{\prime}\Bigr\}; \ \ \ \ \
\alpha=4^{\prime}\equiv \gamma^{\prime}\pi^{\prime}N^{\prime}\Longrightarrow
\gamma^{\prime}\Delta^{\prime}. \eqno(5.8)
$$

and we get

$$
{\cal T}_{\aleph\beta}(E_{\beta})=U_{\aleph\beta}(E_{\beta})+
\sum_{\Lambda=\Delta^{\prime\prime},\pi^{\prime\prime}\Delta^{\prime%
\prime},h^{\prime\prime}N^{\prime\prime}} U_{\aleph\Lambda}(E_{\beta}) {%
\frac{1}{{E_{\beta}-{\cal E}_{\Lambda}(E_{\beta}) } }} {\cal T}_{
\Lambda\beta} (E_{\beta}),\eqno(5.9)
$$

where $\aleph=\Delta^{\prime},\gamma^{\prime}N^{\prime},\pi^{\prime}\Delta^{%
\prime},h^{\prime}N^{\prime},\gamma^{\prime}\Delta^{\prime}=
\Lambda^{\prime},\gamma^{\prime}N^{\prime},\gamma^{\prime}\Delta^{\prime}$.


\vspace{5mm}


\begin{figure}[htb]
\centerline{\epsfysize=225mm\epsfbox{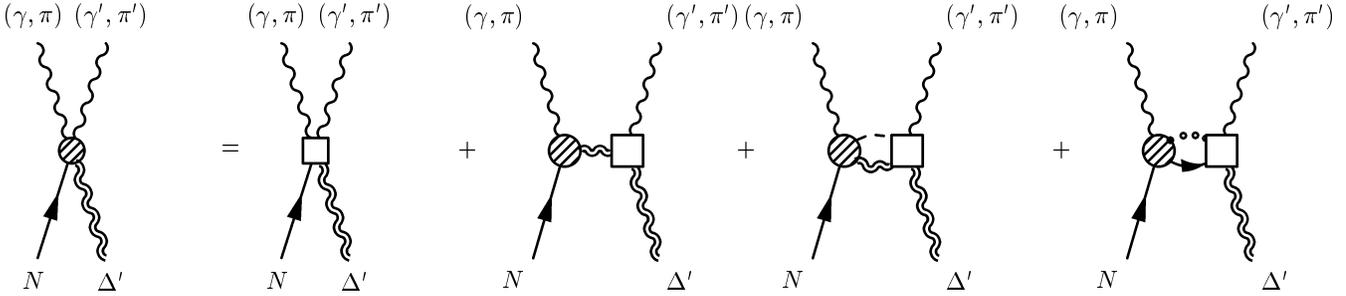}} 
\vspace{-17.0cm} 
\caption{{\protect\footnotesize {\it The graphical representation of
Eq.(5.9) for the $b+N\Longrightarrow a+\Delta^{\prime}$ transition amplitude
with the $\Delta,\ \pi \Delta, h\Delta$ intermediate states. }}}
\label{fig:seven}
\end{figure}
\vspace{5mm}

Substituting the solution of  the two-body equations (5.9) in Eq.(5.6), we
obtain the transition amplitudes into the three-body final states $\alpha
=\pi ^{\prime }\pi ^{\prime }N^{\prime },\gamma ^{\prime }\pi ^{\prime
}N^{\prime }$. Note, that one can obtain similar to Eq.(5.9) equation for $%
F_{\gamma ^{\prime }\pi ^{\prime }N^{\prime };\Delta }^{c}=-<in;{\bf %
p^{\prime }}_{N}{\bf p^{\prime }}_{\pi }|J_{\mu }(0)|\Psi _{{\bf P}_{\Delta
}}^{\Delta }>\Longrightarrow F_{\gamma ^{\prime }\Delta ^{\prime };\Delta
}^{c}=-<{{\bf P}_{\Delta }}|J_{\mu }(0)|\Psi _{{\bf P}_{\Delta }}^{\Delta }>$
from the quadratically-nonlinear, spectral decomposition formula (2.13)

$$
F_{\aleph\beta}=w_{\aleph\beta}+
\sum_{\Lambda=\Delta^{\prime\prime},\pi^{\prime\prime}\Delta^{\prime%
\prime},h^{\prime\prime}N^{\prime\prime}} F_{\aleph\Lambda} {\frac{1}{{%
E_{\beta}-{\cal E}_{\Lambda}(E_{\beta}) } }} F^*_{\beta \Lambda}\eqno(5.10)
$$

which allows us to operate with the $\Delta\to\gamma^{\prime}\Delta^{\prime}$
vertex function.

The structure of the two-body equations (5.9) is illustrated in Fig.7. At
first sight these equation have the same form as the relativistic equation
derived from the Bethe-Salpeter equation in the framework of the Aaron,
Amado and Young (AAY) model with the two-body separable amplitudes \cite
{AAY,GM}. However in our approach only the Green functions (4.2a,b,c) have
the separable form due to the resonance pole. The effective potential $%
U_{\aleph \Lambda }(E)$ consists of the sum of all connected diagrams
(depicted in Fig.2, Fig.3, Fig.4 and Fig.5), that are continued at the
resonance poles. After these continuations, $U_{\aleph \lambda }(E_{\beta })$
with three-body states $\lambda =\pi N,\pi \pi N$ transforms into two-body
form. Thus in Eq.(5.9) the two-body potential $U_{\aleph \lambda }(E_{\beta
})$ is constructed from the three particle potential $U_{\alpha \beta
}(E_{\beta })$ after the projection procedure (4.7a,b) to the resonance pole
position.

\vspace{5mm}

\begin{center}
{\bf 6. Unitarity and gauge invariance}
\end{center}

\vspace{3mm} {\bf A. Unitarity}

\vspace{5mm}

The three-dimensional quantum field-theoretical equations (2.13) have the
form of generalized unitarity conditions. Therefore for these multi-channel
equations the unitarity condition is fulfilled automatically. Moreover, the
equivalent linearized equations (3.4) with the potential $U_{\alpha\beta}(E)$
satisfies the unitarity condition also in the off energy shell region ($%
E=E_{\beta}$). However the unitarity condition for the two-body equations
(5.6), (5.9) and (5.10) with the complex propagators for the intermediate
resonance states need a special consideration. For this aim it is convenient
to rewrite  Eq.(3.4) in the form

$$
{\cal T}_{\alpha\beta}(E)=U_{\alpha\beta}(E)+
\sum_{\gamma=\pi^{\prime\prime}N^{\prime\prime},\pi^{\prime\prime}\pi^{%
\prime\prime}N^{\prime\prime}} {\cal T}_{\alpha\gamma}(E) {\frac{1}{{%
E-P_{\gamma}^o+i\epsilon } }} {\cal T}_{\beta \gamma}(E)^*,\eqno(6.1)
$$

which using the same approximation as by derivation of Eq.(5.6), takes the
form

$$
{\cal T}_{\alpha\beta}(E)\approx U_{\alpha\beta}(E)+
\sum_{\Lambda=\Delta^{\prime\prime},\pi^{\prime\prime}\Delta^{\prime%
\prime},h^{\prime\prime}N^{\prime\prime}} {\cal T}_{\alpha\Lambda}(E) {\frac{%
1}{{E-{\cal E}_{\Lambda}(E) } }} {\cal T}_{\beta \Lambda}(E)^*\eqno(6.2)
$$

From Eq.(6.1) and (6.2) we can  obtain the following condition for the $%
{\cal E}_{\Lambda}(E)$

$$
{\cal T}_{\alpha\beta}(E)- {\cal T}_{\beta\alpha}(E)^*= -2\pi
i\sum_{\gamma=\pi^{\prime\prime}N^{\prime\prime},\pi^{\prime\prime}\pi^{%
\prime\prime}N^{\prime\prime}}{\cal T}_{\alpha\gamma}(E) \delta\Bigl(
E-P_{\gamma}^o\Bigr) {\cal T}_{\beta \gamma}(E)^*\eqno(6.3a)
$$

$$
\approx
\sum_{\Lambda=\Delta^{\prime\prime},\pi^{\prime\prime}\Delta^{\prime%
\prime},h^{\prime\prime}N^{\prime\prime}} {\cal T}_{\alpha\Lambda}(E)\Bigl[ {%
\frac{1}{{E-{\cal E}_{\Lambda}(E) } }} -{\frac{1}{{E-{\cal E}_{\Lambda}(E)^* 
} }}\Bigr] {\cal T}_{\beta\Lambda}(E)^*.\eqno(6.3b)
$$

$$
\approx \sum_{\Lambda=\Delta,\pi\Delta,h
N}\sum_{\Lambda^{\prime}=\Delta^{\prime},\pi^{\prime}\Delta^{\prime},h^{%
\prime}N^{\prime}} {\cal T}_{\alpha\Lambda}(E)\Bigl[ {\frac{1}{{E-{\cal E}%
_{\Lambda}(E) } }} {\cal N}_{\Lambda\Lambda^{\prime}}(E) {\frac{1}{{E-{\cal E%
}_{\Lambda^{\prime}}(E)^* } }}\Bigr] {\cal T}_{\beta\Lambda^{\prime}}(E)^*.%
\eqno(6.3c)
$$

where

$$
{\cal N}_{\Lambda\Lambda^{\prime}}(E)= -2\pi
i\sum_{\gamma=\pi^{\prime\prime}N^{\prime\prime},\pi^{\prime\prime}\pi^{%
\prime\prime}N^{\prime\prime}} {\cal T}_{\gamma \Lambda}(E)^* \delta\Bigl(
E-P_{\gamma}^o\Bigr) {\cal T}_{\gamma\Lambda^{\prime}}(E)\eqno(6.3d)
$$

Equations (6.3c) and (6.3d) provides an simple relations for the
intermediate cluster propagators

$$
\delta_{\Lambda\Lambda^{\prime}}\Bigl[{\frac{1}{{E-{\cal E}_{\Lambda}(E) } }}
-{\frac{1}{{E-{\cal E}_{\Lambda}(E)^* } }}\Bigr] \approx {\frac{1}{{E-{\cal E%
}_{\Lambda}(E) } }} {\cal N}_{\Lambda\Lambda^{\prime}}(E) {\frac{1}{{E-{\cal %
E}_{\Lambda^{\prime}}(E)^* }}} \eqno(6.3e)
$$

which looks like the ``unitarity condition'' for the propagators of the
intermediate resonances. Unfortunately the validity of the condition (6.3e)
is not enough for the validity of the unitarity condition (6.3a) for the
observed amplitudes ${\cal T}_{\alpha \beta }(E)$. In particular, after
comparison of Eq.(6.3a) and Eq.(6.3b) we see that Eq.(6.3a) is given in the
half on energy shell region due to the $\delta \Bigl(E-P_{\gamma }^{o}\Bigr)$
function. Contrary to this in Eq.(6.3b) the intermediate states are
depending on $E$. This means, that the unitarity condition (6.3a) can be
approximately valid only for the resonance energies $E=E_{\Delta ^{\prime
\prime }}^{R}=\sqrt{m_{\Delta }^{2}+{\bf P^{\prime \prime }}_{\Delta }^{2}}$%
, $E=E_{\pi ^{\prime \prime }\Delta ^{\prime \prime }}^{R}\equiv \omega
_{\pi }({\bf p^{\prime \prime }})+\sqrt{m_{\Delta }^{2}+{\bf P^{\prime
\prime }}_{\Delta }^{2}}$ and $E=E_{h^{\prime \prime }N^{\prime \prime
}}^{R}\equiv E_{{\bf p^{\prime \prime }}_{N}}+\sqrt{m_{h}^{2}+{\bf P^{\prime
\prime }}_{h}^{2}}$ (see restrictions (4.8a,b)) and in the neighbor area of
these resonance energies.

For instance, from the (6.3b) we can obtain the following formula

$$
{\cal T}_{\alpha\beta}(E)- {\cal T}_{\beta\alpha}(E)^*\approx -2\pi
i\sum_{\Lambda=\Delta^{\prime\prime},\pi^{\prime\prime}\Delta^{\prime%
\prime},h^{\prime\prime}N^{\prime\prime}} {\cal T}_{\alpha\Lambda}(E)
\delta(E-E^R_{\Lambda}){\cal T}_{\beta\Lambda}(E)^*\eqno(6.4)
$$
if $\Sigma_{\Delta}$ and $\Sigma_{h}$ in Eq.(5.7) tend to zero as 
$$
lim_{E\to E^R}\Sigma_{\Delta}(E,{\bf P^{\prime\prime}}_{\Delta})\Rightarrow
i\epsilon\ \ and\ \ lim_{E\to E^R} \Sigma_{h}(E,{\bf P^{\prime\prime}}%
_{h})\Rightarrow i\epsilon\eqno(6.5)
$$
.

Thus, the unitarity condition (6.4) with the resonance amplitudes ${\cal T}%
_{\alpha\Lambda}(E)$ have the approximate form even in the resonance energy
region $E\approx E_r$  in spite of the conditions (4.8a,b). According to
Eq.(6.5), the approximation of the amplitudes or Green functions (4.2b) by
the resonance part only has an acceptable accuracy for  narrow resonances,
or if $\Gamma_r/E_r\ll 1$.

Unitarity condition (6.4) with the intermediate resonance states can be
improved if one takes into account the nonresonant contributions to the
resonant $\pi N$ and $\pi \pi $ interactions as it was done in Ref..\cite
{Lahiff}, where the contributions of the nonresonant Feynman diagrams for
the $\pi N$ $P33$ partial waves was investigated. Also in Ref..\cite{Blan}
the resonant ($\Delta $) and nonresonant parts of the $\pi N$ interactions
was separated proceeding from the effective three-dimensional Hamiltonian
method and the corresponding Lippmann-Schwinger and Dyson equations were
solved in the framework of the separable potential model. In the
nonrelativistic approach the multichannel equations with the intermediate
resonance and nonresonance parts and the explicit form of the corresponding
unitarity conditions were considered in book of Bohr and Mottelson.\cite
{BohrMot}.

\vspace{5mm} {\bf B. Gauge invariance}

\vspace{5mm}

The choice of {\bf the Coulomb gauge $\nabla^iA^c_i(x)=0$} for the
considered three-dimensional time-ordered formulation allows us  to exclude
the non-physical degrees of freedom of photons. In this way it is easy to
achieve current conservation and gauge invariance for Eq.(2.13) and (3.4)
even for truncated intermediate states.

According to the book of Bjorken and Drell \cite{BD}, in order to redefine
the above equations in the Coulomb gauge one must first redefine the
reduction formula for the $S$-matrix using transversal quantization rules,
transversal photon fields, transversal source operators etc. In particular,
the transversal photon current operator in Eq.(2.2) and (2.4a) can be
constructed from the initial four-dimensional current operator as

$$
J_{\mu=i=1,2,3}(x)\Longrightarrow J_{i}^{tr}(x)=J_{i}(x)- {\frac{{%
\nabla_{i}\partial_{o}}}{{\nabla^2}}}J_{o}(x).\eqno(6.6a)
$$

The conservation of the transverse current $\nabla^{i}J_{i}^{tr}(x)=0$
follows from the conservation of the four-dimensional current $%
\partial^{\mu}J_{\mu}(x)=0$. In the same manner one can construct the photon
field operator  in the Coulomb gauge ${\bf A}^c_i(x)$ from the photon field
operator $A_{\mu}(x)$ in the Lorentz gauge $\partial^{\mu}A_{\mu}=0$ 
$$
A_{\mu}(x)\Longrightarrow {\bf A}^c_{i=1,2,3}(x)=A_{i}(x)- {\frac{{%
\nabla_{i}\partial_{o}}}{{\nabla^2}}}A_{o}(x),\eqno(6.6b)
$$
where the time component of $A_{\mu}$ is determined through $J_{o}$ from the
equation of motion $\partial_{\nu}\partial^{\nu}A_{o}(x)=J_{o}(x)$ and
instead of the Eq.(2.4a) we have $\partial^{\nu}\partial_{\nu}{\bf A}%
_{i}^{c}(x)=J_{i}^{tr}(x)$. The quantization rules of the photon field
operators in the Coulomb gauge are \cite{BD}

$$
\Bigl[\partial_{x_o}{\bf A}^c_{i}(x),{\bf A}^c_{j}(y)\Bigr]_{x_o=y_o}=-i
\delta^{tr}_{ij}({\bf x-y})\eqno(6.6c)
$$

where $\delta^{tr}_{ij}({\bf x})
=(\delta_{ij}-\nabla_i\nabla_j/\nabla^2)\delta({\bf x})$.

The redefined photon current operator (6.6a) and the quantization rules
(6.6c) allow us to rewrite the $S$-matrix reduction formulas and
corresponding equations (2.1) and (2.2) with the $J_{i}^{tr}(x)$ and
equal-time commutators between the photon field operators in the Coulomb
gauge \cite{BD}. The form of the three-dimensional time-ordered equations
(2.13) or (3.4) with the fields in the Coulomb gauge remains the same. Now
one can demonstrate the validity of the current conservation condition and
invariance under the gauge transformation of the photon field operator

$$
{\bf A}^c_{i}(x)={\bf A}^c_{i}(x)+\nabla_i\Lambda(x)\ \ \  or\ \ \
A_{\mu}^{\prime}(x)=A_{\mu}(x)+\partial_{\mu}\Lambda(x) \eqno(6.7a)
$$

or similarly, invariance under the gauge transformation of photon 
polarization vector

$$
{\epsilon^{\eta}_{i}}^{\prime}({\bf k})= \epsilon^{\eta}_{i}({\bf k}%
)+\lambda {\bf k}_{i}\ \ \  or\ \ \ {\epsilon^{\eta}_{\mu}}^{\prime}({\bf k}%
)= \epsilon^{\eta}_{\mu}({\bf k})+\lambda {\bf k}_{\mu} \eqno(6.7b)
$$
for an arbitrary $\Lambda(x)$ and $\lambda$. In particular, due to
three-momentum conservation at every vertex function in the considered
approximation, one can show, that every transition matrix $%
<n|J^{tr}_{\mu}(0)|m>$ with arbitrary $n$ and $m$ $in(out)$ states  in the
effective potential or in the amplitude of Eq.(2.13) or Eq.(3.4)  satisfies
the current conservation condition 
$$
0=<n|i\nabla^j J^{tr}_{j}(0)|m> =({\bf P_N-P_m})^j<n|J^{tr}_{j}(0)|m> ={\bf k%
}_{\gamma}^j<n|J^{tr}_{j}(0)|m>\eqno(6.8)
$$

where according to (6.6a) 
$$
<n|J_{i}^{tr}(0)|m>= <n|J_{i}(0)|m>- {\frac{ \Bigl({\bf P_n-P_m}%
\Bigr)_i(P^o_n-P^o_m) }{{\ \Bigl({\bf P_n-P_m}\Bigr)^2 }}} <n|J_{o}(0)|m>.%
\eqno(6.9)
$$

Thus the three-momentum conservation condition in Eq.(2.13) or Eq.(3.4)
allows us to reduce the action of ${\bf k}_{\gamma}^i$ to the action of $%
i\nabla^i$ on the photon current operator $J_{i}^{tr}(0)$ for the every term
of Eq.(2.13) or (3.4). For example, for the second $u$-channel term in
Eq.(2.9a) with $a=\gamma^{\prime}$ and $b=\gamma$ in virtue of Eq.(6.8) we
have

\[
{\bf k}_{\gamma}^j\sum_{n=N,\pi N,...} <out;{\widetilde \alpha}%
|J^{tr}_{i}(0)|n;in> {\frac{{\delta^{(3)}( {-{\bf k}}_{\gamma}+{\bf P_{{%
\widetilde \alpha}}-P_{n} })}}{{-|{\bf k}_{\gamma}|+ P_{{\widetilde \alpha}%
}^o-P_{n}^o } }} <in;n|J^{tr}_j(0)|{\widetilde \beta};in>
\]
\[
=\sum_{n=N,\pi N...} <out;{\widetilde \alpha}|J^{tr}_{i}(0)|n;in> {\frac{{%
\delta^{(3)}( {-{\bf k}}_{\gamma}+{\bf P_{{\widetilde \alpha}}-P_{n} })}}{{-|%
{\bf k}_{\gamma}|+ P_{{\widetilde \alpha}}^o-P_{n}^o } }} <in;n|i\nabla_{%
{\bf x_j}}J^{tr}_j(0)|{\widetilde \beta};in>=0 
\]


In the same manner for the equal-time commutators we get

$$
{\bf k}_{\gamma}^j <out;{\widetilde \alpha}|\Biggl[J^{tr}_{i}(0),{a}^+_{j}(0)%
\Biggr]
|{\widetilde \beta};in>= <out;{\widetilde \alpha}|\Biggl[
J^{tr}_{i}(0),{\bf k}_{\gamma}^j{a}^+_{j}(0)\Biggr]
|{\widetilde \beta};in>=0.\eqno(6.10)
$$

Thus we have demonstrated that 
\[
{\bf k}_{\gamma }^{j}\Bigl[f_{\gamma ^{\prime }{\widetilde{\alpha }};\gamma {%
\widetilde{\beta }}}\Bigr]_{ij}=-{\bf k}_{\gamma }^{j}<out;{\widetilde{%
\alpha }}|J_{i}^{tr}(0)|{\bf k}_{\gamma }j;{\widetilde{\beta }};in>
\]
$$
={\bf k}_{\gamma }^{j}\Bigl[W_{\gamma ^{\prime }{\widetilde{\alpha }};\gamma 
{\widetilde{\beta }}}+(2\pi )^{3}\sum_{\sigma =1}^{4}f_{\gamma ^{\prime }{%
\widetilde{\alpha }};\sigma }{\frac{{\delta ^{(3)}({\bf k}_{\gamma }+{\bf P_{%
{\widetilde{\beta }}}-P_{\gamma }})}}{{|{\bf k}_{\gamma }|+{P_{{\widetilde{%
\beta }}}^{o}-P_{\gamma }^{o}+i\epsilon }}}}{F^{*}}_{\gamma {\widetilde{%
\beta }};\sigma }\Bigr]_{ij}=0\eqno(6.11a)
$$
for an arbitrary Lagrangian by calculation of the equal-time commutator
(2.15) and any number of the intermediate states in $W_{\alpha \beta }$
(2.9a). Note that equations (2.8a), (2.9a), (2.13), (3.4) etc. are not
depend on the three-momentum ${\bf p^{\prime }}_{a}={\bf k^{\prime }}%
_{\gamma }\ if\ a=\gamma ^{\prime }$. Therefore we can assume, that ${\bf %
k^{\prime }}_{\gamma }={\bf k}_{\gamma }+{\bf P}_{\widetilde{\beta }}-{\bf P}%
_{\widetilde{\alpha }}$. Then in the same way as (6.11a) we obtain

\[
{\bf k^{\prime}}_{\gamma}^i \Bigl[f_{\gamma^{\prime}{\widetilde \alpha}%
;\gamma{\widetilde \beta}}\Bigr]_{ij} =-{\bf k^{\prime}}_{\gamma}^i <out;{%
\widetilde \alpha}|J^{tr}_{i}(0)| {\bf k}_{\gamma} j;{\widetilde \beta}; in>
\]
$$
={\bf k^{\prime}}_{\gamma}^i\Bigl[ W_{\gamma^{\prime}{\widetilde \alpha}%
;\gamma{\widetilde \beta}} +(2\pi)^3\sum_{\sigma=1}^4 f_{\gamma^{\prime}{%
\widetilde \alpha};\sigma}{\frac{{\delta^{(3)}( {\bf k}_{\gamma}+{\bf P_{{%
\widetilde \beta}}-P_{\gamma} })}}{{|{\bf k}_{\gamma}|+{\ P_{{\widetilde
\beta}}^o-P_{\gamma}^o+i\epsilon }} }} {F^{\ast}}_{\gamma {\widetilde \beta}%
; \sigma}\Bigr]_{ij}=0. \eqno(6.11b)
$$

This completes the proof of the current conservation condition of the above
equations in the Coulomb Gauge. It is easy to see that the current
conservation condition (6.11a,b) for the both sides of Eq.(2.13) or Eq.(3.4)
is sufficient for the validity of the invariance of these equation under the
gauge transformation (6.7b) for ${\epsilon _{\mu }^{\eta }}^{\prime }({\bf k}%
)$. The gauge transformation (6.7a) ${\bf A}_{i}^{c}(x)={\bf A}%
_{i}^{c}(x)+\nabla _{i}\Lambda (x)$ leads to the gauge transformation (6.7b)
if we taken into account the conditions $\nabla ^{2}\Lambda (x)=\nabla ^{i}%
{\bf A}_{i}^{c}(x)$ \cite{BD,GrossB}. In the general sense the gauge
invariance means the independence of results on the gauge used. In other
words, if we represent the gauge condition as $n^{\mu }A^{\mu }(x)=0$ with $%
n_{\mu }=k^{\mu }$ for the Lorentz gauge, $n_{\mu }=(0,{\bf k})$ for the
Coulomb gauge, $n_{\mu }=(k_{o},0,0,0)$ for the axial gauge etc., then gauge
invariance reduces to the independence of the observables on the choice of $%
n^{\mu }$ . The proof of this general gauge invariance is out of the scope
of our paper.

\vspace{5mm}

\begin{center}
{\bf 7. Conclusion}
\end{center}

\vspace{5mm}

In this paper we have derived three-dimensional covariant scattering
equations for the coupled system of the amplitudes of the $\pi
N\Longleftrightarrow \gamma N\Longleftrightarrow \pi \pi
N\Longleftrightarrow \gamma \pi N$ reactions. The basis of these three-body
relativistic equations is the standard field-theoretical $S$-matrix
reduction formulas. After decomposition over the complete set of the
asymptotic $^{\prime\prime}in^{\prime\prime}$ states the quadratically
nonlinear three-dimensional equations (2.13) were obtained.

These equations were replaced by the equivalent Lippmann-Schwinger type
equation (3.4). Unlike the three-body Faddeev-type equations, the potentials
of the suggested three-body equations (3.4) consists of the connected parts
only and thereby they have the form of the relativistic Lippmann-Schwinger
type equations with the well defined connected three-body potential. This
sufficient difference follows from the field-theoretical derivation of the
considered equations, where the disconnected parts of the three-body
amplitudes (2.2) $f_{\alpha\beta}=-<out;{\widetilde \alpha}|j_{a}(0)|\beta;
in>$ coincide with the disconnected part from the right side of these
equations,i.e. Eq.(2.2) or Eq.(2.8a) or Eq.(2.13) consist of independent
sets of equations for the connected and disconnected parts of amplitudes. In
particular, if we note, that the potentials of Eq.(3.4) (or Eq.(2.13))
consist of the product of the two renormalized (physical) amplitudes or
vertex functions with the corresponding propagator of the particles , then
it is obvious, that  the graphical method of Taylor, based on the last cut
lemma, does not work in this approach.  The analytical cluster decomposition
leads to independent sets for the disconnected amplitudes in Eq.(2.8a) or in
Eq.(2.13).

The potentials of the suggested equations require as their input the vertex
functions with two on mass shell particles. For the two-body $\pi
N\Longleftrightarrow \gamma N$ reactions these input functions are exactly
the phenomenological one variable  vertex functions that can be determined
from dispersion relations or from two-body observables. The equal-time
commutators (2.15) offer different opportunities to investigate the
off-shell effects resulting from chosen model Lagrangian. Thus the number of
the one off-mass shell particle exchange diagrams and contact (overlapping)
terms in expression (2.15) (see Fig.5) depend on the model Lagrangian.
Certainly, if one includes the higher order derivatives and some nonlocal
effects in the effective Lagrangians, then in the equal-time commutators
(2.15) arise numerous contact terms which will not be easy to take into
account. Usually, in practical calculations simple Lagrangians which
generate a minimal number of contact terms are used. On the other hand these
simple Lagrangians with ``effective'' $\sigma,\rho,\omega,...$-particle
degrees of freedom could help us to estimate the more complicated effects
contained in the much more complicated Lagrangians like the Lagrangian of
the nonlinear $\sigma$ model with vector mesons, some QCD motivated
Lagrangians etc. Thus, if it is possible to describe the connected $\pi
N\Longleftrightarrow \gamma N\Longleftrightarrow \pi \pi
N\Longleftrightarrow \gamma \pi N$  reactions using some simple Lagrangian,
then one can find a number of more complicated Lagrangians which lead to the
description of the same data. On the other hand, one can, in principle,
construct a two-body potential,  coming from the equal-time commutators
using the inverse scattering methods \cite{Ml}. This link between the
effective Lagrangians and the potentials of the solved field-theoretical
equations which arise due to the equal-time commutators, can be considered
as an additional tool for the investigation of the correlations between a
class of Lagrangians and the calculated experimental data.

The present three-dimensional formulation is not more complicated than the
Bethe-Salpeter equations. In the four-dimensional formulation intermediate
particle and anti-particle degrees of freedom are combined in the same
diagram. Therefore on the tree-level approximation Bethe-Salpeter equations
are more simple. But if one wants to use renormalized physical amplitudes
and vertices and if one wants to take into account the rescattering effects,
then the suggested formulation is simpler, because the corresponding
equations are three-dimensional from the beginning and in the amplitudes and
in the vertex functions  two of the external particles are on mass shell.


The important features of the suggested three-body field-theoretical
equations are the following:

1. {\bf Unitarity}. The final three-body equations (3.4) are equivalent to
the nonlinear equations (2.13) which fulfill the off shell unitarity
conditions in the nonrelativistic collision theory \cite{New,Gold}.
Therefore these equations automatically satisfy the unitarity conditions for
the complete set of equations with the infinity intermediate $|n;in>$ states
or for the truncated set of equations with two and three particle
intermediate states $n$. It is important to note, that unlike the
four-dimensional Bethe-Salpeter equations, in the considered approach only
the on-mass shall physical $^{\prime\prime}in(out)^{\prime\prime}$ states
are truncated in the completeness condition $|n;in(out)><in(out);n|={\hat 1}$%
.

2. {\bf Current conservation conditions}. In the considered formulation with
the Coulomb gauge we have demonstrated current conservation and invariance
of Eq. (2.13) and Eq.(3.4) and corresponding potentials under gauge
transformations (6.7a) or (6.7b) for every number of the truncated
intermediate states. In order to achieve gauge invariance in this
formulation it is not necessary to use additional approximations like the
tree approximation with a gauge invariant combination of terms or the
construction of approximate auxiliary gauge-invariance-preserving currents 
or to use the special representation of the off-mass shell $\Delta$
propagator and the corresponding construction of the gauge invariant
electromagnetic $\Delta$ vertex function. The only requirement which is
necessary in the considered approach to attain the invariance under the
gauge transformation (6.7a) or (6.7b) is the existence of the conserved
currents $J_{\mu}(x)$ (2.4a) and the corresponding model Lagrangian.


Another aspect of the considered three-body equations is the intermediate $%
\Delta $ resonance degrees of freedom. We have construct the $\pi
N\Longleftrightarrow \Delta $ wave functions using the extension of the
corresponding Lippmann-Schwinger equations in the complex region for the
observed (physical) $\Delta $ pole position (i.e. for the Breit-Wigner mass $%
m_{\Delta }=1232MeV$ and full width $\Gamma _{\Delta }=120MeV$). We
considered two cases, (i) with the ``on shell $\Delta $'', when $E_{{\bf P}%
_{\Delta }}=\sqrt{{M_{\Delta }}^{2}+{\bf P}_{\Delta }^{2}}$ and $M_{\Delta
}=m_{\Delta }+i\Gamma _{\Delta }/2$ (4.7a) and (ii) ``off shell $\Delta $'',
where $E_{{\bf P}_{\Delta }}=E_{{\bf P}_{\Delta }}^{o}+\Sigma _{\Delta }(E_{%
{\bf P}_{\Delta }},{\bf P}_{\Delta })$ (4.7b). These $\pi
N\Longleftrightarrow \Delta $ wave functions were used for the construction
of the intermediate full $\pi N$ Green function with the $\Delta $ resonance
pole (4.2b,c) and for the equation of motion with intermediate $\Delta $
degrees of freedom (see Eq.(4.10a,b), (5.6) and (5.9)). Moreover, the same
extension procedure generates the $\Delta \gamma -\Delta $ vertex functions
(5.5a,b), that can be considered as the generalization of the $N\gamma
-N^{\prime }$ vertex function. Thus in contrast to other formulations we
have not introduced the effective spin $3/2$ Lagrangian in order to
introduce the intermediate $\Delta ^{\prime }s$, where additional conditions
are necessary in order to determine the actual off-mass shell behavior of
the amplitude.

\begin{center}
{\bf Appendix A: Equivalence of quadratically  nonlinear equations (2.10)
and Lippmann-Schwinger  type equations (3.4) with the linear energy 
depending potential (3.5). }
\end{center}

\bigskip

By solving of the Lippmann-Schwinger type equations (3.4) we can define the
corresponding wave function

$$
<in;{\alpha}|\Psi_{\beta}>= <in;{\alpha}|{\beta};in> + {\frac{1}{{%
E_{\beta}-E_{\alpha}+i\epsilon}}} \sum_{\gamma=1}^4
U_{\alpha\gamma}(E_{\beta}) <in;\gamma|\Psi_{\beta}>.\eqno(A.1)
$$

which satisfies the Schr\"odinger equation

$$
\Bigl(H_o+U(E_{\beta})\Bigr)|\Psi_{\beta}>=E_{\beta}|\Psi_{\beta}> .%
\eqno(A.2)
$$

where in accordance with the definition of the free Hamiltonian, we have $%
H_o|\beta;in>=E_{\beta}|\beta;in>.$

If we taken into account the energy dependence of potential (3.5) $U_{\alpha
\beta }(E_{\beta })=A_{\alpha \beta }+E_{\beta }\ B_{\alpha \beta }$, then
we can rewrite Eq. (A.2) as follows

$$
\Bigl(H_o+A\Bigr)|\Psi_{\beta}>=E_{\beta}\Bigl(1-B\Bigr) |\Psi_{\beta}>.%
\eqno(A.3)
$$

Below we shall assume that exists $(1-B)^{-1}$. i.e. the operator $1-B$ is
smooth enough. This allows us represent Eq. (A.3) in the form of the Schr%
\"{o}dinger equation with Hermitian Hamiltonian $h$

$$
h|\chi_{\beta}>=E_{\beta}|\chi_{\beta}>\eqno(A.4a)
$$

$$
h=(1-B)^{-1/2}(H_o+A)(1-B)^{-1/2}\eqno(A.4b)
$$

$$
|\chi_{\beta}>=(1-B)^{1/2}|\Psi_{\beta}>\eqno(A.4c)
$$

From the hermiticity of the Hamiltonian $h$ there follows the conditions of
the completeness and ortho-normality of its eigenstates $|\chi_{\beta}>$.
Therefore the state vectors $|\Psi_{\beta}>$ satisfy following completeness
conditions

$$
\sum_{\gamma=1}^4 |\Psi_{\gamma}><\Psi_{\gamma}|=(1-B)^{-1}\eqno(A.5a)
$$

and of the ortho-normality

$$
<\Psi_{\alpha}|(1-B)|\Psi_{\beta}>=\delta_{\alpha\beta}\eqno(A.5b)
$$

Now if we insert the identity $U_{\alpha\beta}(E_{\beta})=W_{\alpha\beta}+%
\Bigl(E_{\beta}-E_{\alpha}\Bigr) B_{\alpha\beta}$ in Eq.(A.2) and use
Eq.(3.9) $F_{\alpha\beta}= \sum_{\gamma=1}^4
W_{\alpha\gamma}<in;\gamma|\Psi_{\beta}>$, then after some simple algebraic
transformations we obtain

$$
<in;\alpha|(1-B)|\Psi_{\beta}>= <in;\alpha|\beta;in> + {\frac{1}{{%
E_{\beta}-E_{\alpha}+i\epsilon}}} F_{\alpha\beta},\eqno(A.6a)
$$

and

$$
<\Psi_{\alpha}|(1-B)|\beta;in>= <in;\alpha|\beta;in> - {\frac{1}{{%
E_{\beta}-E_{\alpha}+i\epsilon}}} F_{\beta\alpha}^*.\eqno(A.6b)
$$

Here we multiply relation (A.6b) by $F_{\sigma\alpha}=<in;\sigma|W|\Psi_{%
\alpha}>$ and using the completeness condition (A.5a), after summation over $%
\alpha$ we get

$$
W_{\sigma\beta}=F_{\sigma\beta}- \sum_{\alpha=1}^4 F_{\sigma\alpha}{\frac{1}{%
{E_{ \beta}-E_{\alpha}+i\epsilon } }} F_{\beta \alpha}^{\ast},\eqno(A.7)
$$

Equation (A.7) is identical with Eq.(2.10). The same derivation of the
quadratically-nonlinear integral equations with bound states from the
Lippmann-Schwinger equation  for the one-channel case, was given in \cite{M1}%
.

\medskip

\begin{center}
{\bf Appendix B: The explicit form  of the potentials $w^c_{\alpha\beta}$ 
and $U_{\alpha\beta}(E)$ in the multichannel  equations (2.13) and (3.4)}
\end{center}

\medskip

In order to determine the structure of the potentials $w^c_{\alpha\beta}$
(2.13 or $U_{\alpha\beta}(E)$ (3.4) firstly we must separate the connected
and disconnected parts from $W_{\alpha\beta}$ (2.9a). In the present
formulation this transformation is equivalent to the so called $%
^{\prime\prime}cluster \ decomposition^{\prime\prime}$ procedure \cite
{alf,Ban,M1} which allows us to distinguish the equations for the connected
parts of scattering amplitudes $f^c_{\alpha\beta}$ and for the connected
parts of the effective potential $w^c_{\alpha\beta}$. In the other words, we
will represent $w^c_{\alpha\beta}$ and corresponding $U_{\alpha\beta}(E)$ in
the terms of the connected parts of the transition matrix elements only. In
particular, from Eq.(2.2) after insertion of the complete $%
^{\prime\prime}in^{\prime\prime}$ states we get

\[
f^c_{\alpha\beta}= \biggl\{
-<out;{\widetilde \alpha}|\Bigl[j_{a}(0),a_{b}^+(0)\Bigr]|{\widetilde \beta}%
;in>
\]
$$
+(2\pi)^3 \sum_{n=N,\pi N,\gamma N, \pi\pi N,\pi \gamma N, 3\pi N,...} <out;{%
\widetilde \alpha}|j_{a}(0)|n;in> {\frac{{\delta^{(3)}( {\bf p}_b+{\bf P_{{%
\widetilde \beta}}-P_{n} })}}{{\omega_b({\bf p_b})+{\ P_{{\widetilde \beta}%
}^o-P_{n}^o+i\epsilon }} }} <in;n|j_{b}(0)|{\widetilde \beta};in>\eqno(B.1a)
$$
\[
+(2\pi)^3\sum_{n=N,\pi N,\gamma N, \pi\pi N,\pi\gamma N, 3 \pi N,...} <out;{%
\widetilde \alpha}|j_{b}(0)|n;in>{\frac{{\delta^{(3)}( {-{\bf p}}_b+{\bf P_{{%
\widetilde \alpha}}-P_{n} })}}{{-\omega_b({\bf p_b})+ P_{{\widetilde \alpha}%
}^o-P_{n}^o } }} <in;n|j_{a}(0)|{\widetilde \beta};in>\biggr\}^c
\]
Equations (2.8a) and (2.9a) come out from (B.1a) after separation of the $%
n=\pi N,\gamma N, \gamma \pi N, \pi\pi N$ states. For the amplitude $%
F_{\alpha\beta}$ (2.10) all $^{\prime\prime}out^{\prime\prime}$ states are
replaced by the $^{\prime\prime}in^{\prime\prime}$ states and we have

\[
F^c_{\alpha\beta}=\biggl\{
-<in;{\widetilde \alpha}|\Bigl[j_{a}(0),a_{b}^+(0)\Bigr]|{\widetilde \beta}%
;in>
\]
$$
+(2\pi)^3 \sum_{n=N,\pi N,\gamma N, \pi\pi N,\pi \gamma N, 3\pi N,...} <in;{%
\widetilde \alpha}|j_{a}(0)|n;in> {\frac{{\delta^{(3)}( {\bf p}_b+{\bf P_{{%
\widetilde \beta}}-P_{n} })}}{{\omega_b({\bf p_b})+{\ P_{{\widetilde \beta}%
}^o-P_{n}^o+i\epsilon }} }} <in;n|j_{b}(0)|{\widetilde \beta};in>\eqno(B.1b)
$$
\[
+(2\pi)^3\sum_{n=N,\pi N,\gamma N, \pi\pi N,\pi\gamma N, 3 \pi N,...} <in;{%
\widetilde \alpha}|j_{b}(0)|n;in> {\frac{{\delta^{(3)}( {-{\bf p}}_b+{\bf P_{%
{\widetilde \alpha}}-P_{n} })}}{{-\omega_b({\bf p_b})+ P_{{\widetilde \alpha}%
}^o-P_{n}^o } }} <in;n|j_{a}(0)|{\widetilde \beta};in>\biggr\}^c 
\]

The special case of the disconnected parts of the amplitude $<in;{\widetilde
\alpha}|j_{a}(0)|\beta;in>$  are defined according to Eq.(2.6a,b,c), 
Eq.(2.7a,b) and they are depicted in the Fig.1. The separation of the
disconnected parts for the amplitude $<in;{\widetilde \alpha}%
|j_{a}(0)|n^{\prime\prime};in>$ with asymptotic ${\widetilde \alpha}$ and $N$
states means 
\[
<in;{\widetilde \alpha}|j_{a}(0)|n^{\prime\prime};in>= <in;{\widetilde \alpha%
}|j_{a}(0)|n^{\prime\prime};in>_c+
\]
$$
<in;N^{\prime}|N;in><in;{\widetilde \alpha}_{N^{\prime}}|j_{a}(0)|{%
\widetilde n}^{\prime\prime};in>_c +<in;\pi^{\prime},|{\pi}";in>
<in;N^{\prime}|j_{a}(0)|{\widetilde n}";in>_c \eqno(B.2)
$$
where the subscript $^{\prime\prime}c^{\prime\prime}$ stands for the
connected part of the amplitude, $<in;N^{\prime}|$ denotes the one paticle
nucleon $N^{\prime}$ state and the second term with disconnected in Eq.(B.2)
arise only for the three-particle states, when $<in;\alpha| \equiv
<in;a,\pi^{\prime}N^{\prime}|$, since $a=\pi^{\prime}\ or\ \gamma^{\prime}$
and $|n^{\prime\prime};in>=|{\widetilde n}^{\prime\prime}\pi^{\prime%
\prime};in>$.

The cluster decomposition for the expression (B.1b) is the same as exclusion
of the disconnected parts for the product of the amplitudes according to
Eq.(B.2). Thus, after the separation of the connected parts in the second
and third parts of Eq.(B.1b) according to the (B.2), we obtain the following
representation for the first connected $s$ channel term in (B.1b)

\[
\sum_{n=N,\pi N, \pi\pi N,3\pi N,...} <in;{\widetilde \alpha}|j_{a}(0)|n;in> 
{\frac{{\delta^{(3)}( {\bf P}_{\widetilde \beta}+{\bf p}_b-{\bf P}_{n} )}}{{%
\ \omega_b({\bf p}_b)+ P^o_{\widetilde \beta} -P_{n}^o+i\epsilon } }}
<in;n|j_{b}(0)|{\widetilde \beta};in>
\]
\[
=\sum_{n=N,\pi N,2\pi N,...} <in;{\widetilde \alpha}|j_{a}(0)|n;in>_c {\frac{%
{\delta^{(3)}( {\bf P_{\widetilde b}}+{\bf p}_b-{\bf P}_{n} )}}{{\omega_b(%
{\bf p_b})+P^o_{\widetilde b} -P_{n}^o+i\epsilon }}} <in;n|j_{b}(0)|{%
\widetilde \beta};in>_c+
\]
\[
\sum_{m=mesons,...} <in;{\widetilde \alpha}_{N^{\prime}}|j_{a}(0)|m;in>_c {%
\frac{{\delta^{(3)}( {\bf P}_{\widetilde \beta}+{\bf p}_b -{\bf P}_{m}-{\bf %
p^{\prime}}_N )} }{{\omega_b({\bf p_b})+ P^o_{\widetilde b}+ -P_{m}^o -E_{%
{\bf p^{\prime}}_N}+i\epsilon } }} <in;m,N^{\prime}|j_{b}(0)|{\widetilde
\beta};in>_c
\]
$$
+\ terms\ with\ the\ all\ other\ transposition\ of\ {\widetilde a}%
,N^{\prime}\ and\ {\widetilde b},N\eqno(B.3)
$$
where $p^{\prime}_N=\Bigl(E_{{\bf p^{\prime}}_N}=\sqrt{m_N^2+{{\bf p^{\prime}%
}_N}^2}, {\bf p^{\prime}}_N\Bigr)$ stands for the four-momentum of the
nucleon and $P_{\widetilde \beta}=\Bigl(P^o_{\widetilde \beta}, {\bf P}%
_{\widetilde \beta}\Bigr)$ is the four-momentum of the asymptotic ${%
\widetilde \beta}$ state.

Next we will determine the explicit form of the $w^c_{\alpha\beta}$ in
Eq.(2.13) and the corresponding $U_{\alpha\beta}((E)$ (2.13) separately for
the $2\to 2^{\prime}$, $2\to 3^{\prime}$ and $3\to 3^{\prime}$ reactions

\bigskip

{\bf For the two-body $\pi N\Longleftrightarrow \gamma N$ reactions} 
cluster decomposition is the same as the transposition of the nucleon from
the asymptotic $<{\widetilde \alpha}={\bf p^{\prime}}_N|$ or $|{\widetilde
\beta}={\bf p}_N>$ states using the separation of connected and disconnected
parts in Eq.(B.2) \cite{alf,Ban,M1,MR}. Thus instead of the two $s$ and $u$%
-channel terms (2.6) there will appear eight terms with connected amplitudes

\[
W^c_{a+N^{\prime},b+N}=w^c_{a+N^{\prime},b+N}= -<in;{\bf p^{\prime}}_N|\Bigl[%
j_{a}(0),a_{b}^+(0)\Bigr]|{\bf p}_N;in>
\]
$$
+(2\pi)^3 \sum_{n=N^{\prime\prime}} <in; {\bf p^{\prime}}_N |j_{a}(0)|{\bf p}%
_{N^{\prime\prime}};in> {\frac{ {\delta^{(3)}( {\bf p}_b+{\bf p}_{N}-{\bf p}%
_{N^{\prime\prime}} )}}{{\omega_b({\bf p_b})+ E_{{\bf p}_N}-E_{{\bf p}%
_{N^{\prime\prime}} } } }} <in;{\bf p}_{N^{\prime\prime}}|j_{b}(0)|{\bf p}%
_{N};in>\eqno(B.4a)
$$

$$
+(2\pi)^3 \sum_{m=mesons,N{\overline N}} < 0 |j_{a}(0)|m;in>{\frac{ {%
\delta^{(3)}( {\bf p}_b+{\bf p}_{N}-{\bf P}_{m}-{\bf p^{\prime}}_{N} )}}{{%
\omega_b({\bf p}_b)+ E_{{\bf p}_N}-P_{m}^o-E_{{\bf p^{\prime}}_{N} }
+i\epsilon } }} <in;{\bf p^{\prime}}_{N},m|j_{b}(0)|{\bf p}_{N};in>_c%
\eqno(B.4b)
$$

$$
+(2\pi)^3 \sum_{m=mesons,N{\overline N}} <in;{\bf p^{\prime}}_N |j_{a}(0)|%
{\bf p}_{N},m;in>_c {\frac{ {\delta^{(3)}( {\bf p}_b-{\bf P}_{m} )}}{{%
\omega_b({\bf p_b})-P_{m}^o } }} <in;m|j_{b}(0)|0>\eqno(B.4c)
$$

$$
-(2\pi)^3 \sum_{{\overline N,...}} <0 |j_{a}(0)|{\bf p}_{N}{\bf p}%
_{\overline N};in> {\frac{ {\delta^{(3)}( {\bf p}_b-{\bf p^{\prime}}%
_{N^{\prime}}-{\bf p}_{\overline N} )}}{{\omega_b({\bf p_b})- E_{{\bf %
p^{\prime}}_{N}}-E_{{\bf p}_{\overline N} } } }} <in;{\bf p^{\prime}}_{N},%
{\bf p}_{\overline N}|j_{b}(0)|0>\eqno(B.4d)
$$

$$
+(2\pi)^3
\sum_{n=N^{\prime\prime},\pi^{\prime\prime}N^{\prime\prime},\pi^{\prime%
\prime}\pi^{\prime\prime}N^{\prime\prime},...} <in; {\bf p^{\prime}}_N
|j_{b}(0)|n;in>_c {\frac{ {\delta^{(3)}(-{\bf p}_b+{\bf p^{\prime}}_{N}-{\bf %
P}_{n} )}}{{-\omega_b({\bf p_b})+ E_{{\bf p^{\prime}}_{N}}-P_n^o} }}
<in;n|j_{a}(0)|{\bf p}_{N};in>_c\eqno(B.4e)
$$

$$
+(2\pi)^3 \sum_{m=mesons,N{\overline N}} < 0 |j_{b}(0)|m;in>{\frac{ {%
\delta^{(3)}( -{\bf p}_b-{\bf P}_{m} )}}{{-\omega_b({\bf p}_b)-P_{m}^o } }}
<in;{\bf p^{\prime}}_{N},m|j_{b}(0)|{\bf p}_{N};in>_c\eqno(B.4f)
$$

$$
+(2\pi)^3 \sum_{m=mesons,N{\overline N}} <in;{\bf p^{\prime}}_N |j_{b}(0)|%
{\bf p}_{N},m;in>_c {\frac{ {\delta^{(3)}( -{\bf p}_b-{\bf P}_{m}-{\bf p}%
_{N}+{\bf p^{\prime}}_{N} )}}{{-\omega_b({\bf p_b})-E_{{\bf p}_N}-P_{m}^o+E_{%
{\bf p^{\prime}}_N} } }} <in;m|j_{a}(0)|0>\eqno(B.4g)
$$

$$
-(2\pi)^3 \sum_{{\overline N,...}} <0 |j_{b}(0)|{\bf p}_{N}{\bf p}%
_{\overline N};in> {\frac{ {\delta^{(3)}( -{\bf p}_b-{\bf p}_{N}-{\bf p}%
_{\overline N} )}}{{-\omega_b({\bf p_b})- E_{{\bf p}_{N}}-E_{{\bf p}%
_{\overline N} } } }} <in;{\bf p^{\prime}}_{N}{\bf p}_{\overline
N}|j_{a}(0)|0>,\eqno(B.4h)
$$
where the high order over $e^2$ intermediate terms are omitted. In addition,
we have eliminated the four-particle intermediate states from our
consideration. Therefore from the $s$-channel term (B.4a) are excluded
intermediate $n=3\pi^{\prime\prime}N,...$ states.

The terms (B.4a)-(B.4h) are represented in the Fig.2A-Fig.2H
correspondingly. Unlike to the analogical relations for the $\pi N$
scattering \cite{Ban,MR,M1}, equations (B.4a)-(B.4h) includes in the
asymptotic $\gamma N$ states, because $a=\pi^{\prime}N^{\prime},\gamma^{%
\prime}N^{\prime}\ and \ b=\pi N,\gamma N$, i.e. Eq.(B.4a)-(B.4h) are
derived for the coupled $\pi N\Longleftrightarrow\gamma N$ processes.


In order to transform $w_{\alpha \beta }^{c}$ (B.4a)-(B.4h) in the form of
the Hermitian potential $U(E)_{\alpha \beta }$ (3.5) we will use following
identities for the propagators

$$
{\frac{1}{{\omega_b({\bf p_b})+ E_{{\bf p}_N}-E_{{\bf p}_{N^{\prime\prime}}
} } }}\equiv {\frac{1}{{\omega_b({\bf p_b})+ E_{{\bf p}_N}-E_{{\bf p}%
_{N^{\prime\prime}} } } }} {\frac{ {\Bigl[ \omega_a({\bf p^{\prime}}_a)+ E_{%
{\bf p^{\prime}}_N}\Bigr]-E_{{\bf p}_{N^{\prime\prime}} } } }{{\omega_b({\bf %
p_b})+ E_{{\bf p}_N}-E_{{\bf p}_{N^{\prime\prime}} } } }}\eqno(B.5a)
$$

$$
{\frac{1}{{\omega_b({\bf p}_b)+ E_{{\bf p}_N}-P_{m}^o-E_{{\bf p^{\prime}}%
_{N}} +i\epsilon} }}\equiv {\frac{1}{{\omega_b({\bf p}_b)+ E_{{\bf p}%
_N}-P_{m}^o-E_{{\bf p^{\prime}}_{N}} +i\epsilon} }} {\frac{{\Bigl[\omega_a(%
{\bf p^{\prime}}_a)+E_{{\bf p^{\prime}}_N}\Bigr] -P_{m}^o }-E_{{\bf %
p^{\prime}}_N} }{{\omega_a({\bf p^{\prime}}_a)-P_{m}^o } }} \eqno(B.5b)
$$

$$
{\frac{1}{{\omega_b({\bf p_b})-P_{m}^o } }}\equiv {\frac{1}{{\omega_b({\bf %
p_b})-P_{m}^o } }} {\frac{{\Bigl[\omega_a({\bf p^{\prime}}_a)+ E_{{\bf %
p^{\prime}}_N}\Bigr]-P_{m}^o- E_{{\bf p }_{N}} } }{{\omega_a({\bf p^{\prime}}%
_a)+ E_{{\bf p^{\prime}}_N}-P_{m}^o-E_{{\bf p }_{N}} -i\epsilon} }} %
\eqno(B.5c)
$$

$$
{\frac{1}{{\omega_b({\bf p_b})- E_{{\bf p^{\prime}}_{N}}-E_{{\bf p}%
_{\overline N} } } }}\equiv {\frac{1}{{\omega_b({\bf p_b})- E_{{\bf %
p^{\prime}}_{N}}-E_{{\bf p}_{\overline N}} } }} {\frac{{\Bigl[\omega_a({\bf %
p^{\prime}}_a)+E_{{\bf p^{\prime}}_{N}}\Bigr] -E_{{\bf p^{\prime}}_{N}}- E_{%
{\bf p}_{N}}-E_{{\bf p}_{\overline N}} } }{{\omega_a({\bf p^{\prime}}_a)- E_{%
{\bf p}_{N}}-E_{{\bf p}_{\overline N}} } }} \eqno(B.5d)
$$

$$
{\frac{1}{{-\omega_b({\bf p_b})+ E_{{\bf p^{\prime}}_{N}}-P_n^o} }}\equiv {%
\frac{1}{{-\omega_b({\bf p_b})+ E_{{\bf p^{\prime}}_{N}}-P_n^o} }} {\frac{ {-%
\Bigl[\omega_a({\bf p}_a)+E_{{\bf p^{\prime}}_{N}}\Bigr] +E_{{\bf p^{\prime}}%
_{N}}+ E_{{\bf p}_{N}}-P_n^o} }{{-\omega_a({\bf p^{\prime}}_a)+ E_{{\bf p}%
_{N}}-P_n^o} }} \eqno(B.5e)
$$

$$
{\frac{1}{{-\omega_b({\bf p}_b)-P_{m}^o } }}\equiv {\frac{1}{{-\omega_b({\bf %
p}_b)-P_{m}^o } }} {\frac{{-\Bigl[\omega_a({\bf p^{\prime}}_a)+E_{{\bf %
p^{\prime}}_N}\Bigr]-P_{m}^o+E_{{\bf p}_N} } }{{-\omega_a({\bf p^{\prime}}%
_a)-E_{{\bf p^{\prime}}_N}-P_{m}^o+E_{{\bf p}_N} } }} \eqno(B.5f)
$$

$$
{\frac{1}{{-\omega_b({\bf p_b})-E_{{\bf p}_N}-P_{m}^o+E_{{\bf p^{\prime}}_N} 
} }}\equiv {\frac{1}{{-\omega_b({\bf p_b})-E_{{\bf p}_N}-P_{m}^o+E_{{\bf %
p^{\prime}}_N} } }} {\frac{{-\Bigl[\omega_a({\bf p^{\prime}}_a)+ E_{{\bf %
p^{\prime}}_{N}}\Bigr] + E_{{\bf p^{\prime}}_{N}}-P_{m}^o }}{{-\omega_a({\bf %
p^{\prime}}_a)-P_{m}^o } }} \eqno(B.5g)
$$

$$
{\frac{1}{{-\omega _{b}({\bf p_{b}})-E_{{\bf p}_{N}}-E_{{\bf p}_{\overline{N}%
}}}}}\equiv {\frac{1}{{-\omega _{b}({\bf p_{b}})-E_{{\bf p}_{N}}-E_{{\bf p}_{%
\overline{N}}}}}}{\frac{{-\Bigl[\omega _{a}({\bf p^{\prime }}_{a})+E_{{\bf %
p^{\prime }}_{N}}\Bigr]-E_{{\bf p}_{\overline{N}}}}}{{-\omega _{a}({\bf %
p^{\prime }}_{a})-E_{{\bf p^{\prime }}_{N}}-E_{{\bf p}_{\overline{N}}}}}}%
\eqno(B.5h)
$$
where the propagators (B.5e)-(B.5h) are taken from the corresponding
expressions (B.4a)-(B.4h). Here we wish stress that only the common factor $%
\Bigl[\omega _{a}({\bf p^{\prime }}_{a})+E_{{\bf p^{\prime }}_{N}}\Bigr]$
destroys the hermiticity of the propagators (B.4a)-(B.4h).

All of identities (B.5a)-(B.5h) contain the same expression of the final
state $\alpha=a+N^{\prime}$ energy $E_{\alpha}=\omega_a({\bf p^{\prime}}%
_a)+E_{{\bf p^{\prime}}_N}$ in the square parenthesis. After substitution of
the identities (B.5a)-(B.5h) in the relations (B.4a)-(B.4h) we define the
sum of all terms with $E_{\alpha}=\omega_a({\bf p^{\prime}}_a)+E_{{\bf %
p^{\prime}}_N}$ as the matrix $B_{\alpha\beta}$.

\[
B_{a+N^{\prime},b+N}\equiv B_{a+N^{\prime},b+N}({\bf p^{\prime}}_N {\bf %
p^{\prime}}_a; {\bf p}_N {\bf p}_b)=
\]

$$
+(2\pi)^3 \sum_{n=N^{\prime\prime}} <in; {\bf p^{\prime}}_N |j_{a}(0)|{\bf p}%
_{N^{\prime\prime}};in> {\frac{ {\delta^{(3)}( {\bf p}_b+{\bf p}_{N}-{\bf p}%
_{N^{\prime\prime}} )}}{{\omega_b({\bf p_b})+ E_{{\bf p}_N}-E_{{\bf p}%
_{N^{\prime\prime}} } } }} {\frac{{\ <in;{\bf p}_{N^{\prime\prime}}|j_{b}(0)|%
{\bf p}_{N};in>} }{{\omega_b({\bf p_b})+ E_{{\bf p}_N}-E_{{\bf p}%
_{N^{\prime\prime}} } } }} \eqno(B.6a)
$$
$$
+(2\pi)^3 \sum_{m=mesons,N{\overline N}} <in;{\bf p^{\prime}}_N |j_{a}(0)|%
{\bf p}_{N},m;in>_c {\frac{ {\delta^{(3)}( {\bf p}_b-{\bf P}_{m} )}}{{%
\omega_b({\bf p_b})-P_{m}^o } }} {\frac{{<in;m|j_{b}(0)|0>} }{{\omega_a({\bf %
p^{\prime}}_a)+ E_{{\bf p^{\prime}}_N}-P_{m}^o-E_{{\bf p }_{N}} -i\epsilon} }%
} \eqno(B.6b)
$$

$$
+(2\pi)^3 \sum_{m=mesons,N{\overline N}} < 0 |j_{a}(0)|m;in>{\frac{ {%
\delta^{(3)}( {\bf p}_b+{\bf p}_{N}-{\bf P}_{m}-{\bf p^{\prime}}_{N} )}}{{%
\omega_b({\bf p}_b)+ E_{{\bf p}_N}-P_{m}^o-E_{{\bf p^{\prime}}_{N}}
+i\epsilon} }} {\frac{{<in;{\bf p^{\prime}}_{N},m|j_{b}(0)|{\bf p}_{N};in>_c}
}{{\omega_a({\bf p^{\prime}}_a)-P_{m}^o } }} \eqno(B.6c)
$$

$$
-(2\pi)^3 \sum_{{\overline N},...} <0 |j_{a}(0)|{\bf p}_{N}{\bf p}%
_{\overline N};in> {\frac{ {\delta^{(3)}( {\bf p}_b-{\bf p^{\prime}}%
_{N^{\prime}}-{\bf p}_{\overline N} )}}{{\omega_b({\bf p_b})- E_{{\bf %
p^{\prime}}_{N}}-E_{{\bf p}_{\overline N} } } }} {\frac{{<in;{\bf p^{\prime}}%
_{N},{\bf p}_{\overline N}|j_{b}(0)|0>} }{{\omega_a({\bf p^{\prime}}_a)- E_{%
{\bf p}_{N}}-E_{{\bf p}_{\overline N}} } }} \eqno(B.6d)
$$
\[
\ \ \ \ \ \ \ +\ \ \ a\ and\ b\ crossing\ terms.\ \ \ 
\]
The terms in the left side of the identities (B.4a)-(B.4h) that are not
proportional of $\Bigl[\omega_a({\bf p^{\prime}}_a)+E_{{\bf p^{\prime}}%
_N}\Bigr]$ compose the $A_{a+N^{\prime},b+N}$ matrix

\[
A_{a+N^{\prime},b+N}\equiv A_{a+N^{\prime},b+N}({\bf p^{\prime}}_N {\bf %
p^{\prime}}_a; {\bf p}_N {\bf p}_b)= -<out;N^{\prime}|\Bigl[%
j_{a}(0),a_{b}^+(0)\Bigr]|N;in> 
\]

$$
- \sum_{n=N^{\prime\prime}} E_{{\bf p}_{N"}} <in; {\bf p^{\prime}}_N
|j_{a}(0)|{\bf p}_{N^{\prime\prime}};in> {\frac{ {(2\pi)^3\delta^{(3)}( {\bf %
p}_b+{\bf p}_{N}-{\bf p}_{N^{\prime\prime}} )}}{{\omega_b({\bf p_b})+ E_{%
{\bf p}_N}-E_{{\bf p}_{N^{\prime\prime}} } } }} {\frac{{\ <in;{\bf p}%
_{N^{\prime\prime}}|j_{b}(0)|{\bf p}_{N};in>} }{{\omega_b({\bf p_b})+ E_{%
{\bf p}_N}-E_{{\bf p}_{N^{\prime\prime}} } } }} \eqno(B.7a)
$$

$$
- \sum_{m=mesons} (E_{{\bf p^{\prime}}_{N}}+P_{m}^o) < 0 |j_{a}(0)|m;in>{%
\frac{ {(2\pi)^3\delta^{(3)}( {\bf p}_b+{\bf p}_{N}-{\bf P}_{m}-{\bf %
p^{\prime}}_{N} )}}{{\omega_b({\bf p}_b)+ E_{{\bf p}_N}-P_{m}^o-E_{{\bf %
p^{\prime}}_{N}} +i\epsilon} }} {\frac{{<in;{\bf p^{\prime}}_{N},m|j_{b}(0)|%
{\bf p}_{N};in>_c} }{{\omega_a({\bf p^{\prime}}_a)-P_{m}^o } }} \eqno(B.7b)
$$

$$
- \sum_{m=mesons}(E_{{\bf p}_{N}}+P_{m}^o) <in;{\bf p^{\prime}}_N |j_{a}(0)|%
{\bf p}_{N},m;in>_c {\frac{ {(2\pi)^3\delta^{(3)}( {\bf p}_b-{\bf P}_{m} )}}{%
{\omega_b({\bf p_b})-P_{m}^o } }} {\frac{{<in;m|j_{b}(0)|0>} }{{\omega_a(%
{\bf p^{\prime}}_a)+ E_{{\bf p^{\prime}}_N}-P_{m}^o-E_{{\bf p }_{N}}
-i\epsilon} }} \eqno(B.7c)
$$

$$
+ \sum_{{\overline N},...}(E_{{\bf p^{\prime}}_{N}}+E_{{\bf p}_{N}} +E_{{\bf %
p}_{\overline N}}) <0 |j_{a}(0)|{\bf p}_{N}{\bf p}_{\overline N};in> {\frac{%
(2\pi)^3 {\delta^{(3)}( {\bf p}_b-{\bf p^{\prime}}_{N^{\prime}}-{\bf p}%
_{\overline N} )}}{{\omega_b({\bf p_b})- E_{{\bf p^{\prime}}_{N}}-E_{{\bf p}%
_{\overline N} } } }} {\frac{{<in;{\bf p^{\prime}}_{N},{\bf p}_{\overline
N}|j_{b}(0)|0>} }{{\omega_a({\bf p^{\prime}}_a)- E_{{\bf p}_{N}}-E_{{\bf p}%
_{\overline N}} } }} \eqno(B.7d)
$$
\[
\ \ \ \ \ \ \ +\ \ \ a\ and\ b\ crossing\ terms.\ \ \ 
\]

It easy to observe, that the first and the fourth terms of Eq.(B.7) are
Hermitian, but the second term (B.7b) is Hermitian conjugate of the third
term (B.7c). Note, that these Hermitian conjugate terms (B.7b) and (B.7c)
have the Hermitian conjugate singular propagators. Also the $u$ channel
propagators (B.5f) and (B.5g) are Hermitian conjugate.

We have included in $A_{a+N^{\prime },b+N}$ the equal time commutators
because equal-time commutators produces the Hermitian terms for the case of
the simplest renormalized Lagrangians in $\phi ^{3}$ field theory. The more
complicated types of the equal-time commutators which produces the
contribution in the both matrices $A_{a+N^{\prime },b+N}$ and $%
B_{a+N^{\prime },b+N}$ are considered in our previous papers \cite{M1,M3}.

\bigskip

{\bf For the $3\Longleftarrow 2$ transition amplitude} $<in;{\bf p^{\prime}}%
_{N},{\bf p^{\prime}}_{\pi}|j_{a}(0)|{\bf p}_{N};in>_c$ which corresponds to
the $a+{\pi^{\prime}}+N^{\prime}\Longleftarrow b+N$ reaction with $%
a=\pi^{\prime},\gamma^{\prime}$ and $b=\pi,\gamma$, the cluster
decomposition of the expression (B.1b) generates the following 16 terms.

\[
w^c_{a+{\pi^{\prime}}+N^{\prime},b+N}= -<in;{\bf p^{\prime}}_{\pi}, {\bf %
p^{\prime}}_N|\Bigl[j_{a}(0),a_{b}^+(0)\Bigr]|{\bf p}_N;in>
\]
$$
+\sum_{n=N^{\prime\prime}} <in;{\bf p^{\prime}}_{\pi},{\bf p^{\prime}}_N
|j_{a}(0)|{\bf p}_{N^{\prime\prime}};in> {\frac{ {(2\pi)^3\delta^{(3)}( {\bf %
p}_b+{\bf p}_{N}-{\bf p}_{N^{\prime\prime}} )}}{{\omega_b({\bf p_b})+ E_{%
{\bf p}_N}-E_{{\bf p}_{N^{\prime\prime}} } } }} <in;{\bf p}%
_{N^{\prime\prime}}|j_{b}(0)|{\bf p}_{N};in>\eqno(B.8a)
$$

$$
+\sum_{m=mesons} < {\bf p^{\prime}}_{\pi}|j_{a}(0)|m;in>_c{\frac{ {%
(2\pi)^3\delta^{(3)}( {\bf p}_b+{\bf p}_{N}-{\bf P}_{m}-{\bf p^{\prime}}_{N}
)}}{{\omega_b({\bf p}_b)+ E_{{\bf p}_N}-P_{m}^o-E_{{\bf p^{\prime}}_{N}}
+i\epsilon} }} <in;{\bf p^{\prime}}_{N},m|j_{b}(0)|{\bf p}_{N};in>_c%
\eqno(B.8b)
$$

$$
+\sum_{m=mesons} <in;{\bf p^{\prime}}_{\pi},{\bf p^{\prime}}_N |j_{a}(0)|%
{\bf p}_{N},m;in>_c {\frac{(2\pi)^3 {\delta^{(3)}( {\bf p}_b-{\bf P}_{m} )}}{%
{\omega_b({\bf p_b})-P_{m}^o } }} <in;m|j_{b}(0)|0>\eqno(B.8c)
$$

$$
-\sum_{{\overline N}} <{\bf p^{\prime}}_{\pi} |j_{a}(0)|{\bf p}_{N}{\bf p}%
_{\overline N};in>_c {\frac{ {(2\pi)^3\delta^{(3)}( {\bf p}_b-{\bf p^{\prime}%
}_{N^{\prime}}-{\bf p}_{\overline N} )}}{{\omega_b({\bf p_b})- E_{{\bf %
p^{\prime}}_{N}}-E_{{\bf p}_{\overline N} } } }} <in;{\bf p^{\prime}}_{N},%
{\bf p}_{\overline N}|j_{b}(0)|0>\eqno(B.8d)
$$

$$
+\sum_{n=N^{\prime\prime},\pi"N",...} <in;{\bf p^{\prime}}_N
|j_{a}(0)|n;in>_c {\frac{ {(2\pi)^3\delta^{(3)}( {\bf p}_b+{\bf p}_{N}-{\bf P%
}_{n}-{\bf p^{\prime}}_{\pi} )}}{{\omega_b({\bf p_b})+ E_{{\bf p}_N}-P^o_n-
\omega_{\pi}({\bf p^{\prime}}_{\pi}) +i\epsilon } }} <in;{\bf p^{\prime}}%
_{\pi},n|j_{b}(0)|{\bf p}_{N};in>_c \eqno(B.8e)
$$

$$
+\sum_{m=mesons} < 0|j_{a}(0)|m;in>{\frac{ {(2\pi)^3\delta^{(3)}( {\bf p}_b+%
{\bf p}_{N}-{\bf P}_{m}-{\bf p^{\prime}}_{N} -{\bf p^{\prime}}_{\pi} )}}{{%
\omega_b({\bf p}_b)+ E_{{\bf p}_N}-P_{m}^o-E_{{\bf p^{\prime}}_{N}}
-\omega_{\pi}({\bf p^{\prime}}_{\pi}) +i\epsilon} }} <in;{\bf p^{\prime}}%
_{\pi},{\bf p^{\prime}}_{N},m|j_{b}(0)|{\bf p}_{N};in>\eqno(B.8f)
$$

$$
+\sum_{m=mesons} <in;{\bf p^{\prime}}_N |j_{a}(0)|{\bf p}_{N},m;in>_c {\frac{
{(2\pi)^3\delta^{(3)}( {\bf p}_b-{\bf P}_{m}-{\bf p^{\prime}}_{\pi} )}}{{%
\omega_b({\bf p_b})-P_{m}^o-\omega_{\pi}({\bf p^{\prime}}_{\pi}) } }} <in;%
{\bf p^{\prime}}_{\pi}, m|j_{b}(0)|0>\eqno(B.8g)
$$

$$
-\sum_{{\overline N},...} < 0 |j_{a}(0)|{\bf p}_{N}{\bf p}_{\overline N};in> 
{\frac{ {(2\pi)^3\delta^{(3)}( {\bf p}_b-{\bf p^{\prime}}_{N^{\prime}}-{\bf p%
}_{\overline N} -{\bf p^{\prime}}_{\pi})}}{{\omega_b({\bf p_b})- E_{{\bf %
p^{\prime}}_{N}}-E_{{\bf p}_{\overline N}} -\omega_{\pi}({\bf p^{\prime}}%
_{\pi}) } }} <in;{\bf p^{\prime}}_{\pi}, {\bf p^{\prime}}_{N},{\bf p}%
_{\overline N}|j_{b}(0)|0>\eqno(B.8h)
$$

\[
\ \ \ \ \ \ \ +\ \ \ a\ and\ b\ crossing\ 8\ terms.\ \ \ 
\]

These terms are depicted in Fig. 3A - Fig.3H correspondingly.


The conjugate $2\Longleftarrow 3$ transition amplitude $<in;{\bf p^{\prime}}%
_{N}|j_{a}(0)|{\bf p}_b{\bf p}_{\pi}{\bf p}_{N};in>$ relates to the potential

\[
w^c_{a+N^{\prime},b+{\pi}+N}= -<in;{\bf p^{\prime}}_N|\Bigl[%
j_{a}(0),a_{b}^+(0)\Bigr] |{\bf p}_N{\bf p}_{\pi};in>
\]
$$
+\sum_{n=N^{\prime\prime}} {\frac{{<in;{\bf p^{\prime}}_N |j_{a}(0)|{\bf p}%
_{N^{\prime\prime}};in>_c (2\pi)^3\delta^{(3)}( {\bf p}_b+{\bf p}_{N}+{\bf p}%
_{\pi}-{\bf p}_{N^{\prime\prime}} )}}{{\omega_b({\bf p_b})+ E_{{\bf p}_N}+
\omega_{\pi}({\bf p}_{\pi}) -E_{{\bf p}_{N^{\prime\prime}} } } }} <in;{\bf p}%
_{N^{\prime\prime}}|j_{b}(0)|{\bf p}_{N}{\bf p}_{\pi};in>_c\eqno(B.9a)
$$

$$
+\sum_{m=mesons} {\frac{{<0 |j_{a}(0)|m;in>_c(2\pi)^3\delta^{(3)}( {\bf p}_b+%
{\bf p}_{\pi}+ {\bf p}_{N}-{\bf P}_{m}-{\bf p^{\prime}}_{N} )}}{{\omega_b(%
{\bf p}_b)+ E_{{\bf p}_N}+ \omega_{\pi}({\bf p}_{\pi})-P_{m}^o-E_{{\bf %
p^{\prime}}_{N}} +i\epsilon } }} <in;{\bf p^{\prime}}_{N},m|j_{b}(0)|{\bf p}%
_{\pi}{\bf p}_{N};in>_C\eqno(B.9b)
$$

$$
+\sum_{m=mesons} {\frac{{<in;{\bf p^{\prime}}_N |j_{a}(0)|{\bf p}%
_{N},m;in>_c (2\pi)^3\delta^{(3)}( {\bf p}_b+{\bf p}_{\pi}-{\bf P}_{m} )}}{{%
\omega_b({\bf p_b})+\omega_{\pi}({\bf p}_{\pi}) -P_{m}^o +i\epsilon} }}
<in;m|j_{b}(0)|{\bf p}_{\pi};in>\eqno(B.9c)
$$

$$
-\sum_{{\overline N}} {\frac{{<0|j_{a}(0)|{\bf p}_{N}{\bf p}_{\overline
N};in> (2\pi)^3\delta^{(3)}( {\bf p}_b+{\bf p}_{\pi}-{\bf p^{\prime}}%
_{N^{\prime}}-{\bf p}_{\overline N} )}}{{\omega_b({\bf p_b})+\omega_{\pi}(%
{\bf p}_{\pi}) - E_{{\bf p^{\prime}}_{N}}-E_{{\bf p}_{\overline N} }
+i\epsilon} }} <in;{\bf p^{\prime}}_{N},{\bf p}_{\overline N}|j_{b}(0) |{\bf %
p}_{\pi};in>_c\eqno(B.9d)
$$

$$
+\sum_{n=N^{\prime\prime},\pi^{\prime\prime}N^{\prime\prime},2\pi^{\prime%
\prime}N^{\prime\prime}} {\frac{{<in;{\bf p^{\prime}}_N |j_{a}(0)|{\bf p}%
_{\pi},n;in>_c (2\pi)^3\delta^{(3)}( {\bf p}_b+{\bf p}_{N}-{\bf P}_{n})}}{{%
\omega_b({\bf p_b})+ E_{{\bf p}_N}-P^o_{n}+i\epsilon } }} <in;n|j_{b}(0)|%
{\bf p}_{N};in>_c \eqno(B.9e)
$$

$$
+\sum_{m=mesons} {\frac{{< 0|j_{a}(0)|{\bf p}_{\pi},m;in>_c(2\pi)^3%
\delta^{(3)}( {\bf p}_b+{\bf p}_{N}-{\bf P}_{m}-{\bf p^{\prime}}_{N})}}{{%
\omega_b({\bf p}_b)+ E_{{\bf p}_N}-P_{m}^o-E_{{\bf p^{\prime}}_{N}}
+i\epsilon} }} <in;{\bf p^{\prime}}_{N},m|j_{b}(0)|{\bf p}_{N};in>_c%
\eqno(B.9f)
$$

$$
+\sum_{m=mesons} {\frac{{<in;{\bf p^{\prime}}_N |j_{a}(0)|{\bf p}_{\pi}{\bf p%
}_{N},m;in>_c (2\pi)^3\delta^{(3)}( {\bf p}_b-{\bf P}_{m} )}}{{\omega_b({\bf %
p_b})-P_{m}^o } }} <in; m|j_{b}(0)|0>\eqno(B.9g)
$$

$$
-\sum_{{\overline N}} < 0 |j_{a}(0)|{\bf p}_{\pi}{\bf p}_{N}{\bf p}%
_{\overline N};in> {\frac{ {(2\pi)^3\delta^{(3)}( {\bf p}_b-{\bf p^{\prime}}%
_{N^{\prime}}-{\bf p}_{\overline N})}}{{\omega_b({\bf p_b})- E_{{\bf %
p^{\prime}}_{N}}-E_{{\bf p}_{\overline N}} } }} <in;{\bf p^{\prime}}_{N},%
{\bf p}_{\overline N}|j_{b}(0)|0>\eqno(B.9h)
$$

\[
\ \ \ \ \ \ \ +\ \ \ a\ and\ b\ crossing\ 8\ terms.\ \ \ 
\]

Using the generalization of the identities (B.5a)-(B.5h) and combining
(B.8b), (B.8e) terms with (B.9c), (B.9f) correspondingly, , we obtain

\[
B_{a+{\pi^{\prime}}+N^{\prime},b+N}\equiv B_{a+{\pi^{\prime}}%
+N^{\prime},b+N}({\bf p^{\prime}}_N {\bf p^{\prime}}_a {\bf p^{\prime}}%
_{\pi^{\prime}};{\bf p}_N{\bf p}_b)=
\]

$$
\sum_{n=N^{\prime\prime}} {\frac{ {<in;{\bf p^{\prime}}_{\pi},{\bf p^{\prime}%
}_N |j_{a}(0)|{\bf p}_{N^{\prime\prime}};in>_c (2\pi)^3\delta^{(3)}({\bf p}%
_b+{\bf p}_{N}-{\bf p}_{N^{\prime\prime}} )}}{{\omega_b({\bf p_b})+ E_{{\bf p%
}_N}-E_{{\bf p}_{N^{\prime\prime}} } } }} {\frac{{<in;{\bf p}%
_{N^{\prime\prime}}|j_{b}(0)|{\bf p}_{N};in>_c} }{{\omega_a({\bf p^{\prime}_a%
})+ E_{{\bf p^{\prime}}_N}+ \omega_{\pi}({\bf p^{\prime}}_{\pi})-E_{{\bf p}%
_{N^{\prime\prime}} } } }} \eqno(B.10a)
$$

$$
+\sum_{m=mesons} {\frac{ {< {\bf p^{\prime}}_{\pi}|j_{a}(0)|m;in>_c
(2\pi)^3\delta^{(3)}({\bf p}_b+{\bf p}_{N}-{\bf P}_{m}-{\bf p^{\prime}}_{N} )%
}}{{\omega_b({\bf p}_b)+ E_{{\bf p}_N}-P_{m}^o-E_{{\bf p^{\prime}}_{N}}
+i\epsilon} }} {\frac{{<in;{\bf p^{\prime}}_{N},m|j_{b}(0)|{\bf p}_{N};in>_c}
}{{\omega_a({\bf p^{\prime}}_a)+\omega_{\pi}({\bf p^{\prime}}_{\pi})
-P_{m}^o-i\epsilon } }}\eqno(B.10b)
$$

$$
+\sum_{m=mesons} {\frac{{<in;{\bf p^{\prime}}_{\pi},{\bf p^{\prime}}_N
|j_{a}(0)|{\bf p}_{N},m;in>_c}}{{\omega_b({\bf p_b})-P_{m}^o } }} {\frac{{%
(2\pi)^3\delta^{(3)}({\bf p}_b-{\bf P}_{m} )<in;m|j_{b}(0)|0>} }{{\omega_a(%
{\bf p^{\prime}}_a)+\omega_{\pi}({\bf p^{\prime}}_{\pi}) + E_{{\bf p^{\prime}%
}_N}-P_{m}^o-E_{{\bf p }_{N}} - i\epsilon} }} \eqno(B.10c)
$$

$$
-\sum_{{\overline N}} {\frac{ {<{\bf p^{\prime}}_{\pi} |j_{a}(0)|{\bf p}_{N}%
{\bf p}_{\overline N};in> (2\pi)^3\delta^{(3)}( {\bf p}_b-{\bf p^{\prime}}%
_{N^{\prime}}-{\bf p}_{\overline N} )}}{{\omega_b({\bf p_b})- E_{{\bf %
p^{\prime}}_{N}}-E_{{\bf p}_{\overline N} } } }} {\frac{{<in;{\bf p^{\prime}}%
_{N},{\bf p}_{\overline N}|j_{b}(0)|0>} }{{\omega_a({\bf p^{\prime}}%
_a)+\omega_{\pi}({\bf p^{\prime}}_{\pi })- E_{{\bf p}_{N}}-E_{{\bf %
p^{\prime\prime}}_{\overline N}}-i\epsilon } }} \eqno(B.10d)
$$

$$
+\sum_{n=N^{\prime\prime},\pi^{\prime\prime}N^{\prime\prime},2\pi"N"} {\frac{%
{<in;{\bf p^{\prime}}_N |j_{a}(0)|n;in>_c (2\pi)^3\delta^{(3)}( {\bf p}_b+%
{\bf p}_{N}-{\bf P}_{n}-{\bf p^{\prime}}_{\pi} )}}{{\omega_b({\bf p_b})+ E_{%
{\bf p}_N}-P^o_{n}- \omega_{\pi}( {\bf p^{\prime}}_{\pi}) +i\epsilon } }} {%
\frac{{<in;{\bf p^{\prime}}_{\pi a},n|j_{b}(0)|{\bf p}_{N};in>_c}}{{\omega_a(%
{\bf p^{\prime}_a})+ E_{{\bf p^{\prime}}_N}-P^o_{n} -i\epsilon } }} %
\eqno(B.10e)
$$

$$
+\sum_{m=mesons} {\frac{ {\ < 0|j_{a}(0)|m;in>_c (2\pi)^3\delta^{(3)}( {\bf p%
}_b+{\bf p}_{N}-{\bf P}_{m}-{\bf p^{\prime}}_{N} -{\bf p}_{\pi} )}}{{%
\omega_b({\bf p}_b)+ E_{{\bf p}_N}-P_{m}^o-E_{{\bf p^{\prime}}_{N}}
-\omega_{\pi}({\ {\bf p^{\prime}}_{\pi} }) +i\epsilon} }} {\frac{{<in;{\bf %
p^{\prime}}_{\pi},{\bf p}_{N},m|j_{b}(0)|{\bf p}_{N};in>} }{{\omega_a({\bf %
p^{\prime}_a})-P_{m}^o } }}\eqno(B.10f)
$$

$$
+\sum_{m=mesons} {\frac{ {<in;{\bf p^{\prime}}_N |j_{a}(0)|{\bf p}%
_{N},m;in>_c (2\pi)^3\delta^{(3)}( {\bf p}_b-{\bf P}_{m}-{\bf p}_{\pi} )}}{{%
\omega_b({\bf p_b})-P_{m}^o- \omega_{\pi}({\ {\bf p^{\prime}}_{\pi} }) } }} {%
\frac{{<in;{\bf p^{\prime}}_{\pi}, m|j_{b}(0)|0>}}{{\omega_a({\bf p^{\prime}}%
_a)+ E_{{\bf p^{\prime}}_N}-P_{m}^o-E_{{\bf p}_{N}} -i\epsilon}}}\eqno(B.10g)
$$

$$
-\sum_{{\overline N}} {\frac{ {< 0 |j_{a}(0)|{\bf p}_{N}{\bf p}_{\overline
N};in> (2\pi)^3\delta^{(3)}( {\bf p}_b-{\bf p^{\prime}}_{N^{\prime}}-{\bf p}%
_{\overline N} -{\bf p}_{\pi})}}{{\omega_b({\bf p_b})- E_{{\bf p^{\prime}}%
_{N}}-E_{{\bf p}_{\overline N}} -\omega_{\pi}({\ {\bf p^{\prime}}_{\pi} }) } 
}} {\frac{{<in;{\bf p^{\prime}}_{\pi}, {\bf p^{\prime}}_{N},{\bf p}%
_{\overline N}|j_{b}(0)|0>}}{{\omega_a({\bf p^{\prime}_a})- E_{{\bf p}%
_{N}}-E_{{\bf p}_{\overline N}} } }} \eqno(B.10h)
$$

\[
\ \ \ \ \ \ \ +\ \ \ a\ and\ b\ crossing\ 8\ terms.\ \ \ 
\]

where each term of expression (B.10a)-(B.10h) is nonhermitian. Nevertheless

\[
B_{a+{\pi ^{\prime }}+N^{\prime },b+N}({\bf p^{\prime }}_{N}{\bf p^{\prime }}%
_{a}{\bf p^{\prime }}_{\pi ^{\prime }};{\bf p}_{N}{\bf p}_{b})=B_{b+N^{%
\prime },a+{\pi }+N}^{*}({\bf p}_{N}{\bf p}_{b};{\bf p}_{N}{\bf p}_{a}{\bf p}%
_{\pi }),
\]
i.e. the multichannel matrix $B_{\alpha \beta }$ contains the Hermitian
conjugate non-diagonal parts. For the other part of the $U_{\alpha \beta }(E)
$ potential we have

\[
A_{a+{\pi}^{\prime}+N^{\prime},b+N}\equiv A_{a+{\pi}^{\prime}+N^{%
\prime},b+N}({\bf p^{\prime}}_N {\bf p^{\prime}}_a {\bf p^{\prime}}_{\pi}; 
{\bf p}_N{\bf p}_b)= -<in;{\bf p^{\prime}}_{\pi},{\bf p^{\prime}}_N|\Bigl[%
j_{a}(0),a_{b}^+(0)\Bigr] |{\bf p}_N;in>
\]

$$
-\sum_{n=N^{\prime\prime}}E_{{\bf p}_{N^{\prime\prime}}} {\frac{ {<in;{\bf %
p^{\prime}}_{\pi},{\bf p^{\prime}}_N |j_{a}(0)|{\bf p}_{N^{\prime%
\prime}};in>_c (2\pi)^3\delta^{(3)}({\bf p}_b+{\bf p}_{N}-{\bf p}%
_{N^{\prime\prime}} )}}{{\omega_b({\bf p_b})+ E_{{\bf p}_N}-E_{{\bf p}%
_{N^{\prime\prime}} } } }} {\frac{{<in;{\bf p}_{N^{\prime\prime}}|j_{b}(0)|%
{\bf p}_{N};in>_c} }{{\omega_a({\bf p^{\prime}_a})+ E_{{\bf p^{\prime}}_N}+
\omega_{\pi}({\bf p^{\prime}}_{\pi})-E_{{\bf p}_{N^{\prime\prime}} } } }} %
\eqno(B.11a)
$$

$$
-\sum_{m=mesons}(P_{m}^o+E_{{\bf p^{\prime}}_{N}}) {\frac{ {< {\bf p^{\prime}%
}_{\pi}|j_{a}(0)|m;in>_c (2\pi)^3\delta^{(3)}({\bf p}_b+{\bf p}_{N}-{\bf P}%
_{m}-{\bf p^{\prime}}_{N} )}}{{\omega_b({\bf p}_b)+ E_{{\bf p}_N}-P_{m}^o-E_{%
{\bf p^{\prime}}_{N}} +i\epsilon} }} {\frac{{<in;{\bf p^{\prime}}%
_{N},m|j_{b}(0)|{\bf p}_{N};in>} }{{\omega_a({\bf p^{\prime}}%
_a)+\omega_{\pi}({\bf p^{\prime}}_{\pi}) -P_{m}^o-i\epsilon } }}\eqno(B.11b)
$$

$$
-\sum_{m=mesons}(P_{m}^o+E_{{\bf p }_{N}}) {\frac{{<in;{\bf p^{\prime}}%
_{\pi},{\bf p^{\prime}}_N |j_{a}(0)|{\bf p}_{N},m;in>_c}}{{\omega_b({\bf p_b}%
)-P_{m}^o } }} {\frac{{(2\pi)^3\delta^{(3)}({\bf p}_b-{\bf P}_{m}
)<in;m|j_{b}(0)|0>} }{{\omega_a({\bf p^{\prime}}_a)+\omega_{\pi}({\bf %
p^{\prime}}_{\pi}) + E_{{\bf p^{\prime}}_N}-P_{m}^o-E_{{\bf p }_{N}} -
i\epsilon} }} \eqno(B.11c)
$$

$$
+\sum_{{\overline N}} (E_{{\bf p^{\prime}}_{N}}+E_{{\bf p}_{N}}+E_{{\bf %
p^{\prime\prime}}_{\overline N}}) {\frac{ {<{\bf p^{\prime}}_{\pi} |j_{a}(0)|%
{\bf p}_{N}{\bf p}_{\overline N};in> }}{{\omega_b({\bf p_b})- E_{{\bf %
p^{\prime}}_{N}}-E_{{\bf p}_{\overline N} } } }} {\frac{{(2\pi)^3%
\delta^{(3)}({\bf p}_b-{\bf p^{\prime}}_{N^{\prime}}-{\bf p}_{\overline N} )
<in;{\bf p^{\prime}}_{N},{\bf p}_{\overline N}|j_{b}(0)|0>} }{{\omega_a({\bf %
p^{\prime}}_a)+\omega_{\pi}({\bf p^{\prime}}_{\pi})- E_{{\bf p}_{N}}-E_{{\bf %
p^{\prime\prime}}_{\overline N}}-i\epsilon } }} \eqno(B.11d)
$$

\[
-\sum_{n=N^{\prime\prime},\pi^{\prime\prime}N^{\prime\prime},2\pi^{\prime%
\prime}N^{\prime\prime}}(P^o_{n}+\omega_{\pi}({\bf p^{\prime}}_{\pi} )) {%
\frac{{<in;{\bf p^{\prime}}_N |j_{a}(0)|n;in>_c (2\pi)^3\delta^{(3)}({\bf p}%
_b+{\bf p}_{N}-{\bf P}_{n}-{\bf p^{\prime}}_{\pi} )} }{{\omega_b({\bf p_b})+
E_{{\bf p}_N}-P^o_{n}- \omega_{\pi}( {\bf p^{\prime}}_{\pi}) +i\epsilon } }}
\]
$$
{\frac{{<in;{\bf p^{\prime}}_{\pi},n|j_{b}(0)|{\bf p}_{N};in>_c}}{{\omega_a(%
{\bf p^{\prime}_a})+ E_{{\bf p^{\prime}}_N}-P^o_{n} -i\epsilon } }} %
\eqno(B.11e)
$$

\[
-\sum_{m=mesons} ( P_{m}^o+E_{{\bf p^{\prime}}_N}+\omega_{\pi}({\ {\bf %
p^{\prime}}_{\pi} })) {\frac{ {\ < 0|j_{a}(0)|m;in>_c (2\pi)^3\delta^{(3)}( 
{\bf p}_b+{\bf p}_{N}-{\bf P}_{m}-{\bf p^{\prime}}_{N} -{\bf p^{\prime}}%
_{\pi} )}}{{\omega_b({\bf p}_b)+ E_{{\bf p}_N}-P_{m}^o-E_{{\bf p^{\prime}}%
_{N}} -\omega_{\pi}({\ {\bf p^{\prime}}_{\pi} }) +i\epsilon} }}
\]
$$
{\frac{{<in;{\bf p^{\prime}}_{\pi},{\bf p}_{N},m|j_{b}(0)|{\bf p}_{N};in>} }{%
{\omega_a({\bf p^{\prime}_a})-P_{m}^o } }}\eqno(B.11f)
$$

\[
-\sum_{m=mesons} (P_{m}^o+E_{{\bf p}_{N}}+\omega_{\pi}({\ {\bf p^{\prime}}%
_{\pi} }) ) {\frac{ {<in;{\bf p^{\prime}}_N |j_{a}(0)|{\bf p}_{N},m;in>_c
(2\pi)^3\delta^{(3)}({\bf p}_b-{\bf P}_{m}-{\bf p^{\prime}}_{\pi} )}}{{%
\omega_b({\bf p_b})-P_{m}^o- \omega_{\pi}({\ {\bf p^{\prime}}_{\pi} }) } }}
\]
$$
{\frac{{<in;{\bf p^{\prime}}_{\pi}, m|j_{b}(0)|0>}}{{\omega_a({\bf p^{\prime}%
}_a)+ E_{{\bf p^{\prime}}_N}-P_{m}^o-E_{{\bf p}_{N}} -i\epsilon}}}%
\eqno(B.11g)
$$

\[
+\sum_{{\overline N}} (E_{{\bf p^{\prime}}_{N}}+E_{{\bf p}_{N}}-E_{{\bf p}%
_{\overline N}} +\omega_{\pi}({\ {\bf p^{\prime}}_{\pi} } ) ) {\frac{ {< 0
|j_{a}(0)|{\bf p}_{N}{\bf p}_{\overline N};in> (2\pi)^3\delta^{(3)}( {\bf p}%
_b-{\bf p^{\prime}}_{N^{\prime}}-{\bf p}_{\overline N} -{\bf p}_{\pi})}}{{%
\omega_b({\bf p_b})- E_{{\bf p^{\prime}}_{N}}-E_{{\bf p}_{\overline N}}
-\omega_{\pi}({\ {\bf p^{\prime}}_{\pi} }) } }}
\]
$$
{\frac{{<in;{\bf p^{\prime}}_{\pi}, {\bf p^{\prime}}_{N},{\bf p}_{\overline
N}|j_{b}(0)|0>}}{{\omega_a({\bf p^{\prime}_a})- E_{{\bf p}_{N}}-E_{{\bf p}%
_{\overline N}} } }} \eqno(B.11h)
$$

\[
\ \ \ \ \ \ \ +\ \ \ a\ and\ b\ crossing\ 8\ terms.\ \ \ 
\]


{\bf The pure three-body amplitude $<in;p^{\prime}_{\pi}p^{%
\prime}_{N}|j_{a}(0) |p_b p_{\pi}p_{N};in>_c$ that describes the $a+{%
\pi^{\prime}}+N^{\prime}\Longleftarrow b+\pi+N$} reaction with $%
a=\pi^{\prime}\ or\ \gamma^{\prime}$ and $b=\pi\ or\ \gamma$ relates to the
three-body potential $w_{a+{\pi^{\prime}}+N^{\prime},b+\pi+N}$ (2.13), that
contains the following 48 potential terms after the cluster decomposition of
the expression (B.1b)

\[
w^c_{a+{\pi^{\prime}}+N^{\prime},b+\pi+N}= -<in;{\bf p^{\prime}}_{\pi}, {\bf %
p^{\prime}}_N|\Bigl[j_{a}(0),a_{b}^+(0)\Bigr]|{\bf p}_{\pi}{\bf p}_N;in>
\]
$$
+\biggl\{\sum_{n=N^{\prime\prime}} <in;{\bf p^{\prime}}_{\pi},{\bf p^{\prime}%
}_N |j_{a}(0)|{\bf p}_{N^{\prime\prime}};in> {\frac{ {(2\pi)^3\delta^{(3)}( 
{\bf p}_b+{\bf p}_{\pi}+{\bf p}_{N}-{\bf p}_{N^{\prime\prime}} )}}{{\omega_b(%
{\bf p_b})+ \omega_{\pi}({\bf p}_{\pi})+ E_{{\bf p}_N}-E_{{\bf p}%
_{N^{\prime\prime}} } } }} <in;{\bf p}_{N^{\prime\prime}}|j_{b}(0)|{\bf p}%
_{N};in>\eqno(B.12a)
$$

\[
+\sum_{m=mesons} < {\bf p^{\prime}}_{\pi}|j_{a}(0)|m;in>_c{\frac{ {%
(2\pi)^3\delta^{(3)}({\bf p}_{\pi}+ {\bf p}_b+{\bf p}_{\pi}+ {\bf p}_{N}-%
{\bf P}_{m}-{\bf p^{\prime}}_{N} )}}{{\omega_b({\bf p}_b)+ \omega_{\pi}({\bf %
p}_{\pi}) + E_{{\bf p}_N}-P_{m}^o-E_{{\bf p^{\prime}}_{N}} +i\epsilon} }}
\]
$$
<in;{\bf p^{\prime}}_{N},m|j_{b}(0)|{\bf p}_{\pi}{\bf p}_{N};in>_c%
\eqno(B.12b)
$$

$$
+\sum_{m=mesons} <in;{\bf p^{\prime}}_{\pi},{\bf p^{\prime}}_N |j_{a}(0)|%
{\bf p}_{N},m;in>_c {\frac{(2\pi)^3 {\delta^{(3)}({\bf p}_{\pi}+ {\bf p}_b-%
{\bf P}_{m} )}}{{\omega_b({\bf p_b}) +\omega_{\pi}({\bf p}_{\pi}) -P_{m}^o } 
}} <in;m|j_{b}(0)|{\bf p}_{\pi}>_c\eqno(B.12c)
$$

$$
-\sum_{{\overline N}} <{\bf p^{\prime}}_{\pi} |j_{a}(0)|{\bf p}_{N}{\bf p}%
_{\overline N};in>_c {\frac{ {(2\pi)^3\delta^{(3)}({\bf p}_{\pi}+ {\bf p}_b-%
{\bf p^{\prime}}_{N^{\prime}}-{\bf p}_{\overline N} )}}{{\omega_b({\bf p_b})
+\omega_{\pi}({\bf p}_{\pi}) - E_{{\bf p^{\prime}}_{N}}-E_{{\bf p}%
_{\overline N} } } }} <in;{\bf p^{\prime}}_{N},{\bf p}_{\overline
N}|j_{b}(0)|{\bf p}_{\pi}>\eqno(B.12d)
$$

\[
+\sum_{n=N^{\prime\prime},\pi"N"} <in;{\bf p^{\prime}}_N |j_{a}(0)|n;in>_c {%
\frac{ {(2\pi)^3\delta^{(3)}( {\bf p}_b+{\bf p}_{\pi}+{\bf p}_{N}-{\bf P}%
_{n}-{\bf p^{\prime}}_{\pi} )}}{{\omega_b({\bf p_b}) +\omega_{\pi}({\bf p}%
_{\pi}) + E_{{\bf p}_N}-P^o_n- \omega_{\pi}({\bf p^{\prime}}_{\pi})
+i\epsilon } }}
\]
$$
<in;{\bf p^{\prime}}_{\pi},n|j_{b}(0)|{\bf p}_{\pi}{\bf p}_{N};in>_c %
\eqno(B.12e)
$$

\[
+\sum_{m=mesons} <in;{\bf p^{\prime}}_N |j_{a}(0)|{\bf p}_{N},m;in>_c {\frac{
{(2\pi)^3\delta^{(3)}( {\bf p}_b+{\bf p}_{\pi} -{\bf P}_{m}-{\bf p^{\prime}}%
_{\pi} )}}{{\omega_b({\bf p_b}) +\omega_{\pi}({\bf p}_{\pi})
-P_{m}^o-\omega_{\pi}({\bf p^{\prime}}_{\pi})+i\epsilon } }}
\]
$$
<in;{\bf p^{\prime}}_{\pi}, m|j_{b}(0)|{\bf p}_{\pi}>_c\eqno(B.12f)
$$

\[
+\sum_{m=mesons} < 0|j_{a}(0)|m;in>{\frac{ {(2\pi)^3\delta^{(3)}({\bf p}%
_{\pi}+ {\bf p}_b+{\bf p}_{N}-{\bf P}_{m}-{\bf p^{\prime}}_{N} -{\bf %
p^{\prime}}_{\pi} )}}{{\omega_b({\bf p}_b) +\omega_{\pi}({\bf p}_{\pi}) + E_{%
{\bf p}_N}-P_{m}^o-E_{{\bf p^{\prime}}_{N}} -\omega_{\pi}({\bf p^{\prime}}%
_{\pi}) +i\epsilon} }}
\]
$$
<in;{\bf p^{\prime}}_{\pi},{\bf p}_{N},m|j_{b}(0)|{\bf p}_{\pi}{\bf p}%
_{N};in>_c \eqno(B.12g)
$$

$$
-\sum_{{\overline N}} < 0 |j_{a}(0)|{\bf p}_{N}{\bf p}_{\overline N};in> {%
\frac{ {(2\pi)^3\delta^{(3)}({\bf p}_{\pi}+ {\bf p}_b-{\bf p^{\prime}}%
_{N^{\prime}}-{\bf p}_{\overline N}- {\bf p^{\prime}}_{\pi})}}{{\omega_b(%
{\bf p_b}) +\omega_{\pi}({\bf p}_{\pi}) - E_{{\bf p^{\prime}}_{N}}-E_{{\bf p}%
_{\overline N}}-\omega_{\pi}({\bf p^{\prime}}_{\pi})} }} <in;{\bf p^{\prime}}%
_{\pi}, {\bf p^{\prime}}_{N},{\bf p}_{\overline N}|j_{b}(0)|{\bf p}_{\pi}>%
\eqno(B.12h)
$$

\[
\ \ \ + \ 8\ terms\ with\ \pi\ transposition\ and\ +8\ terms\ with\ the\
both\ pion\ transposition.\ \ \ \biggr\}
\]

\[
\ \ \ \ \ \ \ + \ \ a\ and\ b\ crossing\ 24\ terms.\ \ \ 
\]

Expressions (B.12a)-(B.12h) are depicted in Fig.4A-Fig.4H correspondingly.
Using the three-body generalization of the identities (B.5a)-(B.5h) we obtain

\[
B_{a+{\pi^{\prime}}+N^{\prime},b+\pi+N}\equiv B_{a+{\pi^{\prime}}%
+N^{\prime},b+\pi+N}({\bf p^{\prime}}_a {\bf p^{\prime}}_{\pi} {\bf %
p^{\prime}}_{N};{\bf p}_b{\bf p}_{\pi}{\bf p}_N)=
\]
$$
\sum_{n=N^{\prime\prime}} {\frac{ {<in;{\bf p^{\prime}}_{\pi},{\bf p^{\prime}%
}_N |j_{a}(0)|{\bf p}_{N^{\prime\prime}};in>_c (2\pi)^3\delta^{(3)}({\bf p}%
_b+ {\bf p}_{\pi}+ {\bf p}_{N}-{\bf p}_{N^{\prime\prime}} )}}{{\omega_b({\bf %
p_b})+ \omega_{\pi}({\bf p}_{\pi})+ E_{{\bf p}_N}-E_{{\bf p}%
_{N^{\prime\prime}} } } }} {\frac{{<in;{\bf p}_{N^{\prime\prime}}|j_{b}(0)|%
{\bf p}_{\pi}{\bf p}_{N};in>_c} }{{\omega_a({\bf p^{\prime}_a})+ E_{{\bf %
p^{\prime}}_N}+ \omega_{\pi}({\bf p^{\prime}}_{\pi})-E_{{\bf p}%
_{N^{\prime\prime}} } } }} \eqno(B.13a)
$$

$$
+\sum_{m=mesons} {\frac{ {< {\bf p^{\prime}}_{\pi}|j_{a}(0)|m;in>_c
(2\pi)^3\delta^{(3)}({\bf p}_b+ {\bf p}_{\pi}+ {\bf p}_{N}-{\bf P}_{m}-{\bf %
p^{\prime}}_{N} )}}{{\omega_b({\bf p}_b)+ \omega_{\pi}({\bf p^{\prime}}%
_{\pi})+ E_{{\bf p}_N}-P_{m}^o-E_{{\bf p^{\prime}}_{N}} +i\epsilon} }} {%
\frac{{<in;{\bf p^{\prime}}_{N},m|j_{b}(0)|{\bf p}_{\pi}{\bf p}_{N};in>_c} }{%
{\omega_a({\bf p^{\prime}}_a)+\omega_{\pi}({\bf p^{\prime}}_{\pi}) -P_{m}^o }
}}\eqno(B.13b)
$$

$$
+\sum_{m=mesons} {\frac{{<in;{\bf p^{\prime}}_{\pi},{\bf p^{\prime}}_N
|j_{a}(0)|{\bf p}_{N},m;in>_c}}{{\omega_b({\bf p_b}) +\omega_{\pi}({\bf p}%
_{\pi}) -P_{m}^o } }} {\frac{{(2\pi)^3\delta^{(3)}({\bf p}_b-{\bf P}_{m} )
<in;m|j_{b}(0)|{\bf p}_{\pi}>_c} }{{\omega_a({\bf p^{\prime}}%
_a)+\omega_{\pi}({\bf p^{\prime}}_{\pi}) + E_{{\bf p^{\prime}}_N}-P_{m}^o-E_{%
{\bf p }_{N}} - i\epsilon} }} \eqno(B.13c)
$$

$$
-\sum_{{\overline N}} {\frac{ {<{\bf p^{\prime}}_{\pi} |j_{a}(0)|{\bf p}_{N}%
{\bf p}_{\overline N};in> (2\pi)^3\delta^{(3)}( {\bf p}_b+ {\bf p}_{\pi}-%
{\bf p^{\prime}}_{N^{\prime}}-{\bf p}_{\overline N} )}}{{\omega_b({\bf p_b}%
)+ \omega_{\pi}({\bf p}_{\pi}) - E_{{\bf p^{\prime}}_{N}}-E_{{\bf p}%
_{\overline N} } } }} {\frac{{<in;{\bf p^{\prime}}_{N},{\bf p}_{\overline
N}|j_{b}(0)|{\bf p}_{\pi}>_c} }{{\omega_a({\bf p^{\prime}}_a)+\omega_{\pi}(%
{\bf p^{\prime}}_{\pi })- E_{{\bf p}_{N}}-E_{{\bf p^{\prime\prime}}%
_{\overline N}} } }} \eqno(B.13d)
$$

\[
+\sum_{n=N^{\prime\prime},\pi^{\prime\prime}N^{\prime\prime},2\pi"N"} {\frac{%
{<in;{\bf p^{\prime}}_N |j_{a}(0)|n;in>_c (2\pi)^3\delta^{(3)}( {\bf p}_b+%
{\bf p}_{\pi}+ {\bf p}_{N}-{\bf P}_{n}-{\bf p^{\prime}}_{\pi} )}}{{\omega_b(%
{\bf p_b})+ \omega_{\pi}({\bf p}_{\pi})+ E_{{\bf p}_N}-P^o_{n}-
\omega_{\pi}( {\bf p^{\prime}}_{\pi}) +i\epsilon } }}
\]
$$
{\frac{{<in;{\bf p^{\prime}}_{\pi a},n|j_{b}(0)|{\bf p}_{\pi}{\bf p}%
_{N};in>_c}}{{\omega_a({\bf p^{\prime}_a})+ \omega_{\pi}({\bf p^{\prime}}%
_{\pi})+ E_{{\bf p^{\prime}}_N}-P^o_{n}-\omega_{\pi}({\bf p}_{\pi})
-i\epsilon } }} \eqno(B.13e)
$$

\[
+\sum_{m=mesons} {\frac{ {\ < 0|j_{a}(0)|m;in>_c (2\pi)^3\delta^{(3)}({\bf p}%
_b+{\bf p}_{\pi}+ {\bf p}_{N}-{\bf P}_{m}-{\bf p^{\prime}}_{N} -{\bf p}%
_{\pi} )}}{{\omega_b({\bf p}_b)+\omega_{\pi}({\bf p}_{\pi})
-P_{m}^o-\omega_{\pi}({\ {\bf p^{\prime}}_{\pi} }) +i\epsilon} }}
\]
$$
{\frac{{<in;{\bf p^{\prime}}_{\pi},{\bf p}_{N},m|j_{b}(0)|{\bf p}_{\pi}{\bf p%
}_{N};in>_c} }{{\omega_a({\bf p^{\prime}_a})+\omega_{\pi}({\bf p^{\prime}}%
_{\pi}) + E_{{\bf p^{\prime}}_N}-E_{{\bf p}_{N}} -P_{m}^o-\omega_{\pi}({\bf p%
}_{\pi}) -i\epsilon} }}\eqno(B.13f)
$$

\[
+\sum_{m=mesons} {\frac{ {<in;{\bf p^{\prime}}_N |j_{a}(0)|{\bf p}%
_{N},m;in>_c (2\pi)^3\delta^{(3)}( {\bf p}_b+{\bf p}_{\pi}+{\bf p}_{N} -{\bf %
P}_{m}-{\bf p^{\prime}}_{\pi}-{\bf p^{\prime}}_{N} )}}{{\omega_b({\bf p_b}%
)+\omega_{\pi}({\bf p}_{\pi})+E_{{\bf p}_{N}} -P_{m}^o-E_{{\bf p^{\prime}}%
_N}-\omega_{\pi}( {\bf p^{\prime}}_{\pi} ) } }}
\]
$$
{\frac{{<in;{\bf p^{\prime}}_{\pi}, m|j_{b}(0)|{\bf p}_{\pi}>_c}}{{\omega_a(%
{\bf p^{\prime}}_a)+\omega_{\pi}( {\bf p}_{\pi} ) -P_{m}^o -\omega_{\pi}(%
{\bf p^{\prime}}_{\pi})-i\epsilon} }} \eqno(B.13g) 
$$

\[
-\sum_{{\overline N}} {\frac{ {< 0 |j_{a}(0)|{\bf p}_{N}{\bf p}_{\overline
N};in> (2\pi)^3\delta^{(3)}( {\bf p}_b+{\bf p}_{\pi} -{\bf p^{\prime}}%
_{N^{\prime}}-{\bf p}_{\overline N} -{\bf p^{\prime}}_{\pi})}}{{\omega_b(%
{\bf p_b})+\omega_{\pi}({\bf p}_{\pi}) - E_{{\bf p^{\prime}}_{N}}-E_{{\bf p}%
_{\overline N}} -\omega_{\pi}({\ {\bf p^{\prime}}_{\pi} }) } }}
\]
$$
{\frac{{<in;{\bf p^{\prime}}_{\pi} {\bf p^{\prime}}_{N},{\bf p}_{\overline
N}|j_{b}(0)|{\bf p}_{\pi}>_c}}{{\omega_a({\bf p^{\prime}_a})+\omega_{\pi}(%
{\bf p^{\prime}}_{\pi}) - E_{{\bf p}_{N}}-E_{{\bf p}_{\overline N}}
-\omega_{\pi}({\ {\bf p^{\prime}}_{\pi} }) } }} \eqno(B.13h)
$$

\[
\ \ \ + \ 8\ terms\ with\ \pi\ transposition\ and\ +8\ terms\ with\ the\
both\ pion\ transposition.\ \ \ \biggr\}
\]

\[
\ \ \ \ \ \ \ +\ \ a\ and\ b\ crossing\ 24\ terms.\ \ \ 
\]

Here expressions (B.13a),(B.13d),(B.13e) and (B.13h) are Hermitian.. And the
expressions (B.13b), (B.13f) are Hermitian conjugate of (B.13c),(B.13g)
correspondingly. Therefore the complete potential $B_{a+{\pi }^{\prime
}+N^{\prime },b+\pi +N}$ is Hermitian..

\[
A_{a+{\pi}^{\prime}+N^{\prime},b+\pi+N}\equiv A_{a+{\pi}^{\prime}+N^{%
\prime},b+\pi+N}({\bf p^{\prime}}_N {\bf p^{\prime}}_a {\bf p^{\prime}}%
_{\pi}; {\bf p}_{b} {\bf p}_{\pi}{\bf p}_N)= -<in;{\bf p^{\prime}}_{\pi},%
{\bf p^{\prime}}_N|\Bigl[j_{a}(0),a_{b}^+(0)\Bigr] |{\bf p}_N;in>
\]

\[
\biggl\{-\sum_{n=N^{\prime\prime}}E_{{\bf p}_{N^{\prime\prime}}} {\frac{ {%
<in;{\bf p^{\prime}}_{\pi},{\bf p^{\prime}}_N |j_{a}(0)|{\bf p}%
_{N^{\prime\prime}};in>_c (2\pi)^3\delta^{(3)}({\bf p}_b +{\bf p}_{\pi} +%
{\bf p}_{N}-{\bf p}_{N^{\prime\prime}} )}}{{\omega_b({\bf p_b})
+\omega_{\pi}({\bf p}_{\pi}) + E_{{\bf p}_N}-E_{{\bf p}_{N^{\prime\prime}} } 
} }}
\]
$$
{\frac{{<in;{\bf p}_{N^{\prime\prime}}|j_{b}(0)|{\bf p}_{\pi}{\bf p}%
_{N};in>_c} }{{\omega_a({\bf p^{\prime}_a})+ E_{{\bf p^{\prime}}_N}+
\omega_{\pi}({\bf p^{\prime}}_{\pi})-E_{{\bf p}_{N^{\prime\prime}} } } }} %
\eqno(B.14a)
$$

\[
-\sum_{m=mesons}\Bigl[P_{m}^o+E_{{\bf p^{\prime}}_{N}}\Bigr] {\frac{ {< {\bf %
p^{\prime}}_{\pi}|j_{a}(0)|m;in>_c (2\pi)^3\delta^{(3)}({\bf p}_b+ {\bf p}%
_{\pi}+ {\bf p}_{N}-{\bf P}_{m}-{\bf p^{\prime}}_{N} )}}{{\omega_b({\bf p}%
_b)+ \omega_{\pi}({\bf p^{\prime}}_{\pi})+ E_{{\bf p}_N}-P_{m}^o-E_{{\bf %
p^{\prime}}_{N}} +i\epsilon} }}
\]
$$
{\frac{{<in;{\bf p^{\prime}}_{N},m|j_{b}(0)|{\bf p}_{\pi}{\bf p}_{N};in>_c} 
}{{\omega_a({\bf p^{\prime}}_a)+\omega_{\pi}({\bf p^{\prime}}_{\pi})
-P_{m}^o } }}\eqno(B.14b)
$$

\[
-\sum_{m=mesons}\Bigl[P_{m}^o+E_{{\bf p}_{N}}\Bigr] {\frac{{<in;{\bf %
p^{\prime}}_{\pi},{\bf p^{\prime}}_N |j_{a}(0)|{\bf p}_{N},m;in>_c}}{{%
\omega_b({\bf p_b}) +\omega_{\pi}({\bf p}_{\pi}) -P_{m}^o } }}
\]
$$
{\frac{{(2\pi)^3\delta^{(3)}({\bf p}_b-{\bf P}_{m} ) <in;m|j_{b}(0)|{\bf p}%
_{\pi}>_c} }{{\omega_a({\bf p^{\prime}}_a)+\omega_{\pi}({\bf p^{\prime}}%
_{\pi}) + E_{{\bf p^{\prime}}_N}-P_{m}^o-E_{{\bf p }_{N}} - i\epsilon} }} %
\eqno(B.14c)
$$

\[
+\sum_{{\overline N}} \Bigl[ E_{{\bf p^{\prime}}_{N}}+E_{{\bf p}_{N}}+E_{%
{\bf p^{\prime\prime}}_{\overline N}} \Bigr] {\frac{ {<{\bf p^{\prime}}%
_{\pi} |j_{a}(0)|{\bf p}_{N}{\bf p}_{\overline N};in> (2\pi)^3\delta^{(3)}( 
{\bf p}_b+ {\bf p}_{\pi}-{\bf p^{\prime}}_{N^{\prime}}-{\bf p}_{\overline N}
)}}{{\omega_b({\bf p_b})+ \omega_{\pi}({\bf p}_{\pi}) - E_{{\bf p^{\prime}}%
_{N}}-E_{{\bf p}_{\overline N} } } }}
\]
$$
{\frac{{<in;{\bf p^{\prime}}_{N},{\bf p}_{\overline N}|j_{b}(0)|{\bf p}%
_{\pi}>_c} }{{\omega_a({\bf p^{\prime}}_a)+\omega_{\pi}({\bf p^{\prime}}%
_{\pi })- E_{{\bf p}_{N}}-E_{{\bf p^{\prime\prime}}_{\overline N}} } }} %
\eqno(B.14d)
$$

\[
-\sum_{n=N^{\prime\prime},\pi^{\prime\prime}N^{\prime\prime},2\pi"N"} \Bigl[
P^o_{n}+\omega_{\pi}({\bf p}_{\pi}) \Bigr] {\frac{{<in;{\bf p^{\prime}}_N
|j_{a}(0)|n;in>_c (2\pi)^3\delta^{(3)}( {\bf p}_b+{\bf p}_{\pi}+ {\bf p}_{N}-%
{\bf P}_{n}-{\bf p^{\prime}}_{\pi} )}}{{\omega_b({\bf p_b})+ \omega_{\pi}(%
{\bf p}_{\pi})+E_{{\bf p}_N}-P^o_{n}- \omega_{\pi}( {\bf p^{\prime}}_{\pi})
+i\epsilon } }}
\]
$$
{\frac{{<in;{\bf p^{\prime}}_{\pi a},n|j_{b}(0)|{\bf p}_{\pi}{\bf p}%
_{N};in>_c}}{{\omega_a({\bf p^{\prime}_a})+ \omega_{\pi}({\bf p^{\prime}}%
_{\pi}) E_{{\bf p^{\prime}}_N}-P^o_{n}-\omega_{\pi}({\bf p}_{\pi})
-i\epsilon } }} \eqno(B.14e)
$$

\[
-\sum_{m=mesons} \Bigl[E_{{\bf p}_{N}}+ P^o_{m}+\omega_{\pi}({\bf p}_{\pi})
\Bigr] {\frac{ {\ < 0|j_{a}(0)|m;in>_c (2\pi)^3\delta^{(3)}({\bf p}_b+{\bf p}%
_{\pi}+ {\bf p}_{N}-{\bf P}_{m}-{\bf p^{\prime}}_{N} -{\bf p}_{\pi} )}}{{%
\omega_b({\bf p}_b)+\omega_{\pi}({\bf p}_{\pi}) -P_{m}^o-\omega_{\pi}({\ 
{\bf p^{\prime}}_{\pi} }) +i\epsilon} }}
\]
$$
{\frac{{<in;{\bf p^{\prime}}_{\pi},{\bf p}_{N},m|j_{b}(0)|{\bf p}_{\pi}{\bf p%
}_{N};in>_c} }{{\omega_a({\bf p^{\prime}_a})+\omega_{\pi}({\bf p^{\prime}}%
_{\pi}) + E_{{\bf p^{\prime}}_N}-E_{{\bf p}_{N}} -P_{m}^o-\omega_{\pi}({\bf p%
}_{\pi}) -i\epsilon} }}\eqno(B.14f)
$$

\[
-\sum_{m=mesons} \Bigl[E_{{\bf p^{\prime}}_{N}}+ P^o_{m}+\omega_{\pi}({\bf p}%
_{\pi}) \Bigr] {\frac{ {<in;{\bf p^{\prime}}_N |j_{a}(0)|{\bf p}_{N},m;in>_c
(2\pi)^3\delta^{(3)}( {\bf p}_b+{\bf p}_{\pi}+{\bf p}_{N} -{\bf P}_{m}-{\bf %
p^{\prime}}_{\pi}-{\bf p^{\prime}}_{N} )}}{{\omega_b({\bf p_b})+\omega_{\pi}(%
{\bf p}_{\pi})+E_{{\bf p}_{N}} -P_{m}^o-E_{{\bf p^{\prime}}_N}-\omega_{\pi}( 
{\bf p^{\prime}}_{\pi} ) } }}
\]
$$
{\frac{{<in;{\bf p^{\prime}}_{\pi}, m|j_{b}(0)|{\bf p}_{\pi}>_c}}{{\omega_a(%
{\bf p^{\prime}}_a)+\omega_{\pi}( {\bf p}_{\pi} ) -P_{m}^o -\omega_{\pi}(%
{\bf p^{\prime}}_{\pi})-i\epsilon} }} \eqno(B.13g)
$$

\[
+\sum_{{\overline N}} \Bigl[E_{{\bf p^{\prime}}_{N}}+ E_{{\bf p}_{N}}+E_{%
{\bf p}_{\overline N}}+ \omega_{\pi}({\bf p}_{\pi}) \Bigr] {\frac{ {< 0
|j_{a}(0)|{\bf p}_{N}{\bf p}_{\overline N};in> (2\pi)^3\delta^{(3)}( {\bf p}%
_b+{\bf p}_{\pi} -{\bf p^{\prime}}_{N^{\prime}}-{\bf p}_{\overline N} -{\bf %
p^{\prime}}_{\pi})}}{{\omega_b({\bf p_b})+\omega_{\pi}({\bf p}_{\pi}) - E_{%
{\bf p^{\prime}}_{N}}-E_{{\bf p}_{\overline N}} -\omega_{\pi}({\ {\bf %
p^{\prime}}_{\pi} }) } }}
\]
$$
{\frac{{<in;{\bf p^{\prime}}_{\pi} {\bf p^{\prime}}_{N},{\bf p}_{\overline
N}|j_{b}(0)|{\bf p}_{\pi}>_c}}{{\omega_a({\bf p^{\prime}_a})+\omega_{\pi}(%
{\bf p^{\prime}}_{\pi}) - E_{{\bf p}_{N}}-E_{{\bf p}_{\overline N}}
-\omega_{\pi}({\ {\bf p^{\prime}}_{\pi} }) } }} \eqno(B.13h)
$$

\[
\ \ \ + \ 8\ terms\ with\ \pi\ transposition\ and\ +8\ terms\ with\ the\
both\ pion\ transposition.\ \ \ \biggr\}
\]

\[
\ \ \ \ \ \ \ +\ \ \ a\ and\ b\ crossing\ 24\ terms,\ \ \ 
\]

where $A_{a+{\pi}^{\prime}+N^{\prime},b+\pi+N}$ has the same structure as $%
B_{a+{\pi}^{\prime}+N^{\prime},b+\pi+N}$. But there one must taken into
account an additional combinations  expressions
(B.14e),(B.14f),(B.14g),(B.14h) wit other 4 terms, where the initial pion $%
\pi$ is transposed instead of the final pion $\pi^{\prime}$.



\end{document}